\documentclass[a4paper]{article}
\usepackage[T1]{fontenc}
\usepackage[utf8]{inputenc}
\usepackage{color}
\usepackage{float}
\usepackage{amsmath}
\usepackage{amssymb}
\usepackage{graphicx}
\usepackage[unicode=true,
 bookmarks=false,
 breaklinks=false,pdfborder={0 0 0},backref=section,colorlinks=false]
 {hyperref}

\makeatletter

\pdfpageheight\paperheight
\pdfpagewidth\paperwidth

\newcommand{\lyxaddress}[1]{
\par {\raggedright #1
\vspace{1.4em}
\noindent\par}
}

\usepackage{multirow}
\usepackage{a4wide}
\usepackage{amsfonts}
\usepackage{mathrsfs}
\usepackage{wasysym}
\rmfamily
\usepackage{listings}
\usepackage{tikz}

\usepackage{setspace}

\usepackage{mathrsfs}
\usepackage{color}
 \definecolor{hellgelb}{rgb}{1,1,0.85}
 \definecolor{colKeys}{rgb}{0,0,1}
 \definecolor{colIdentifier}{rgb}{0,0,0}
 \definecolor{colComments}{rgb}{1,0,0}
 \definecolor{colString}{rgb}{0,0.5,0}
\makeatother

\usepackage{listings}

\begin{document}

\title{Causality studied in reconstructed state space. \\
Examples of uni-directionally connected chaotic systems}

\author{Anna Krakovská, Jozef Jakubík, Hana Budáčová, Mária Holecyová}

\maketitle

\lyxaddress{Institute of Measurement Science, Slovak Academy of Sciences, }

\lyxaddress{Dúbravská Cesta 9, 842 19 Bratislava, Slovakia}

\lyxaddress{E-mail address: krakovska@savba.sk}

\tableofcontents{}\newpage{}
\begin{abstract}
Three state-space based methods were tested in relation to the ability
to detect unidirectional coupling and synchronization of interconnected
dynamical systems. The first method, based on measure named M, was
introduced by Andrzejak et al. in 2003 \cite{andrz2003}. The second
one, based on measure L, was described in 2009 by Chicharro et al.
\cite{Chich2009}. The third method, called convergent cross-mapping,
came from Sugihara et al., 2012 \cite{sugih2012}. 

The methods were compared on 9 test examples of uni-directionally
connected chaotic systems of Hénon, Rössler and Lorenz type. The tested
systems were selected from previously published causality studies.
Matlab code for the three methods is provided.

The results show that each of the three examined state-space methods
managed to reveal the presence and the direction of couplings and
also to detect the onset of full synchronization. 

\medskip{}

\end{abstract}

\section{Introduction}

Nowadays, the study of drive-response relationships between dynamical
systems is a topic of increasing interest. Applications are found,
among others, in domains as economics, climatology, electrical activity
of brain, or cardio-respiratory relations. 

In this paper examples of the next type of uni-directionally coupled
bivariate dynamical systems are studied:

\smallskip{}

$\dot{x}(t)=F(x(t))$ 

$\dot{y}(t)=G(y(t),x(t))$ 

\smallskip{}

where $x$ and $y$ are state vectors of driving system $X$ and the
driven response $Y$. If the following relation $y(t)=\text{\ensuremath{\Psi}}(x(t))$
applies for some smooth and invertible function $\text{\ensuremath{\Psi}}$
then there is said to be a generalized synchronization between $X$
and $Y$. If $\Psi$ is an identity, the synchronization is called
identical. After some definitions, $\Psi$ does not need to be smooth.
E.g., Pyragas defines as strong and weak synchronizations the cases
of smooth and non-smooth transformations, respectively \cite{pyrag1996}. 

The direction of coupling can only be uncovered when the coupling
is weaker than the threshold for an emergence of synchronization.
Once the systems are synchronized, the future states of the driver
$X$ can be predicted from the response $Y$ equally well as vice
versa. 

A first mathematical approach to detect causal relationships has been
proposed in 1969 by Clive Granger \cite{grang1969}. The method was
based on the premiss that causality could be reflected by measuring
the ability of predicting the future values of a time series using
past values of the driving time series. More specifically, a time
series $X$ is said to cause $Y$ if it can be shown, usually through
statistical hypothesis tests, that past $X$ values provide statistically
significant information about future values of $Y$.

To be able to consider Granger causality (GC), separability is required.
Namely, information about the causative factor $X$ is expected to
be available as an explicit variable. This requirement can be problematic
in a case of variables which are dynamically linked and sharing the
same manifold in the state space. 

Moreover, the initially linear concept requires generalizations to
enable investigation of complex nonlinear processes. Therefore, new
approaches were proposed, including nonlinear Granger causality, transfer
entropy, cross predictability, conditional mutual information, measures
evaluating distances of conditioned neighbors in reconstructed state
spaces, etc. 

In this study, we are going to focus on testing uni-directionally
coupled chaotic systems by state-space methods. 

The possibility to study synchronization in chaotic systems was discussed
already in 1990 \cite{pecor1990}. The problem with chaos is the long-term
unpredictable behavior. Two identical chaotic systems starting at
very close initial conditions soon diverge from each other, while
remaining on the same attractor. However, although it may sound surprising,
it is possible to ``lock'' one chaotic system to the other to get
them to synchronize. Our test data will include several examples of
such synchronized chaos.

In the case of the state-space methods, efforts to reveal causal links
are based on the following idea. When trajectories of driving and
response systems are strongly connected, then two close states in
the state space of the response system correspond to two close states
in the space of the driving system. Already in 1995, this approach
was used by Rulkov et al. to explore systems where an observable of
response system $y(t)$ is driven with the output of an autonomous
driving system $x(t)$, but there is no feedback to the driver \cite{rulko1995}.
To investigate the existence of the synchronization the authors introduced
the idea of mutual false nearest neighbors to determine when closeness
in response space implies closeness in driving space. Similar idea
was applied in a variety of modifications many times since then. Some
of them will be presented below.

The paper is organized as follows. 

In Section 2, the methods of causality detection are presented. 

Section 3 describes the nine examples of uni-directionally coupled
chaotic systems. At the same place the results of the causality detection
are given for each case.

In Section 4, the findings are summarized. 

In Appendix, Matlab code used for the detection of causality is provided.
The program includes all three testing methods used in this study,
namely measure $M$, measure $L$, and the cross-mapping.

\section{Data and methods }

\subsection{Data}

Our data set consists of 9 examples of chaotic systems of Hénon, Rössler
and Lorenz type that are coupled with variable coupling strengths.
Details on systems are provided in section \ref{sec:Causality-detection}.

Next, we briefly introduce the existing causality methods:

\subsection{Granger test}

When looking for causality, as first the test of Granger causality
is usually applied \cite{grang1969}. For this purpose, freely available
Matlab function written by Chandler Lutz \cite{lutz2009} can be used,
for example. However, in cases of non-separable non-linear dynamic
systems, the Granger test fails to reliably detect and characterize
the causal relations.

\subsection{Transfer entropy and conditional mutual information }

Although we will concentrate on state-space based approaches, for
completeness let us mention the methods that originate in information
theory. However, they will not be tested in the present study.

\subsubsection*{Transfer entropy (TE)}

Transfer entropy was introduced by Schreiber \cite{schre2000b} in
2000 as an information theoretic measure which shares some of the
desired properties of mutual information but takes the dynamics of
information transport into account. The definition is as follows

\begin{align*}
TE_{Y\to X} & =\sum_{x_{t},x_{t-\tau}^{(k)},y_{t-\tau}^{(l)}}p(x_{t},x_{t-\tau}^{(k)},y_{t-\tau}^{(l)})\log\frac{p(x_{t}|x_{t-\tau}^{(k)},y_{t-\tau}^{(l)})}{p(x_{t}|x_{t-\tau}^{(k)})}\\
 & =H(x_{t},x_{t-\tau}^{(k)})-H(x_{t}|y_{t-\tau}^{(l)},x_{t-\tau}^{(k)})
\end{align*}
where $H$ is the Shannon entropy, $t$ indicates a given point in
time, $\tau$ is a time lag (usually the same time lag is used in
both $X$ and $Y$) and $k$ and $l$ are the block lengths of past
values in $X$ and $Y$. 

\begin{align*}
x_{t-\tau}^{(k)} & =\{y_{t-\tau-k+1},y_{t-\tau-k+2},...,y_{t-\tau}\}\\
y_{t-\tau}^{(l)} & =\{y_{t-\tau-l+1},y_{t-\tau-l+2},...,y_{t-\tau}\}
\end{align*}

\begin{flushleft}
Mostly, $l=k=1$, so we get 
\begin{align*}
x_{t-\tau}^{(1)} & =\{y_{t-\tau-1+1}\}\\
y_{t-\tau}^{(1)} & =\{y_{t-\tau-1+1}\}\\
TE_{Y\to X} & =\sum_{x_{t},x_{t-\tau},y_{t-\tau}}p(x_{t},x_{t-\tau},y_{t-\tau})\log\frac{p(x_{t}|x_{t-\tau},y_{t-\tau})}{p(x_{t}|x_{t-\tau})}\\
 & =H(x_{t},x_{t-\tau})-H(x_{t}|y_{t-\tau},x_{t-\tau}).
\end{align*}

\par\end{flushleft}

\subsubsection*{Conditional mutual information}

Transfer entropy has been independently formulated as a conditional
mutual information by Paluš et al. \cite{palus2007}. The joint entropy
is defined as $H(X,Y)=-\sum\sum p(x,y)\, log\, p(x,y)$ and the conditional
entropy is $H(Y|X)=-\sum\sum p(x,y)\, log\, p(y|x)$. Then the mutual
information $I(X;Y)$ is defined as $I(X;Y)=H(X)+H(Y)-H(X,Y)$. The
conditional mutual information of the variables $X,Y$, given the
variable $Z$ is defined as the reduction in the uncertainty of $X$
due to knowledge of $Y$ when $Z$ is given:

\begin{flushleft}
\[
I(X;Y|Z)=H(X|Z)+H(Y|Z)-H(X,Y|Z).
\]
For $Z$ independent of $X$ and $Y$: 
\par\end{flushleft}

\begin{center}
$I(X;Y|Z)=I(X;Y)=I(X;Y;Z)-I(X;Z)-I(Y;Z)$.
\par\end{center}

\begin{flushleft}
After the next substitution
\begin{align*}
x_{t-\tau}^{(k)} & =Z\\
x_{t} & =X\\
y_{t-\tau}^{(l)} & =Y,
\end{align*}
and following simple probability relations it turns out that the conditional
mutual information $I(X;Y|Z)$ and the transfer entropy $TE$ are
the same: 
\begin{align*}
TE_{Y\to X} & =H(X,Z)-H(X|Y,Z)=\sum_{X,Y,Z}p(X,Y,Z)\log\frac{p(X|Y,Z)}{p(X|Y)}
\end{align*}

\par\end{flushleft}

\begin{align*}
TE_{Y\to X} & =\sum_{X,Y,Z}p(X,Y,Z)\log\frac{p(X|Y,Z)}{p(X|Z)}\frac{p(Y|Z)}{p(Y|Z)}=\sum_{X,Y,Z}p(X,Y,Z)\log\frac{p(X,Y|Z)}{p(X|Z)p(Y|Z)}\\
 & =\sum_{X,Y,Z}p(X,Y,Z)\log p(X,Y|Z)-\sum_{X,Y,Z}p(X,Y,Z)\log p(X|Z)\\
 & -\sum_{X,Y,Z}p(X,Y,Z)\log p(Y|Z)\\
 & =\sum_{X,Y,Z}p(X,Y,Z)\log p(X,Y|Z)-\sum_{X,Z}p(X,Z)\log p(X|Z)-\sum_{Y,Z}p(Y,Z)\log p(Y|Z)\\
 & =-H(X,Y|Z)+H(X|Z)+H(Y|Z)=I(X;Y|Z)
\end{align*}

In \cite{barne2009} Barnett et al. have shown that for Gaussian variables,
Granger causality and transfer entropy (conditional mutual information)
are equivalent.

\subsection{Correlation dimension}

In this study, we tested two aspects of causality. First of all we
tested the ability to detect the presence and direction of the causal
link for a particular value of coupling. Secondly, we would like to
be able to detect a possible onset of synchronization following an
increase of coupling above a certain value. Therefore, we need to
know which levels of coupling lead to synchronization for our artificial
chaotic systems. Typically, to this end, the Lyapunov exponents are
evaluated, as it was shown that the synchronization takes place when
all of the conditional Lyapunov exponents of the response subsystem
become negative. 

However, in this study, similarly as in \cite{krako2013} we use a
different complexity measure, namely the correlation dimension, computed
after Grassberger-Proccacia algorithm \cite{grass1983} to reveal
the emergence of synchronization. The idea behind this approach is
as follows.

Suppose we have a driving system $X$ and response $Y$ with a unidirectional
coupling. Let us create $X+Y$ combining state vectors of $X$ and
$Y$. Then, we have the next expectations with regard to the coupling
effects on the correlation dimension: 

- for uncoupled $X$ and $Y$ the correlation dimension of the combined
system are equal to the sum of the dimensions of $X$ and $Y$,

- for coupled but not synchronized case, the correlation dimension
of $X+Y$ is higher than the dimension of the driver,

- once the coupling reaches the synchronization level, the dimension
of the attractor of the combined system saturates to the dimension
of the driving systems attractor.

\subsection{State space based causality measures}

\subsubsection*{\textcolor{black}{State space reconstruction}}

\textcolor{black}{State space reconstruction is usually the first
step in the analysis of a time series in terms of dynamical systems
theory.} Suppose that we have a single time series $y(t)$ presumably
generated by a $d$-dimensional deterministic dynamical system. Then,
the usual choice for a reconstruction is a matrix of time shifts of
one variable, as supported by Takens theorem from 1981 \cite{taken1981}.
The time-delayed versions $[y(t),y(t-\tau),y(t-2\tau),\ldots,y(t-2m\tau)]$
of the known observable $y(t)$ form an embedding from the original
$m$-dimensional manifold into $R^{2m+1}$ (where $2m+1$ is the so
called embedding dimension, and $\tau$ is the time lag between consecutive
states). The reconstructed \textcolor{black}{state space is, in the
sense of diffeomorphism, equivalent to the original state space. T}o
select the embedding parameters, namely the size of the space of the
reconstruction and the $\tau$ , many competing approaches have been
proposed. \textcolor{black}{T}he most common practice is to take the
delay as the first minimum of the mutual information between the delayed
components. Then, the minimal embedding dimension is estimated, usually
by the false near neighbor test \cite{kenne1992}. In this study,
the setting of the embedding parameters is not an issue, since we
work with huge number of clean data from exactly described low-dimensional
systems. However, in real data the search for optimal embedding parameters
can be quite challenging (see \cite{krako2015} and references therein).

\subsubsection{Measures S, H, N, M, L}

In the last two decades, there have been several attempts to infer
causal relationships for complex systems in state spaces. To clarify
the relevant approaches, suppose we have two systems $X$ and $Y$
reconstructed from two observed time series. Let the arrays of the
delay vectors are ($x_{1},$ $x_{2},$\ldots{}$,x_{N}$) and ($y_{1},$
$y_{2},$\ldots{}$,y_{N}$ ). Suppose that $X$ causally influences
$Y$. A fundamental signature of such (non-synchronizing) causal connection
is that close states of $Y$ are mapped to close states of $X$ with
a higher probability if compared to uncoupled systems. However, an
increased probability of the opposite mapping, i.e., close states
of $X$ are mapped to close states of $Y$, also holds, although with
lower probability. Therefore, we have to examine both directions and
evaluate the difference in the results.

We denote $x_{r_{n,1}},$ $x_{r_{n,2}},$ \ldots{}$x_{r_{n,k}}$
the $k$ nearest neighbors to the point $x_{n}$, where $r_{n,1},$
$r_{n,2},$ \ldots{}$r_{n,k}$ denote the indices of first, second,
\ldots{}$k-$th nearest neighbor of point $x_{n}.$ Let us at first
define the average distance of the point $x_{n}$ to its $k$ nearest
neighbors as:

\[
R_{n}^{(k)}(X)=\frac{1}{k}\sum_{j=1}^{k}(x_{n}-x_{r_{n,j}})^{2}.
\]

We denote $y_{s_{n,1}},$ $y_{s_{n,2}},$ \ldots{}$y_{s_{n,k}}$
the $k$ nearest neighbors to the point $y_{n}$ , where $s_{n,1},$
$s_{n,2},$ \ldots{}$s_{n,k}$ denote the indices of first, second,
\ldots{}$k-$th nearest neighbor of point $y_{n}.$ We define the
average distance of the point $x_{n}$ to the $k$ points in $X$,
which correspond to the $k$ nearest neighbors of $y_{n}$ as:

\[
R_{n}^{(k)}(X|Y)=\frac{1}{k}\sum_{j=1}^{k}(x_{n}-x_{s_{n,j}})^{2}.
\]

For simplicity we denote the average distance of point $x_{n}$ from
all other points in $X$ as

\[
R_{n}(X)=\frac{1}{N-1}\sum_{j=1}^{N}(x_{n}-x_{j})^{2}.
\]

Then, starting from the formula $R_{n}(X)$ based on the computations
of distances, several interdependence measures can be proposed:

\begin{eqnarray*}
S^{(k)}(X|Y) & = & \frac{1}{N}\sum_{n=1}^{N}\frac{R_{n}^{k}(X)}{R_{n}^{k}(X|Y)}\\
H^{(k)}(X|Y) & = & \frac{1}{N}\sum_{n=1}^{N}\log\frac{R_{n}(X)}{R_{n}^{k}(X|Y)}\\
N^{(k)}(X|Y) & = & \frac{1}{N}\sum_{n=1}^{N}\frac{R_{n}(X)-R_{n}^{k}(X|Y)}{R_{n}(X)}\\
M^{(k)}(X|Y) & = & \frac{1}{N}\sum_{n=1}^{N}\frac{R_{n}(X)-R_{n}^{k}(X|Y)}{R_{n}(X)-R_{n}^{k}(X)}
\end{eqnarray*}

$S$ and $H$ were introduced in \cite{arnho1999}. In $H$ geometric
averages are used because, in general, they are considered to be more
robust and easier to interpret than the arithmetic averages. The asymmetry
under the exchange $X\leftrightarrow Y$ is the main difference between
$H$ and mutual information. $H$ is more sensitive to weak dependencies
and it should be easier to estimate than the mutual information. 

Quiroga et al. proposed a new measure $N$, similar to $H$ but using
arithmetic averaging and normalized. Both $H$ and $N$ can be slightly
negative \cite{quiro2002}. However, $N$ is equal to $1$ only if
$R_{n}^{(k)}(Y|X)=0$, where $R_{n}^{(k)}(Y|X)\geq R_{n}^{(k)}(Y)$.
On the other hand, $R_{n}^{(k)}(Y)=0$ only for periodic process.
In consequence, even in the case of identical synchronization $N$
is smaller than $1$. Therefore, Andrzejak et al. proposed the measure
$M$ \cite{andrz2003}. Occasional negative values are replaced by
$0$. Then $M$ falls into interval $<0,1>$.

The measure $L$ is not based on computations of average distances,
but instead we use only ranks - for each point $x_{n}$ we sort the
other points with respect to distances and apply similar formula as
before. It is obvious, that the average rank of the $k$ nearest neighbors
is 
\[
G_{n}^{k}(X)=\frac{1}{k}\frac{k(k+1)}{2}=\frac{k+1}{2}
\]
and the average rank of all the neighbors is 
\[
G_{n}(X)=\frac{1}{N-1}\frac{N(N-1)}{2}=\frac{N}{2}\cdot
\]

To obtain the average conditional rank for $x_{n}$ we compute the
average rank of the points in $X$ that correspond to the k-nearest
neighbors of $y_{n}$ in $Y.$ We denote $g_{i,j}$ the rank of the
distance of $x_{i}$ and $x_{j}$ among the distances of $x_{i}$
from all other points in ascending order. Then 
\[
G_{n}^{(k)}(X|Y)=\frac{1}{k}\sum_{j=1}^{k}g_{n,s_{n,j}},
\]
and we define an interdependence measure similar to $M$ as follows:
\[
L^{(k)}(X|Y)=\frac{1}{N}\sum_{n=1}^{N}\frac{G_{n}(X)-G_{n}^{k}(X|Y)}{G_{n}(X)-G_{n}^{k}(X)}\cdot
\]

Since the more recent methods overcome some problems of the early
ones, we used only the last two measures, $M$ and $L$, in this study.

\subsubsection{Convergent cross-mapping}

In 2012 Sugihara et al. introduced yet another method based on state
space reconstruction \cite{sugih2012}. The method called convergent
cross-mapping (CCM) tests for causation between systems $X$ and $Y$
by measuring the extent to which the historical record of $Y$ values
can reliably estimate states of $X$. 

The algorithm for CCM is the following: 

Consider two time series and the corresponding lagged-coordinate vectors
($x_{1},$ $x_{2},$\ldots{}$,x_{L}$) and ($y_{1},$ $y_{2},$\ldots{}$,y_{L}$
) in $E$-dimensional reconstructed manifolds $M_{X}$ and $M_{Y}$
respectively. 

To generate a cross-mapped estimate of point $y(t)$, locate the contemporaneous
vector on $M_{X}$, $x(t)$, and find its $E+1$ nearest neighbors. 

Denote the time indices (from closest to farthest) of the nearest
neighbors of $x(t)$ by $t_{1},t_{2},\ldots,t_{E+1}$. These time
indices of nearest neighbors to $x(t)$ on $M_{X}$ are used to identify
points (neighbors) in $M_{Y}$ to estimate $y(t)$ from a locally
weighted mean of the $E+1$ $y(t_{i})$ values. 

The difference between values estimated and the actual values is evaluated
by the Pearson correlation coefficient. 

For more details on the algorithm see \cite{sugih2012}.

If $X$ and $Y$ are dynamically coupled, the nearest neighbors on
$M_{X}$ should identify the time indices of corresponding nearest
neighbors on $M_{Y}$. As $L$ increases, the attractor manifold fills
in and the distances among the nearest neighbors shrink. Consequently,
the estimates of $Y$ based on $M_{X}$ should converge to the true
values of $Y$ and the estimates of $X$ based on $M_{Y}$ should
converge to the true $X$. In this way, the convergence is used to
test whether there is a correspondence between states on $M_{X}$
and states on $M_{Y}$. 

Consider that a system $X$ is driving the system $Y$, but the reverse
is not true. The forcing variable $X$ contains no information about
the dynamics of $Y$, although there may be significant predictability
for $Y$ using $M_{X}$ that depends on the conditional probability.
However, this predictability will not converge with increasing $L$.
Cross-mapping that converges in only one direction is the criterion
for unidirectional causality. 

The authors of the CCM method emphasize that convergence is a key
property that distinguishes causality from possible correlation. 

However, in examples used in this paper the possibility of correlation
instead of causation is excluded. Therefore, we do not use the aspect
of convergence, we just compare effectiveness of the cross-mapping
(CM) evaluated by the correlation coefficient with the effectiveness
of measures $M$ and $L$ described above.

\newpage{}

\section{\label{sec:Causality-detection}Causality detection between uni-directionally
coupled chaotic systems}

\subsection{Hénon $0.3$ $\rightarrow$ Hénon $0.3$}

As our first example we will use two uni-directionally coupled identical
Hénon maps. The first two lines correspond to the driver system and
the last two equations describe the response system:

\begin{align}
x_{1}(n+1) & =1.4-x_{1}^{2}(n)+0.3x_{2}(n)\nonumber \\
x_{2}(n+1) & =x_{1}(n)\label{eq:hen33}\\
y_{1}(n+1) & =1.4-\left(Cx_{1}(n)y_{1}(n)+(1-C)y_{1}^{2}(n)\right)+0.3y_{2}(n)\nonumber \\
y_{2}(n+1) & =y_{1}(n)\nonumber 
\end{align}

\begin{figure}[H]
\begin{centering}
\includegraphics[bb=10bp 100bp 600bp 700bp,clip,width=13cm,height=15cm]{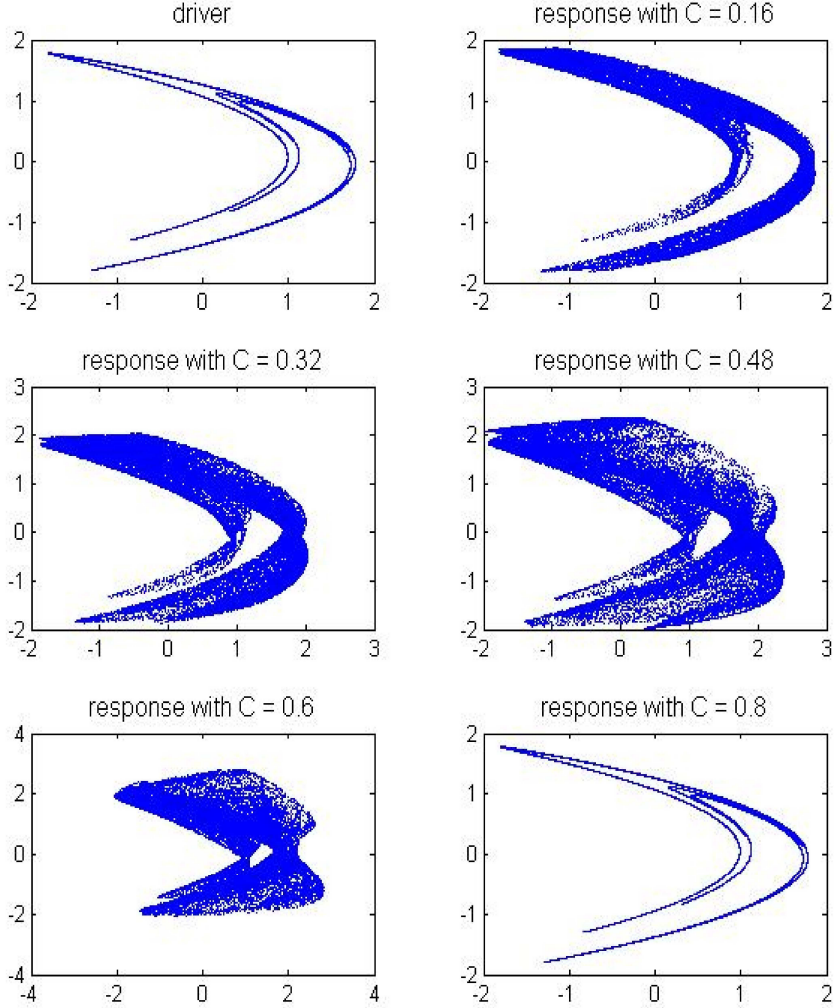}
\par\end{centering}

\protect\caption{Attractors of driver and response Hénon maps (\ref{eq:hen33}) for
various couplings.}
\end{figure}

For each coupling strength a total number of $100000$ were generated
by iterative method. The coupling strengths were chosen from $0$
to $0.8$ with the step $0.04.$ The starting point was $[0.7,0,0.7,0].$
First $1000$ data points were thrown away. 

The same Hénon-Hénon system has been studied in \cite{schif1996},
\cite{quiro2000}, \cite{palus2001}, \cite{stam2002}, \cite{kreuz2007},
\cite{palus2007}, \cite{roman2007}, \cite{janja2008}, \cite{Vlach2010}. 

The variables of the coupled systems can be arranged into an interaction
graph, which is a set of nodes connected by directed edges wherever
one variable directly drives another. Based on definitions in \cite{cummi2015},
in a system of ordinary differential equations, a variable $x$ directly
drives $y$ if it appears non-trivially on the right-hand side of
the equation for the derivative of $y$. Our two connected Hénon systems
represent distinct dynamical subsystems coupled through one-way driving
relationship between variables $x_{1}$ and $y_{1}$. See Figure 2.
This causal link is what we would like to recover.

\usetikzlibrary{arrows,positioning} \newdimen\nodeDist \nodeDist=2cm \newcommand\edgeAngel{20}
\begin{figure}[H] \centering \begin{tikzpicture}[->,>=latex,thick,node distance=\nodeDist,main node/.style={circle,draw} ]
\node[main node] (x1) {$x_1$};   \node[main node] (x2) [right= of x1] {$x_2$};   \node[main node] (y1) [below= of x1] {$y_1$};   \node[main node] (y2) [right= of y1] {$y_2$};
  \path[every node]     (x1) edge node {}  (y1)         edge [bend left=\edgeAngel ] node {} (x2)     (x2) edge [bend left=\edgeAngel ] node {} (x1)     (y1) edge [bend left=\edgeAngel ] node {} (y2)     (y2) edge [bend left=\edgeAngel ] node {} (y1); \end{tikzpicture}\caption{Interaction graph for the coupling of two Henon systems.} \end{figure}

\begin{figure}[H]
\begin{centering}
\includegraphics[width=10cm]{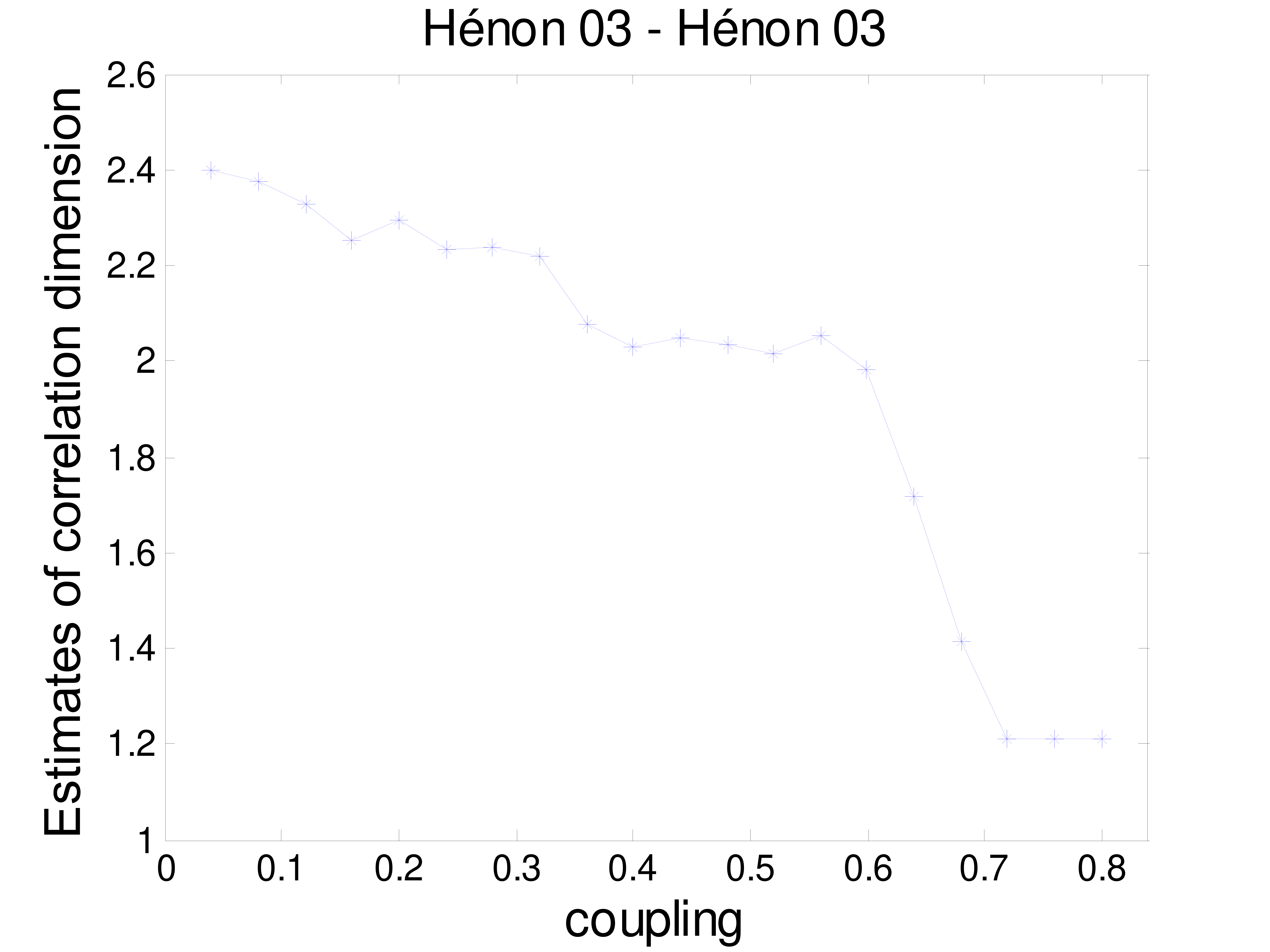}\protect\caption{Correlation dimension estimates for two identical Hénon systems (\ref{eq:hen33})
connected with different coupling strengths.}
\label{D2hen33}
\par\end{centering}

\end{figure}

Estimates of correlation dimension of the combined Hénon-Hénon maps
(driver + response), computed for $100000$ numerically generated
data lead to values of dimension below $2.44$. The estimate of the
dimension of the driving system is about $D_{2}=1.22$ for the same
amount of data. $D_{2}$ estimates around the coupling threshold of
$0.7$ clearly reveal the onset of synchronization by drop to the
value of $1.22$ (the dimension of the driving system) (Figure \ref{D2hen33}).
The same result was indicated by the analysis of the conditional Lyapunov
exponent \cite{schif1996}.

\begin{figure}[H]
\begin{centering}
\includegraphics[width=10cm,height=6.7cm]{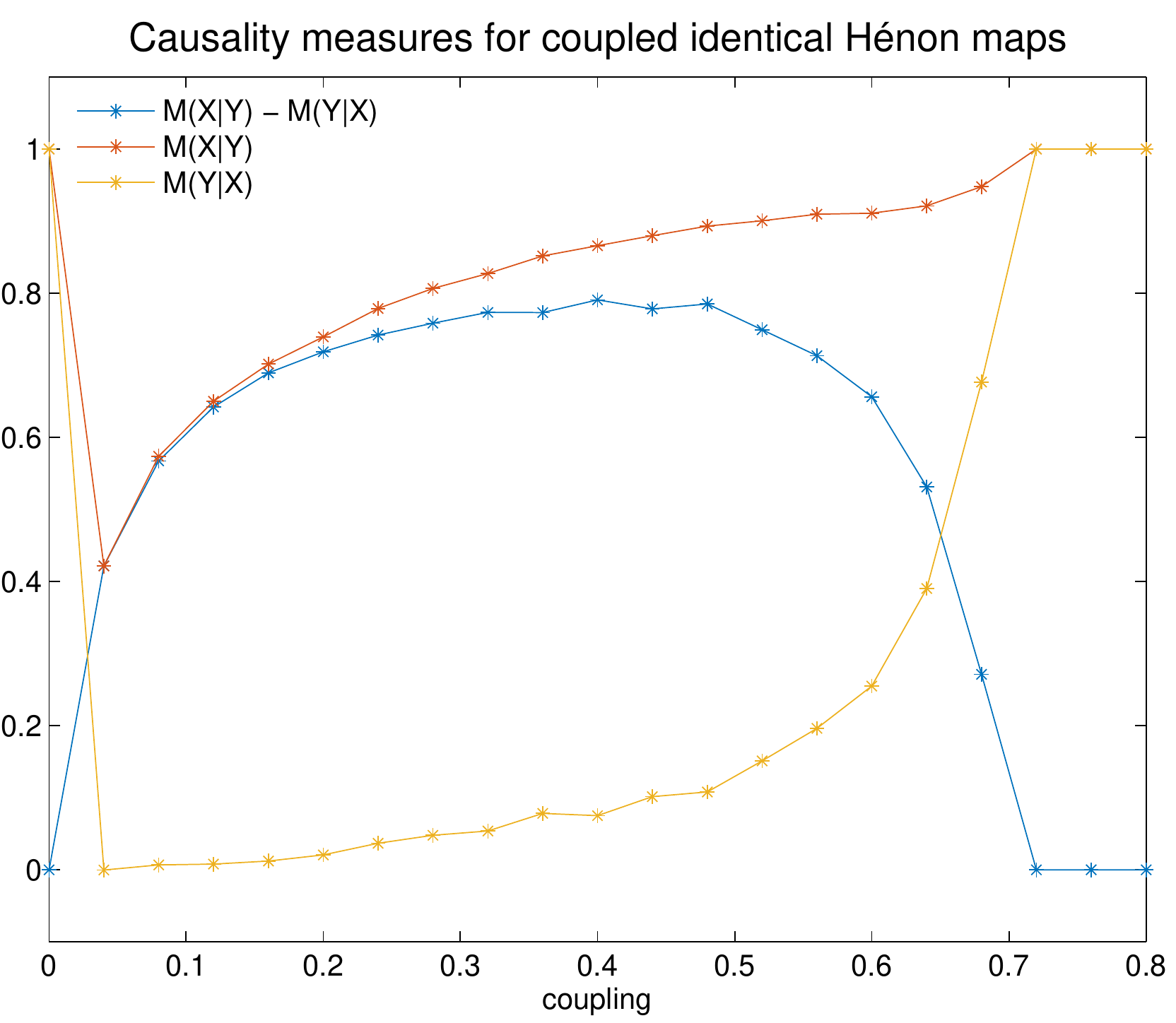}
\par\end{centering}

\medskip{}

\begin{centering}
\includegraphics[width=10cm,height=6.7cm]{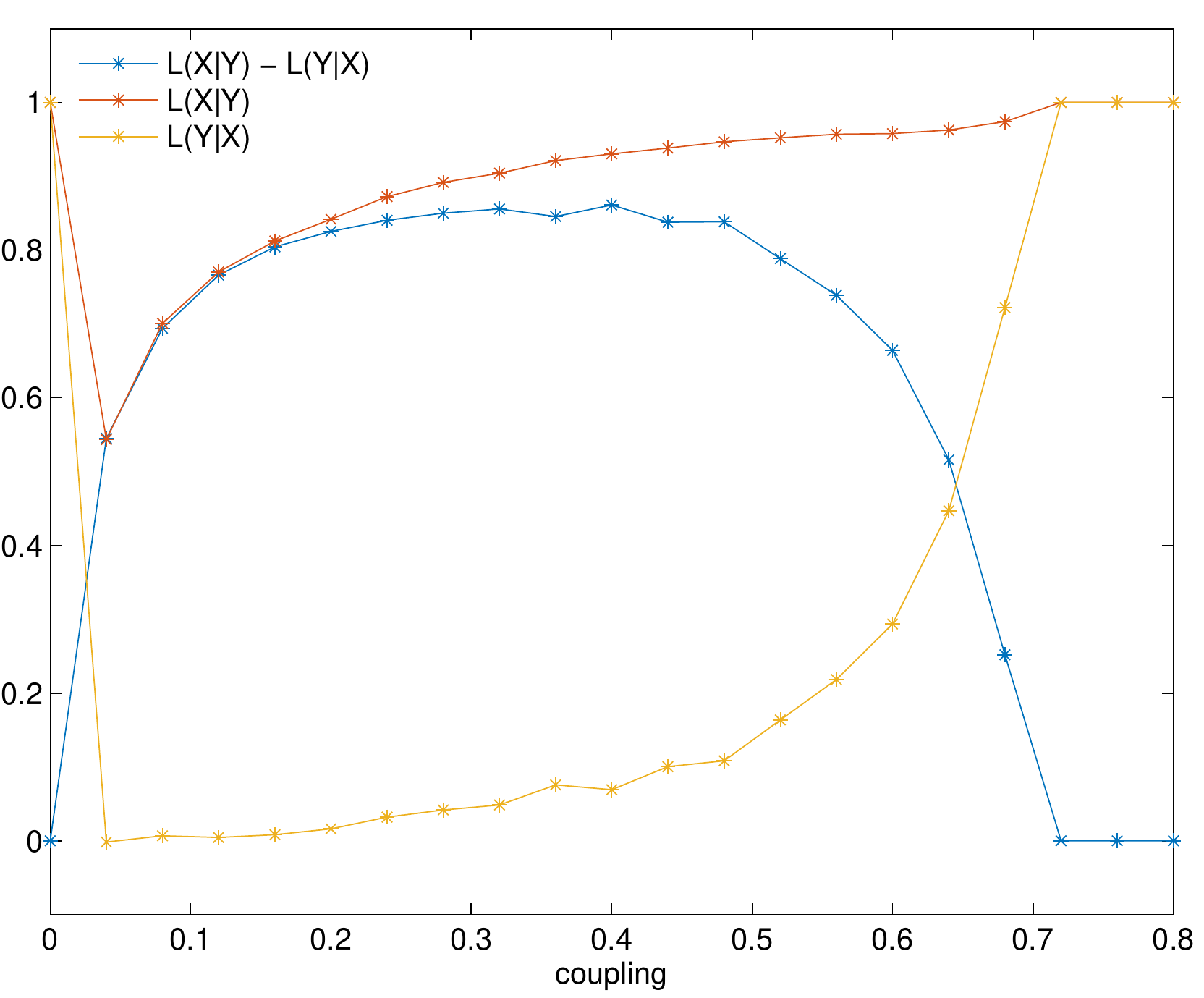}
\par\end{centering}

\medskip{}

\begin{centering}
\includegraphics[width=10cm,height=6.7cm]{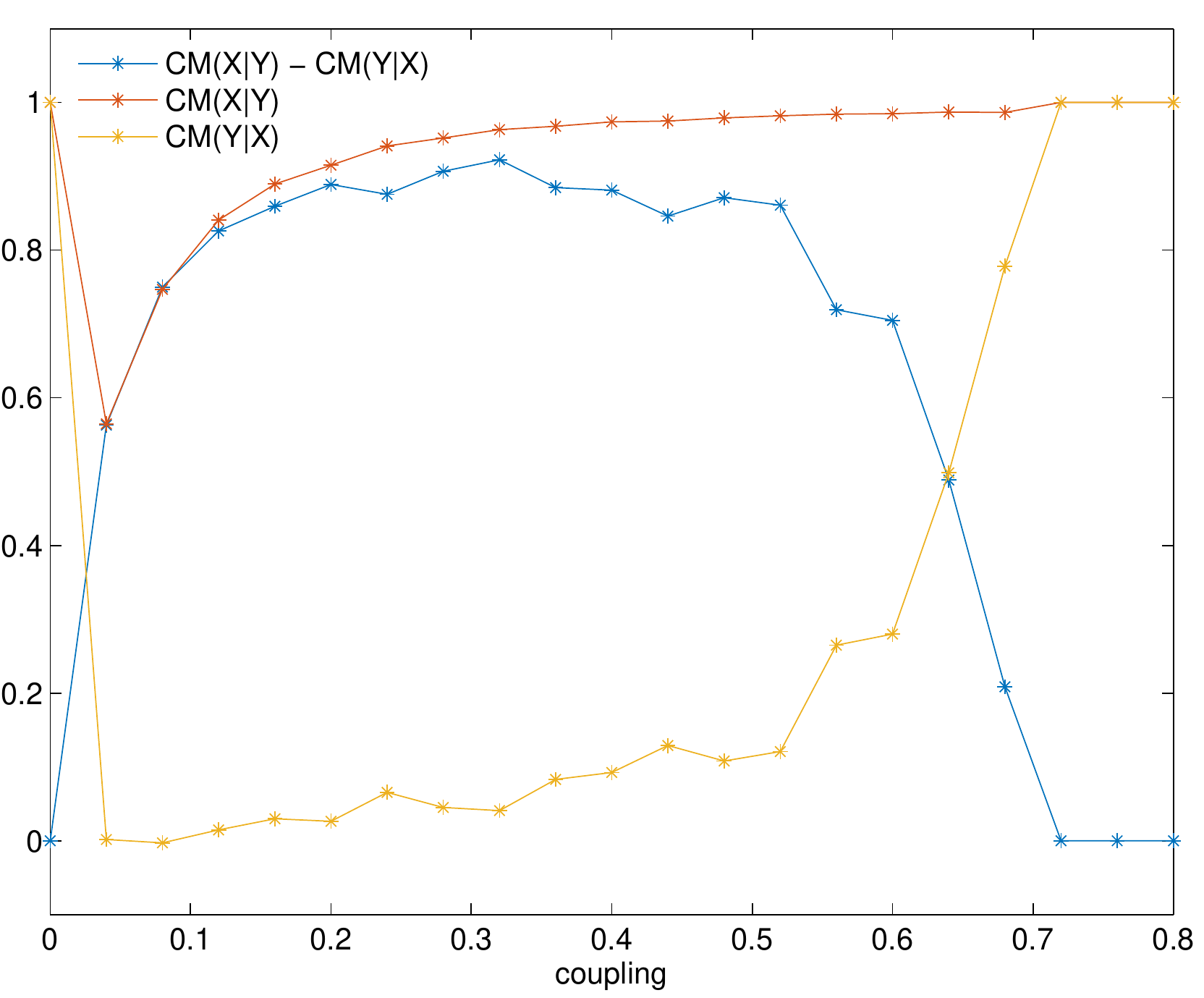}
\par\end{centering}

\protect\caption{Measures $M$, $L$, and $CM$ computed for uni-directionally coupled
identical Hénon systems (\ref{eq:hen33}). The measures show that
$X$ drives $Y$ until the onset of synchronization around the coupling
threshold of $0.7$.}
\end{figure}

\subsubsection*{Results of causality detection using reconstructed manifolds}

Suppose that we only know $10000$ data-points of variable $x_{1}$
of the driving system and variable $y_{1}$ of the response system
from (\ref{eq:hen33}) and we would like to know whether there is
a causal relationship between the two systems. 

One orbit of the attractor has no more than hundred points. In order
to use the state-space based methods of search for causality we used
delay coordinates with the delay equal to $1$ to reconstruct state
portraits of the dynamics in $5-$dimensional state spaces. For the
methods $6$ nearest neighbors were taken. 

In the following, let us denote the direction from $X$ to $Y$ by
$X|Y$ and the direction from $Y$ to $X$ by $Y|X$ . Take, for example
the measure $M$. If $X$ drives $Y$, the measure $M(X|Y)$ is expected
to be higher than $M(Y|X)$. In figures, $M(X|Y)$ is displayed in
red, $M(Y|X)$ in yellow and their difference $\text{\ensuremath{\Delta}}M(X|Y)=M(X|Y)-M(Y|X)$
is shown in blue.

\newpage{}

\subsection{Hénon $0.3$ $\rightarrow$ Hénon $0.1$}

The second example is formed by uni-directionally coupled nonidentical
Hénon maps. Variables $x_{1}$, $x_{2}$ correspond to the driver
system and $y_{1}$, $y_{2}$ are the variables of the response system:

\begin{eqnarray}
x_{1}(n+1) & = & 1.4-x_{1}^{2}(n)+0.3x_{2}(n)\nonumber \\
x_{2}(n+1) & = & x_{1}(n)\label{eq:hen31}\\
y_{1}(n+1) & = & 1.4-\left(Cx_{1}(n)y_{1}(n)+(1-C)y_{1}^{2}(n)\right)+0.1y_{2}(n)\nonumber \\
y_{2}(n+1) & = & y_{1}(n)\nonumber 
\end{eqnarray}

The data were generated by iterative method. The coupling strength
was chosen from $0$ to $1.4$ with the step $0.04.$ The starting
point was $[0.7,0,0.7,0].$ First $1000$ data points were thrown
away. The total number of obtained data was $100000.$ This system
was investigated in \cite{schif1996}, \cite{quiro2000}, \cite{stam2002}.
The interaction graph for this connection is the same as in the previous
case (Figure 2). 

\begin{figure}[H]
\centering{}\includegraphics[width=13cm,height=15cm]{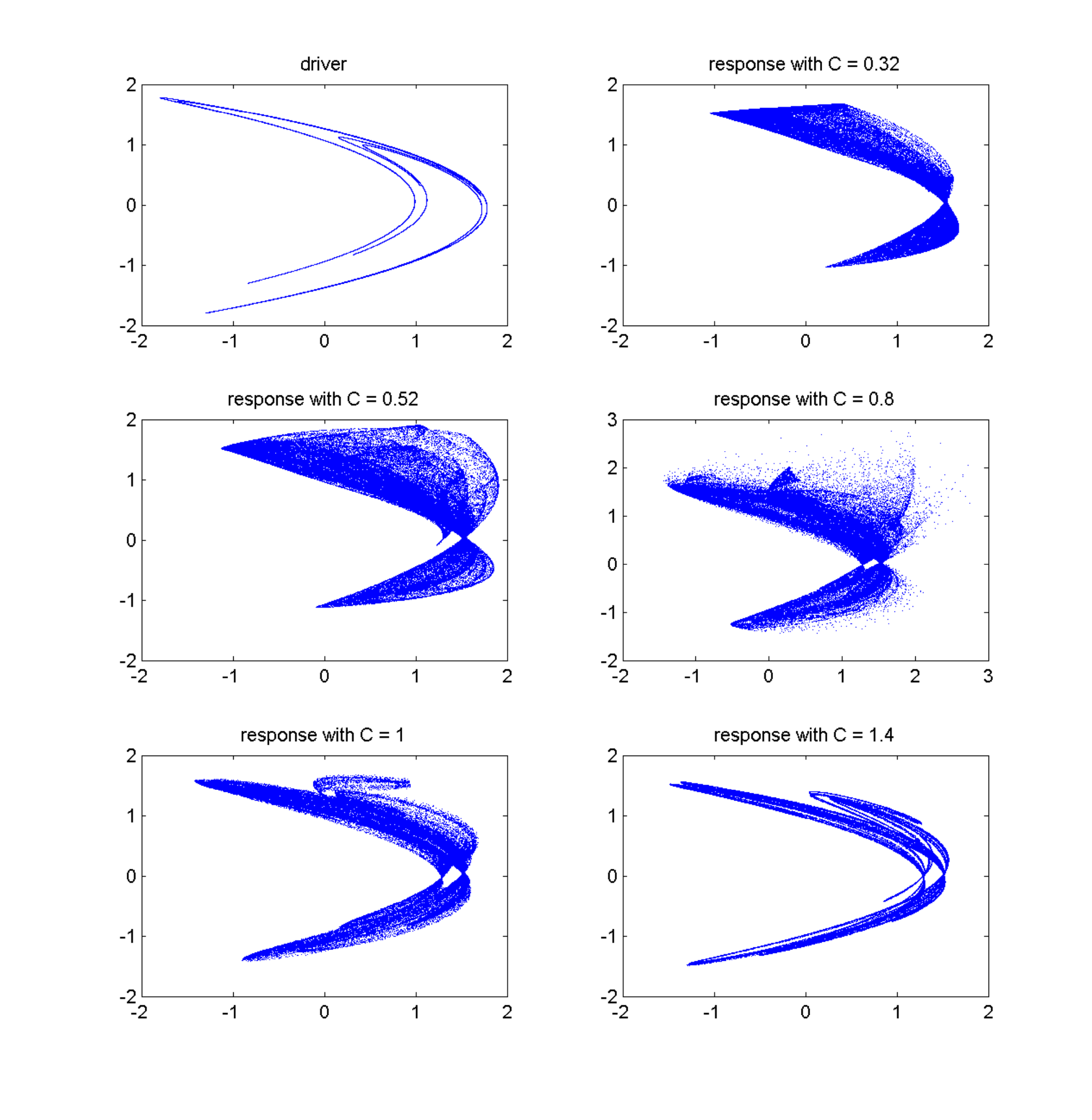}\protect\caption{Coupled non-identical Hénon systems (\ref{eq:hen31}). 2-dimensional
plots of attractors of driver and response system for various couplings.}
\end{figure}

\begin{figure}[H]
\centering{}\includegraphics[bb=0bp 300bp 595bp 400bp,scale=0.4,angle=-90,origin=rB]{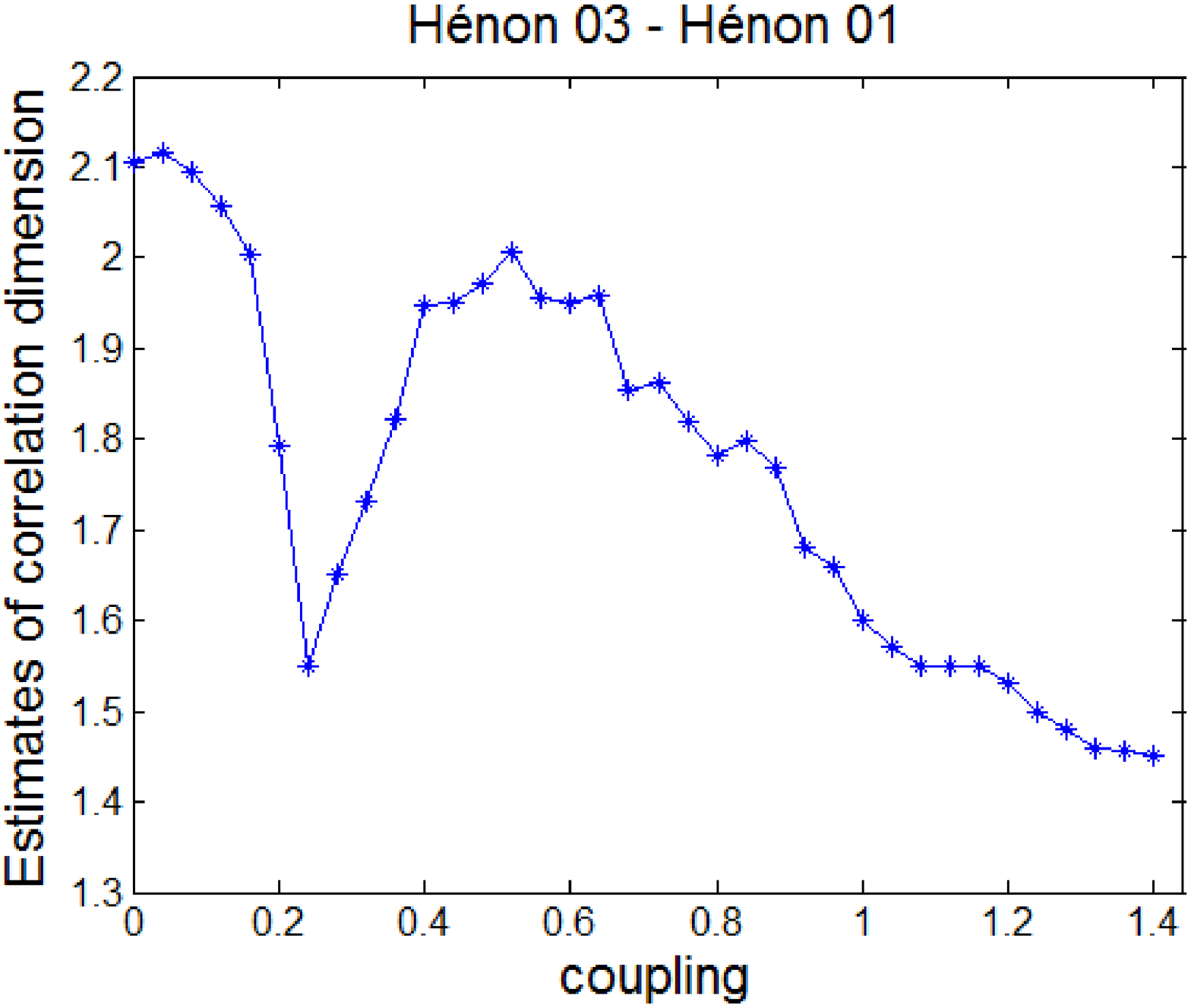}\protect\caption{Correlation dimension estimates for two non-identical Hénon systems
(\ref{eq:hen31}) connected with different coupling strengths.}
\label{D2hen31}
\end{figure}

Maximum Lyapunov exponent of the response system turns negative near
the coupling $0.2$ and rises to positive values around the couplings
$0.4-0.5.$ Then it falls again to negative values showing generalized
(nonidentical) synchronization \cite{quiro2000}. Similar behavior
is presented by our estimates of correlation dimension of the $X+Y$
(see Figure \ref{D2hen31}). The dimension of the attractor of the
combined system saturates to the value which remains relatively unchanging
for couplings somewhat higher than $1$.

\subsubsection*{Results of causality detection using reconstructed manifolds}

Suppose that we only know $10000$ data-points of variable $x_{1}$
of the driving system and variable $y_{1}$ of the response system
and we would like to know whether there is a causal relationship between
the two systems. In order to use the state-space based methods of
search for causality we used delay coordinates with the delay equal
to $1$ to reconstruct state portraits of the dynamics in $5-$dimensional
state spaces. For each method $6$ nearest neighbors were taken. 

\begin{figure}
\begin{centering}
\includegraphics[bb=50bp 190bp 580bp 620bp,clip,width=10cm,height=6.7cm]{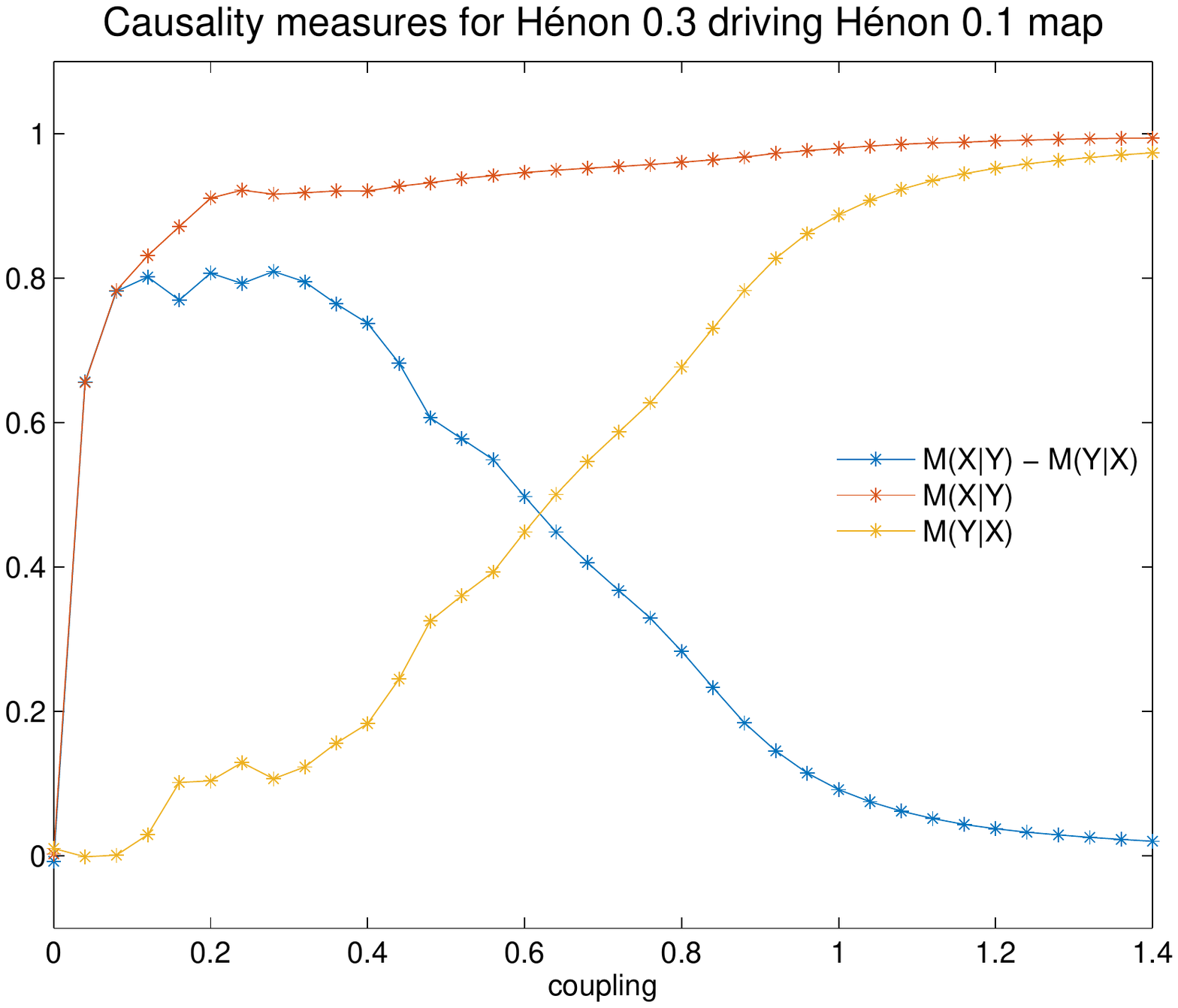}
\par\end{centering}

\medskip{}

\begin{centering}
\includegraphics[bb=50bp 190bp 580bp 620bp,clip,width=10cm,height=6.7cm]{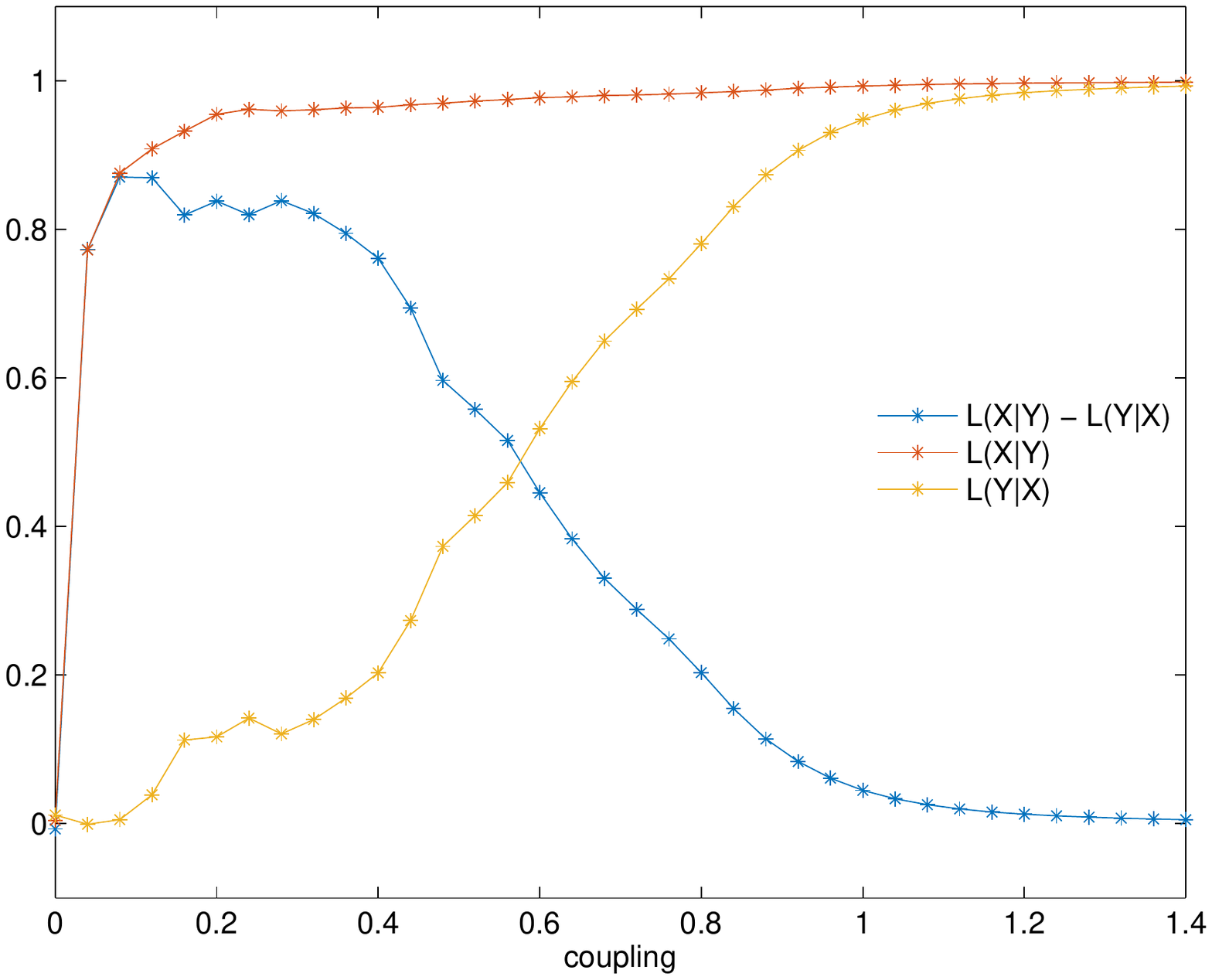}
\par\end{centering}

\medskip{}

\begin{centering}
\includegraphics[bb=50bp 190bp 580bp 620bp,clip,width=10cm,height=6.7cm]{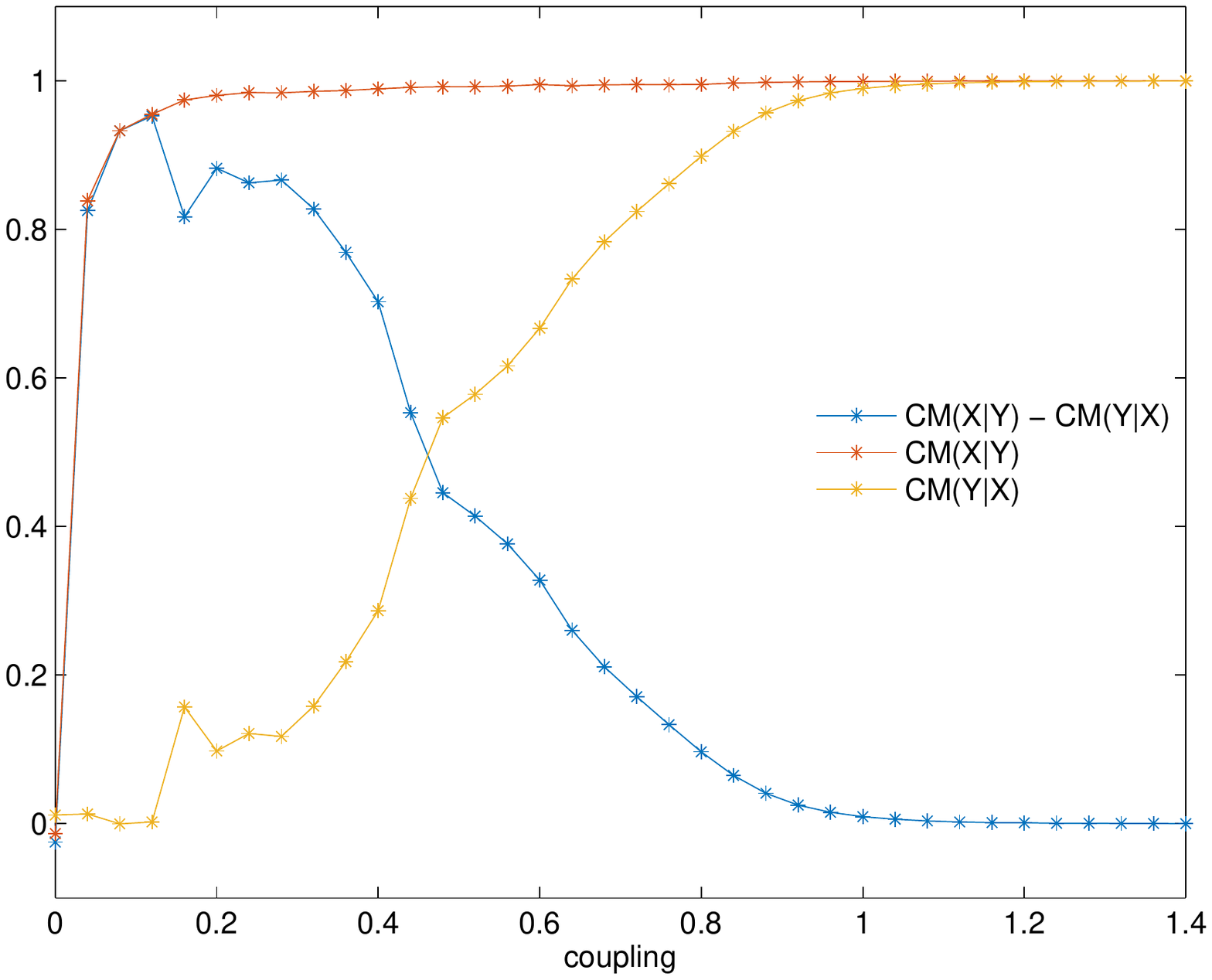}
\par\end{centering}

\protect\caption{Measures $M$, $L$, and $CM$ computed for uni-directionally coupled
non-identical systems (Hénon $0.3$ $\rightarrow$ Hénon $0.1$).
The measures show that $X$ drives $Y$ until the onset of synchronization
around the coupling threshold of about $1$.}
\end{figure}

\newpage{}

\subsection{Hénon $0.1$ $\rightarrow$ Hénon $0.3$}

In the next example, the previous two Hénon maps change roles. Now
the map with parameter $0.1$ is the driver system and the map with
parameter $0.3$ is the response system:

\begin{eqnarray}
x_{1}(n+1) & = & 1.4-x_{1}^{2}(n)+0.1x_{2}(n)\nonumber \\
x_{2}(n+1) & = & x_{1}(n)\label{eq:hen13}\\
y_{1}(n+1) & = & 1.4-\left(Cx_{1}(n)y_{1}(n)+(1-C)y_{1}^{2}(n)\right)+0.3y_{2}(n)\nonumber \\
y_{2}(n+1) & = & y_{1}(n)\nonumber 
\end{eqnarray}
 The data were generated by iterative method. The coupling strength
was chosen from $0$ to $1.4$ with the step $0.04.$ The starting
point was $[0.7,0,0.7,0].$ First $1000$ data points were thrown
out. The total number of obtained data was $100000.$ The same Hénon-Hénon
system was used in \cite{quiro2000}, \cite{palus2001}, and \cite{roman2007}. 

\begin{figure}[H]
\centering{}\includegraphics[width=13cm,height=15cm]{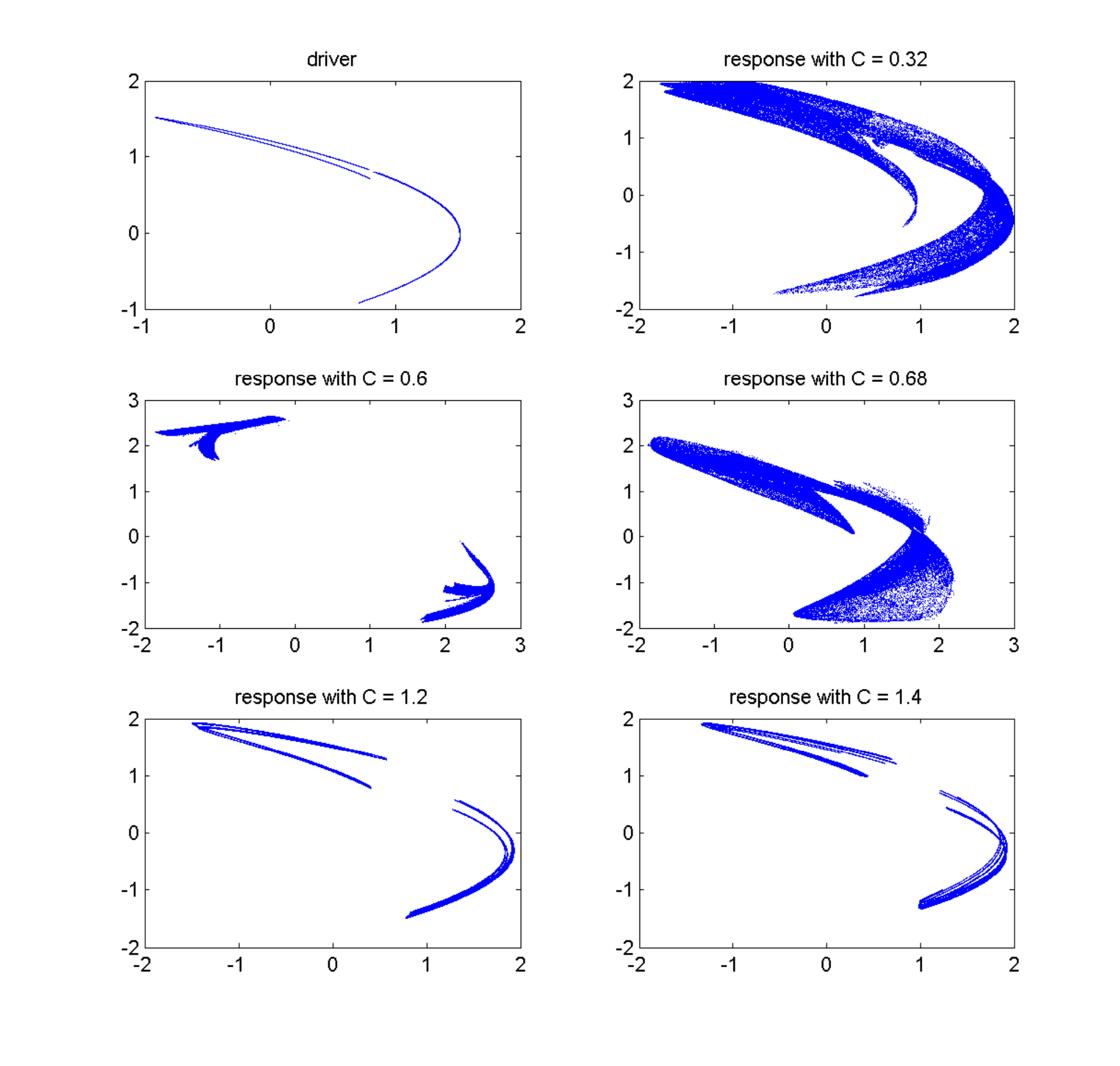}\protect\caption{Coupled non-identical Hénon systems (\ref{eq:hen13}). 2-dimensional
plots of attractors of driver and response system for various couplings.}
\end{figure}

Also in this case variables of the coupled systems can be arranged
into the interaction graph shown in Figure 2. It means that the two
connected Hénon systems represent distinct dynamical subsystems coupled
through one-way driving relationship between variables $x_{1}$ and
$y_{1}$. This causal link is what we would like to recover.

In this example, the correlation dimension of the driving system (estimates
about $1.02$) is lower than the dimension of the response system
(estimates about $1.22$).

The largest Lyapunov exponent of the response decreases with increasing
coupling and becomes negative at $0.38$. After coupling of $0.6$
it rises and touches zero around $0.62$ and then it falls again to
negative values, which indicate generalized synchronization of two
nonidentical systems \cite{palus2001}. 

\begin{figure}[H]
\centering{}\includegraphics[bb=0bp 400bp 595bp 842bp,clip,width=10cm]{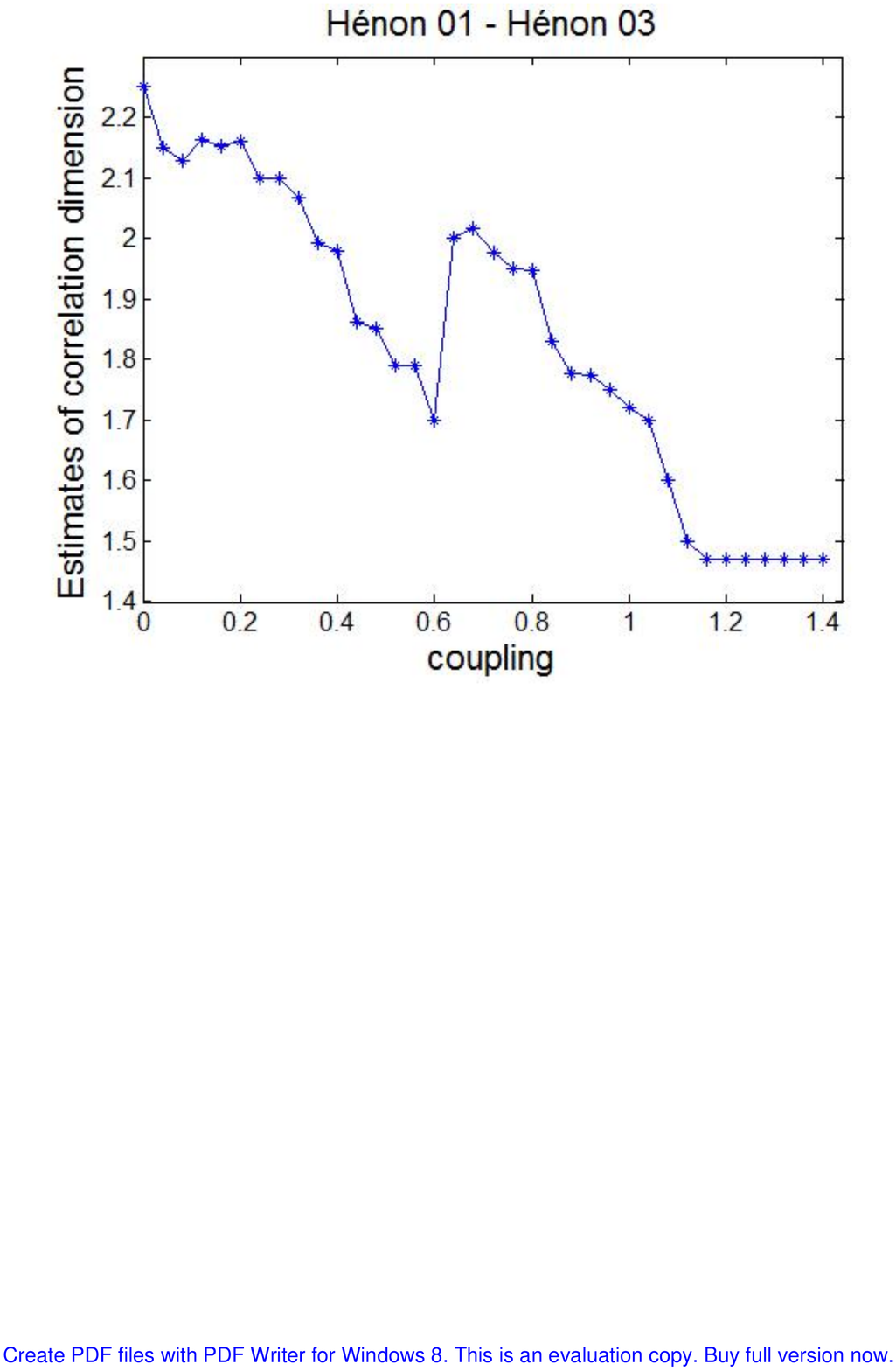}\protect\caption{Correlation dimension estimates for two non-identical Hénon systems
(\ref{eq:hen13}). The individual values correspond to different coupling
strengths.}
\label{D2hen13}
\end{figure}

Our estimates of correlation dimension of the $X+Y$ system (Figure
\ref{D2hen13}) show similar declines and risings to finally settle
for couplings higher than $1.1$.

\subsubsection*{Results of causality detection using reconstructed manifolds}

Suppose that we only know $10000$ data-points of variable $x_{1}$
of the driving system and variable $y_{1}$ of the response system
and we would like to know whether there is a causal relationship between
the two systems. In order to use the state-space based methods of
search for causality we used delay coordinates with the delay equal
to $1$ to reconstruct state portraits of the dynamics in $5-$dimensional
state spaces. For each method $6$ nearest neighbors were taken. 

\begin{figure}[H]
\begin{centering}
\includegraphics[bb=50bp 190bp 580bp 620bp,clip,width=10cm,height=6.7cm]{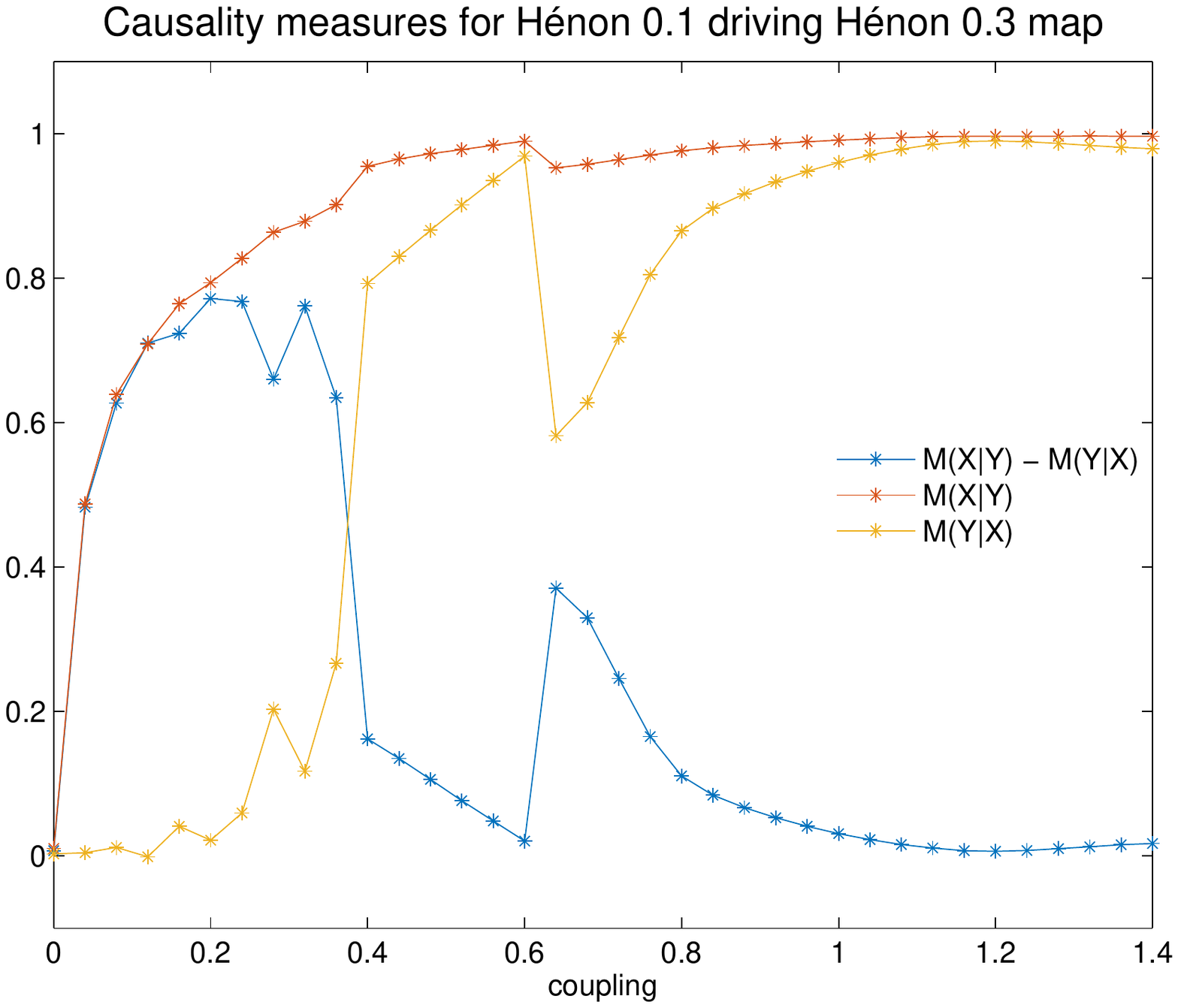}
\par\end{centering}

\medskip{}

\begin{centering}
\includegraphics[bb=50bp 190bp 580bp 620bp,clip,width=10cm,height=6.7cm]{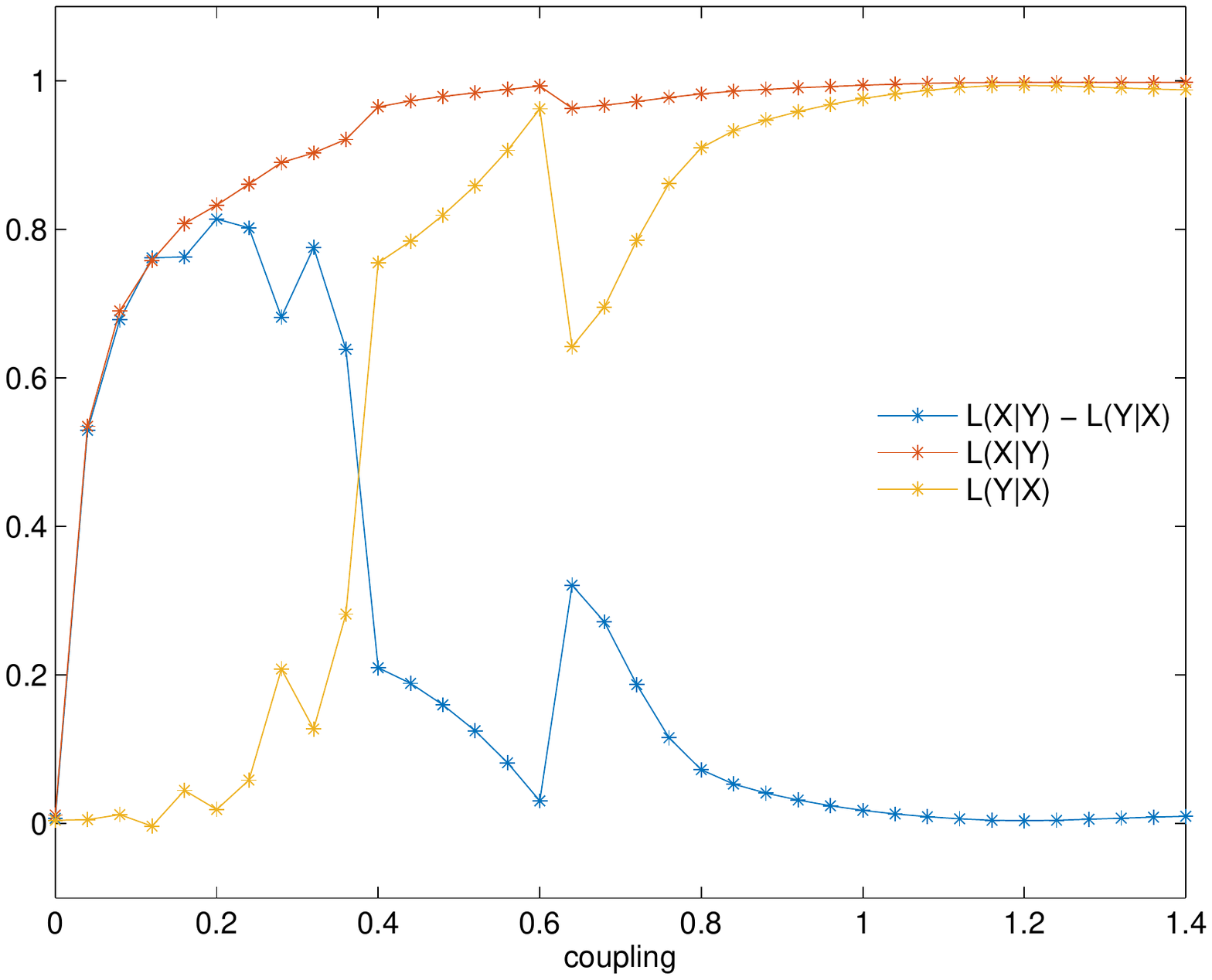}
\par\end{centering}

\medskip{}

\begin{centering}
\includegraphics[bb=50bp 190bp 580bp 620bp,clip,width=10cm,height=6.7cm]{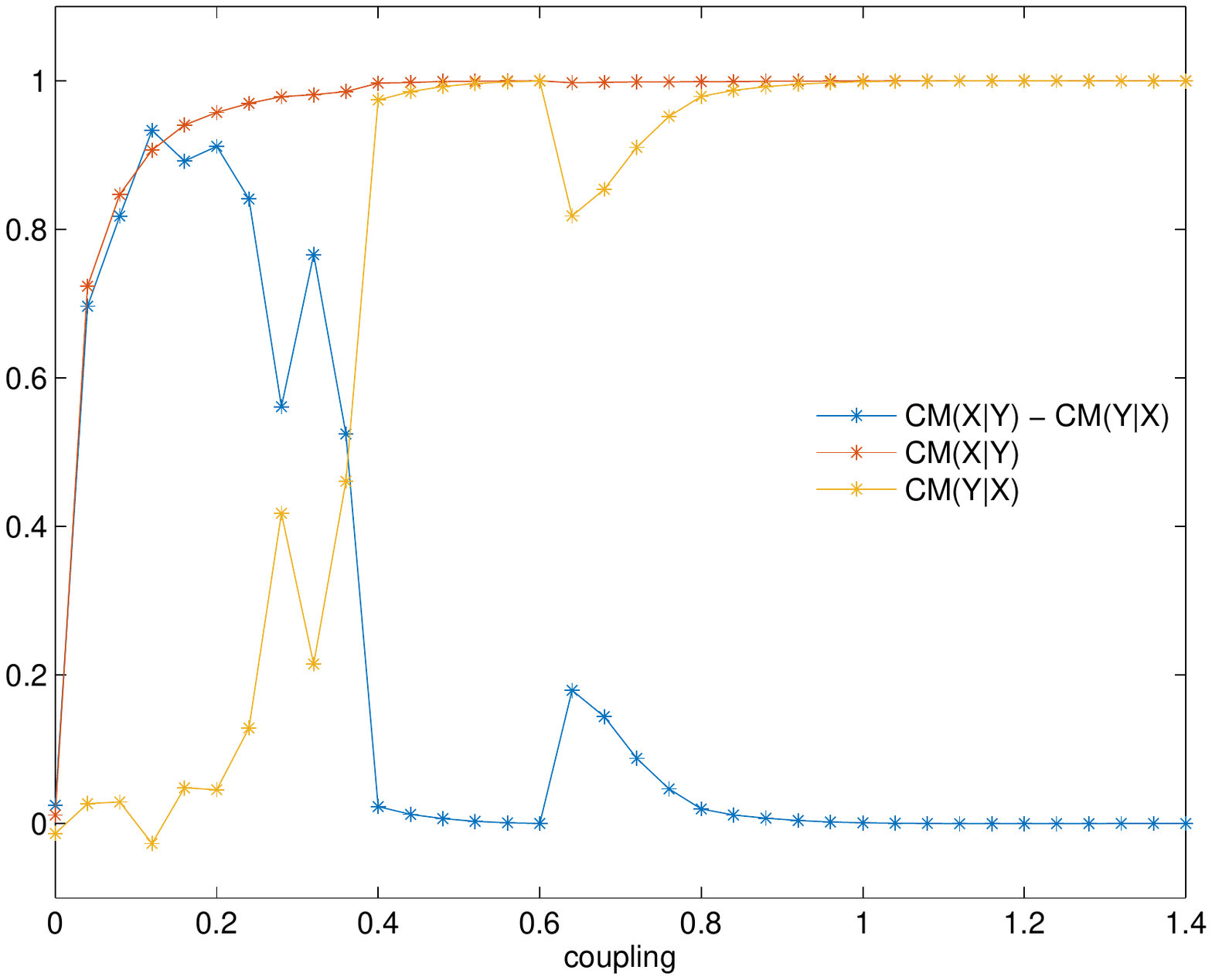}
\par\end{centering}

\protect\caption{Measures $M$, $L$, and $CM$ computed for uni-directionally coupled
non-identical systems (Hénon $0.1$ $\rightarrow$ Hénon $0.3$).
The measures show that $X$ drives $Y$ until the onset of synchronization
around the coupling threshold of about $1.1$.}
\end{figure}

\newpage{}

\subsection{Rössler $\rightarrow$ Lorenz}

In this example Rössler system ($x_{1}$, $x_{2}$, $x_{3}$) drives
the Lorenz system ($y_{1}$, $y_{2}$, $y_{3}$): 
\begin{eqnarray}
\dot{x}_{1} & = & -6(x_{2}+x_{3})\nonumber \\
\dot{x}_{2} & = & 6(x_{1}+0.2x_{2})\nonumber \\
\dot{x}_{3} & = & 6\left(0.2+x_{3}(x_{1}-5.7)\right)\label{eq:roslor}\\
\dot{y}_{1} & = & 10(-y_{1}+y_{2})\nonumber \\
\dot{y}_{2} & = & 28y_{1}-y_{2}-y_{1}y_{3}+Cx_{2}^{2}\nonumber \\
\dot{y}_{3} & = & y_{1}y_{2}-\frac{8}{3}y_{3}\nonumber 
\end{eqnarray}
A total number of $100000$ data were obtained from Matlab solver
of ordinary differential equations ode45 which is based on explicit
Runge-Kutta formula. The coupling strength was chosen from $0$ to
$5$ with the step $0.1$. The starting point was $[0,0,0.4,0.3,0.3,0.3].$
First $1000$ data points were thrown away. The same system was studied
in \cite{pyrag1996}, \cite{levan1999}, \cite{quiro2000}, \cite{palus2001},
\cite{andrz2003} and \cite{palus2007}. 

\begin{figure}[H]
\begin{centering}
\includegraphics[bb=30bp 130bp 560bp 740bp,clip,width=12cm,height=14cm]{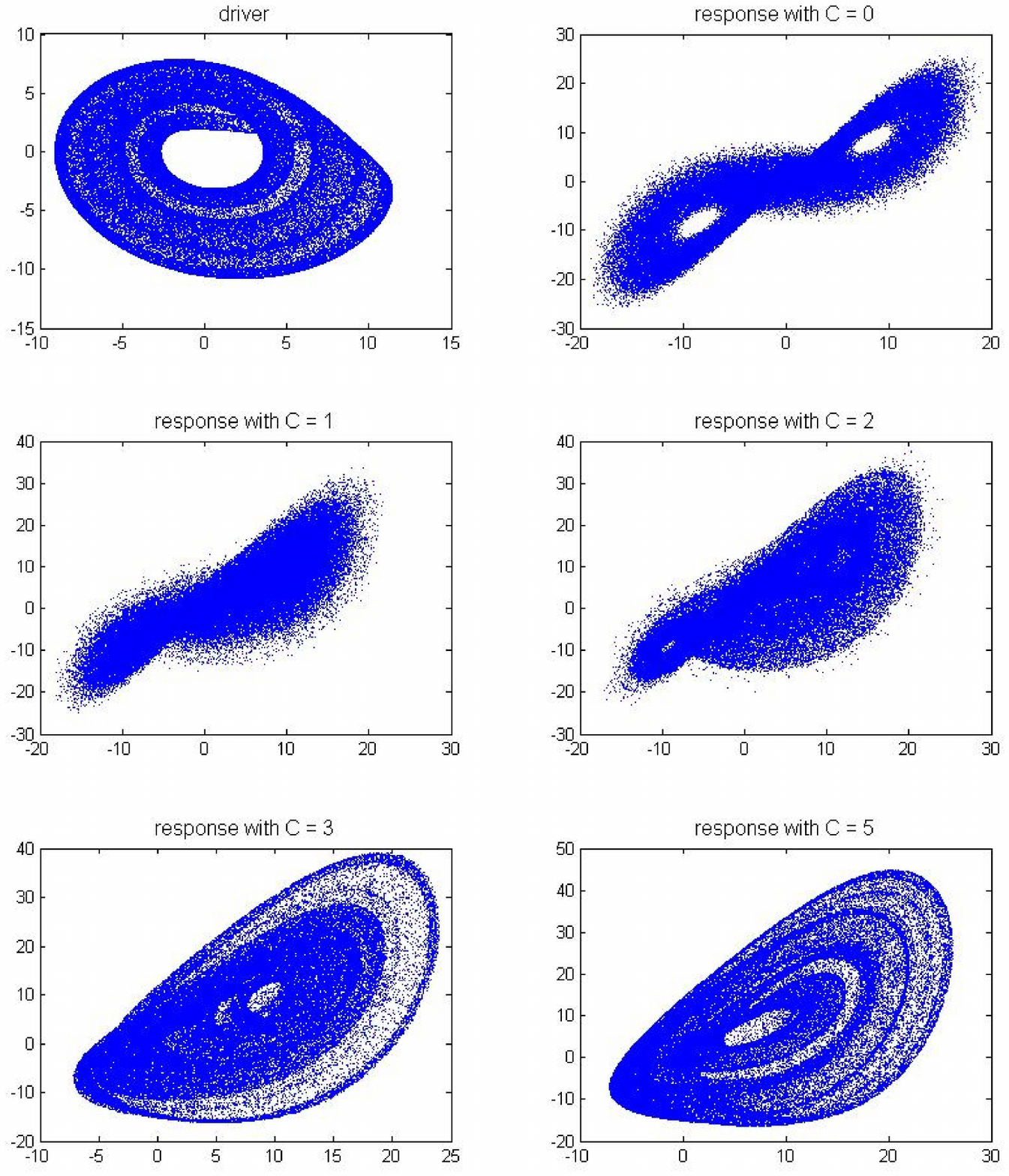}
\par\end{centering}

\protect\caption{Rössler system driving Lorenz system. 2-dimensional plots of attractors
of driver and response system for various couplings.}

\end{figure}

\begin{figure}[H]
\centering{}\includegraphics[bb=30bp 190bp 570bp 700bp,clip,width=13cm,height=7cm]{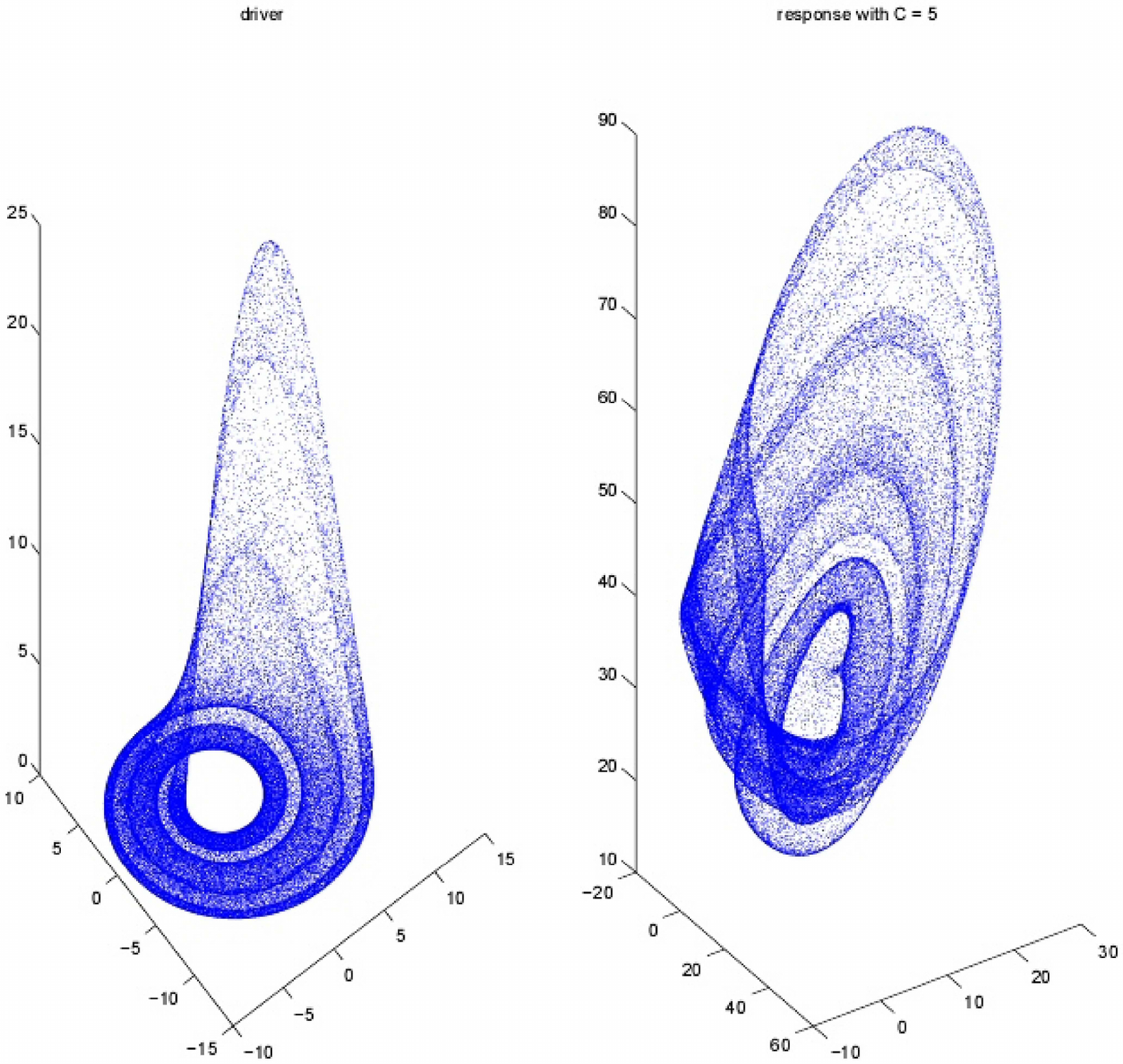}\protect\caption{Rössler system driving Lorenz system. 3-dimensional plots of attractors
of driver and response system for coupling 5.}
\end{figure}

The interaction graph in Figure 13 shows that the Rössler and the
Lorenz system are coupled through one-way driving relationship between
variables $x_{2}$ and $y_{2}$. This causal link is what we would
like to recover.

\usetikzlibrary{arrows,positioning} \newdimen\nodeDist \nodeDist=2cm
\begin{figure}[H] \centering \begin{tikzpicture}[->,>=latex,thick,node distance=\nodeDist,main node/.style={circle,draw} ]
\node[main node] (x2) {$x_2$};   \node[main node] (x1) [right= of x2] {$x_1$};     \node[main node] (x3) [right= of x1] {$x_3$};   \node[main node] (y2) [below= of x2] {$y_2$};   \node[main node] (y1) [right= of y2] {$y_1$};   \node[main node] (y3) [below right= sqrt(3)/2*\nodeDist  and .5*\nodeDist  of y2] {$y_3$};      \path[every node]     (x1) edge [bend left=\edgeAngel ] node {} (x2)     edge [bend left=\edgeAngel ] node {} (x3)     (x2) edge node {}  (y2)     edge [bend left=\edgeAngel ] node {} (x1)     (x3) edge [bend left=\edgeAngel ] node {} (x1)     (y1) edge [bend left=\edgeAngel ] node {} (y2)     edge node {} (y3)     (y2) edge [bend left=\edgeAngel ] node {} (y1)     edge [bend left=\edgeAngel ] node {} (y3)     (y3) edge [bend left=\edgeAngel ] node {} (y2); \end{tikzpicture}\caption{Interaction graph for the coupling of Rössler system and Lorenz system.} \end{figure}
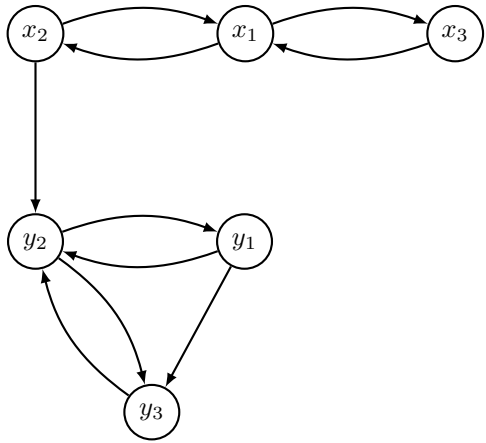

\begin{figure}[H]
\centering{}\includegraphics[width=10cm]{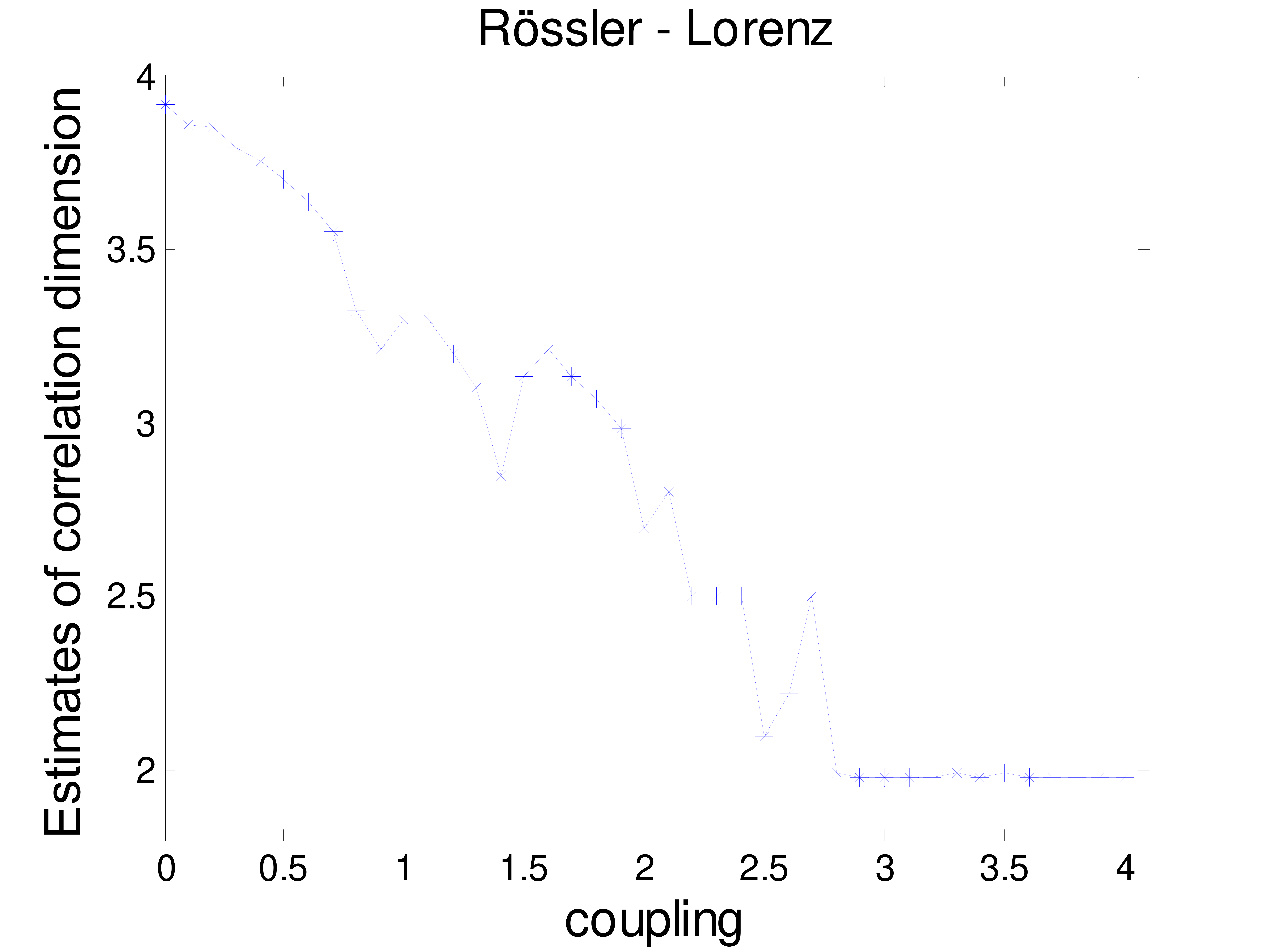}\protect\caption{Correlation dimension estimates for Rössler and Lorenz systems connected
with different coupling strengths.}
\label{D2roslor}
\end{figure}

For these coupled systems only weak synchronization is considered
\cite{quiro2000}. Lyapunov exponents show the synchronization takes
place between the coupling strengths $2$ and $3$. The same is indicated
by our estimates of correlation dimensions (Figure \ref{D2roslor}).

\subsubsection*{Results of causality detection using reconstructed manifolds}

Suppose that we know $50000$ data-points of variable $x_{2}$ of
the driving Rössler system and variable $y_{2}$ of the responsive
Lorenz system and we would like to know whether there is a causal
relationship between the two systems. In order to use the state-space
based methods of search for causality we made reconstructions of the
state portraits. We used time-delayed vectors of $x_{2}$ and $y_{2}$
with time delay equal to $1$ and embedding dimension of $7$. Methods
for all three measures used $8$ nearest neighbors.

\begin{figure}[H]
\begin{centering}
\includegraphics[bb=50bp 190bp 580bp 620bp,clip,width=10cm,height=6.7cm]{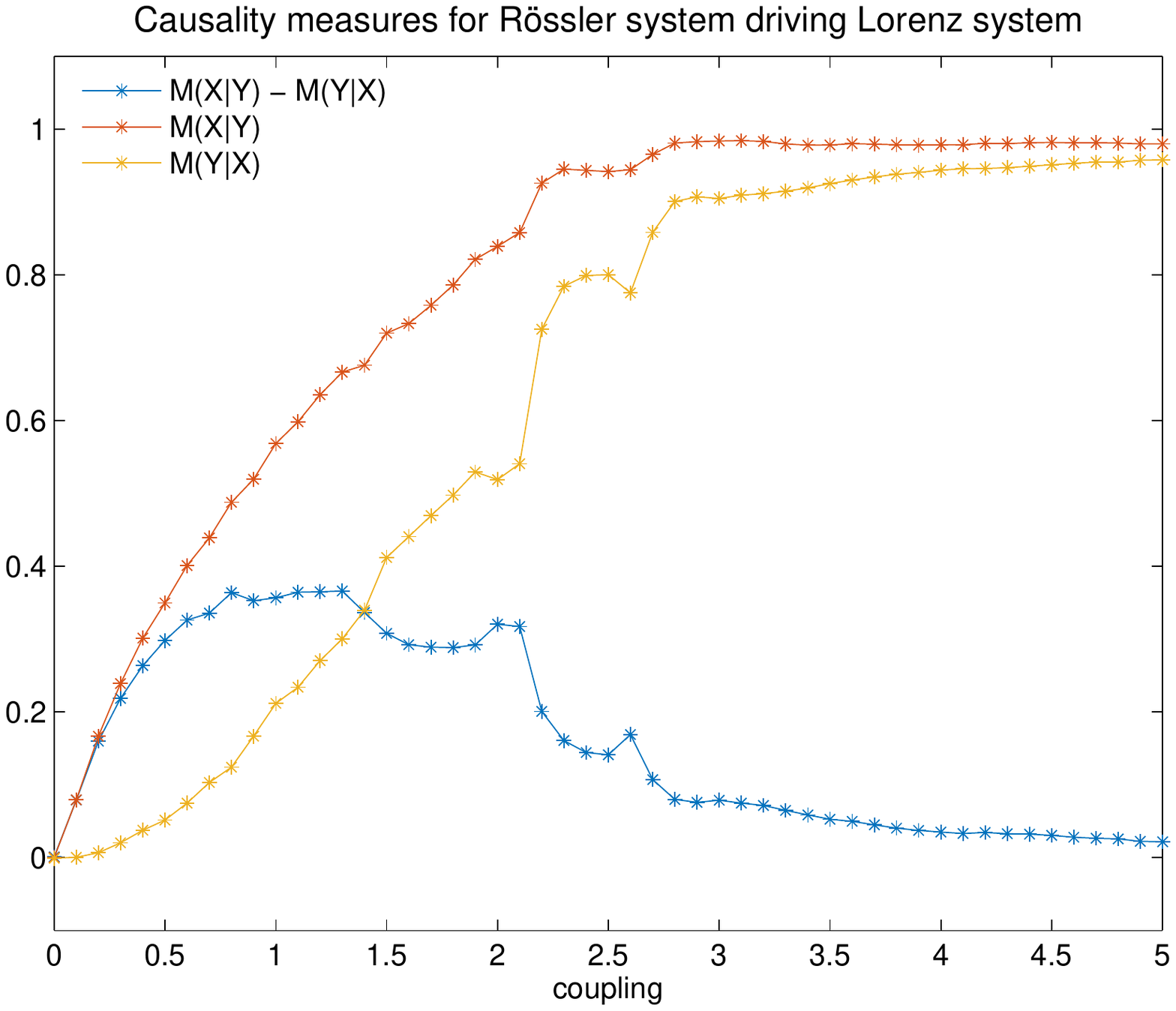}
\par\end{centering}

\medskip{}

\begin{centering}
\includegraphics[bb=50bp 190bp 580bp 620bp,clip,width=10cm,height=6.7cm]{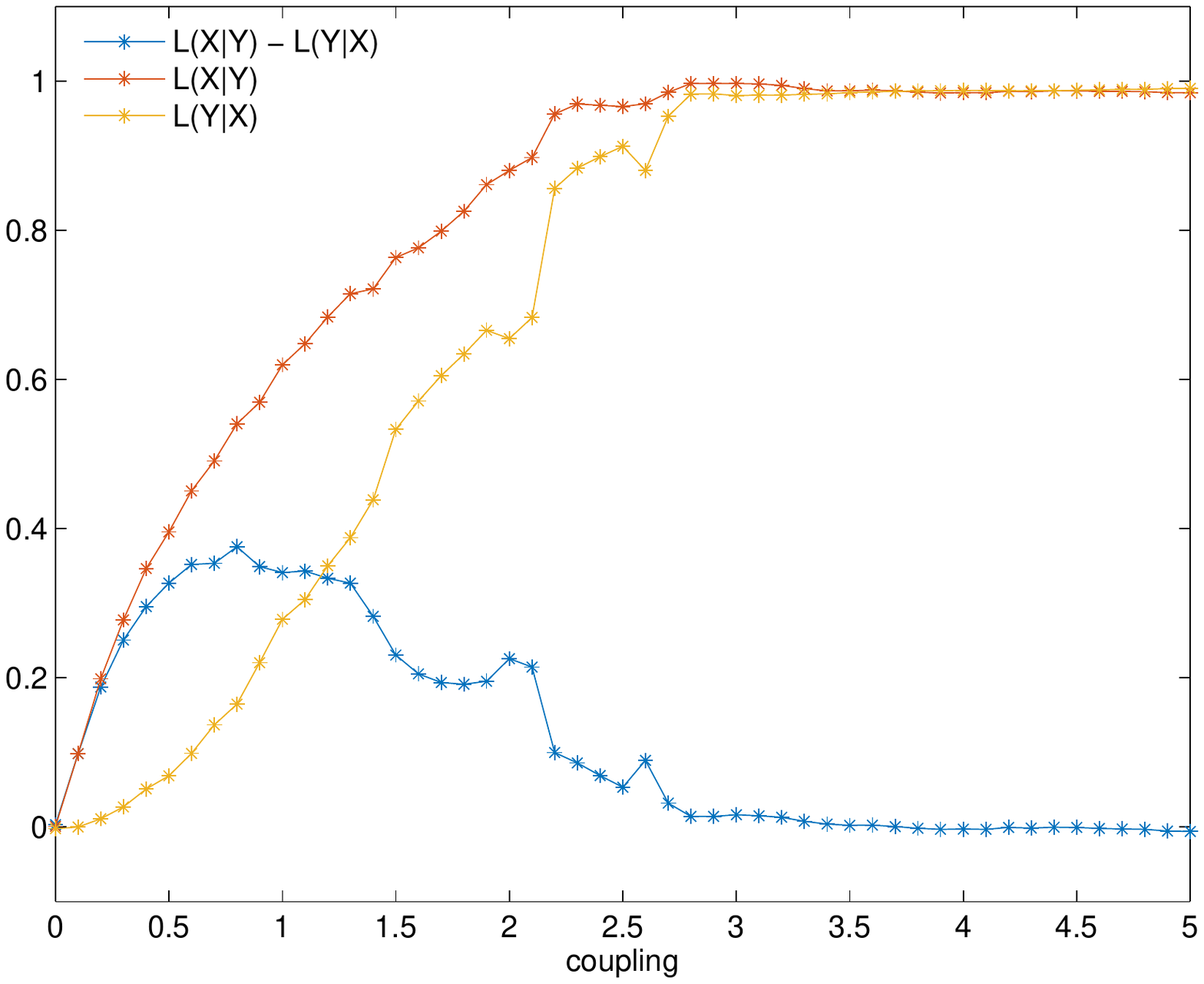}
\par\end{centering}

\medskip{}

\begin{centering}
\includegraphics[bb=50bp 190bp 580bp 620bp,clip,width=10cm,height=6.7cm]{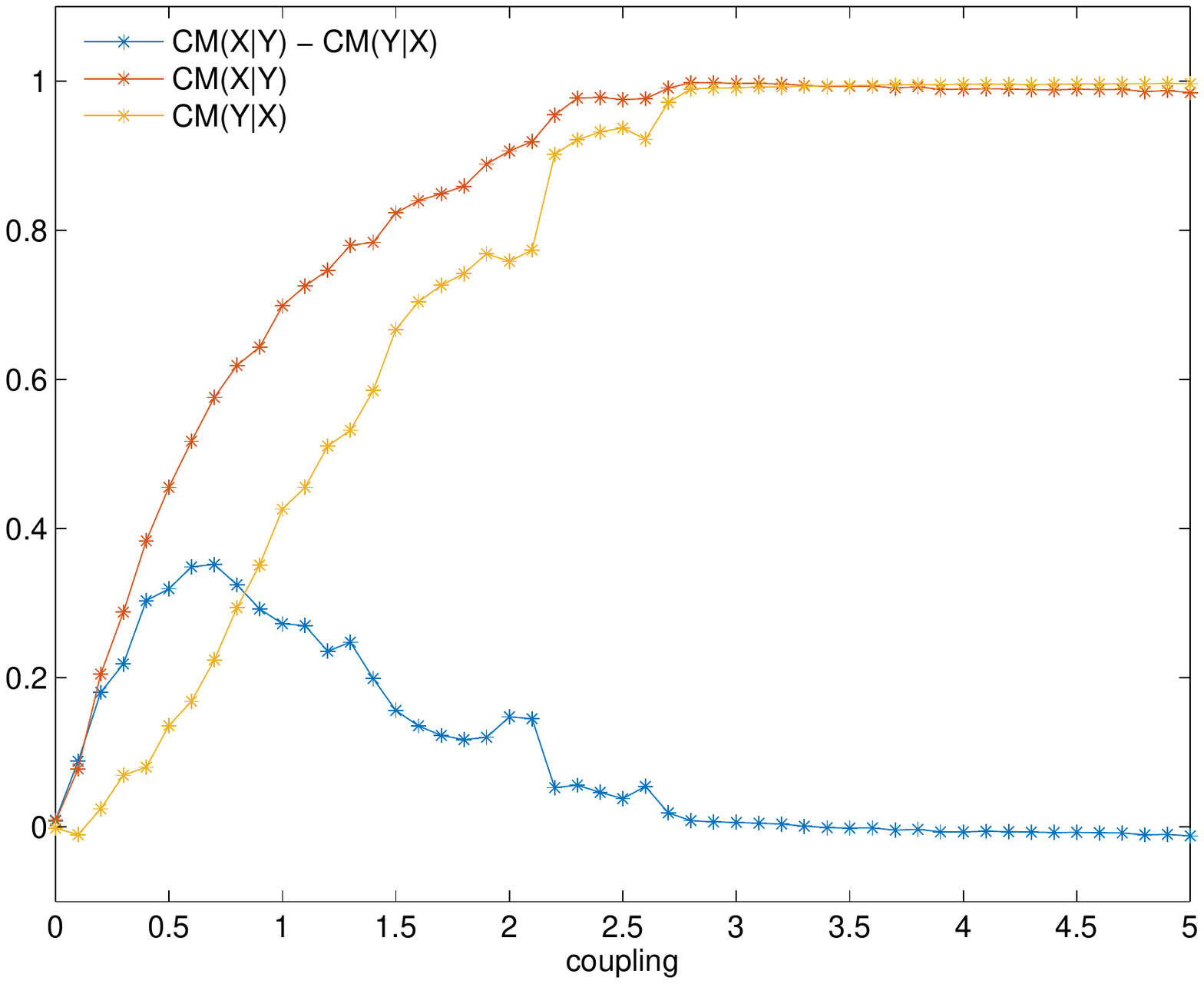}
\par\end{centering}

\protect\caption{Measures $M$, $L$, and $CM$ computed for uni-directionally coupled
Rössler and Lorenz systems. The measures show that $X$ drives $Y$
until the onset of synchronization between the couplings $2$ and
$3$.}
\end{figure}

\newpage{}

\subsection{Rössler $1.015$ $\rightarrow$ Rössler $0.985$}

The fifth data-set comes from coupling of two Rössler systems: 
\begin{eqnarray}
\dot{x}_{1} & = & -\omega_{1}x_{2}-x_{3}\nonumber \\
\dot{x}_{2} & = & \omega_{1}x_{1}+0.15x_{2}\nonumber \\
\dot{x}_{3} & = & 0.2+x_{3}(x_{1}-10)\label{eq:rosros}\\
\dot{y}_{1} & = & -\omega_{2}y_{2}-y_{3}+C(x_{1}-y_{1})\nonumber \\
\dot{y}_{2} & = & \omega_{2}y_{1}+0.15y_{2}\nonumber \\
\dot{y}_{3} & = & 0.2+y_{3}(y_{1}-10)\nonumber 
\end{eqnarray}

The parameters $\omega_{1}$, $\omega_{2}$ were set to: 
\[
\omega_{1}=1.015,\quad\omega_{2}=0.985.
\]

\begin{figure}[H]
\begin{centering}
\includegraphics[width=11.6cm,height=4.2cm]{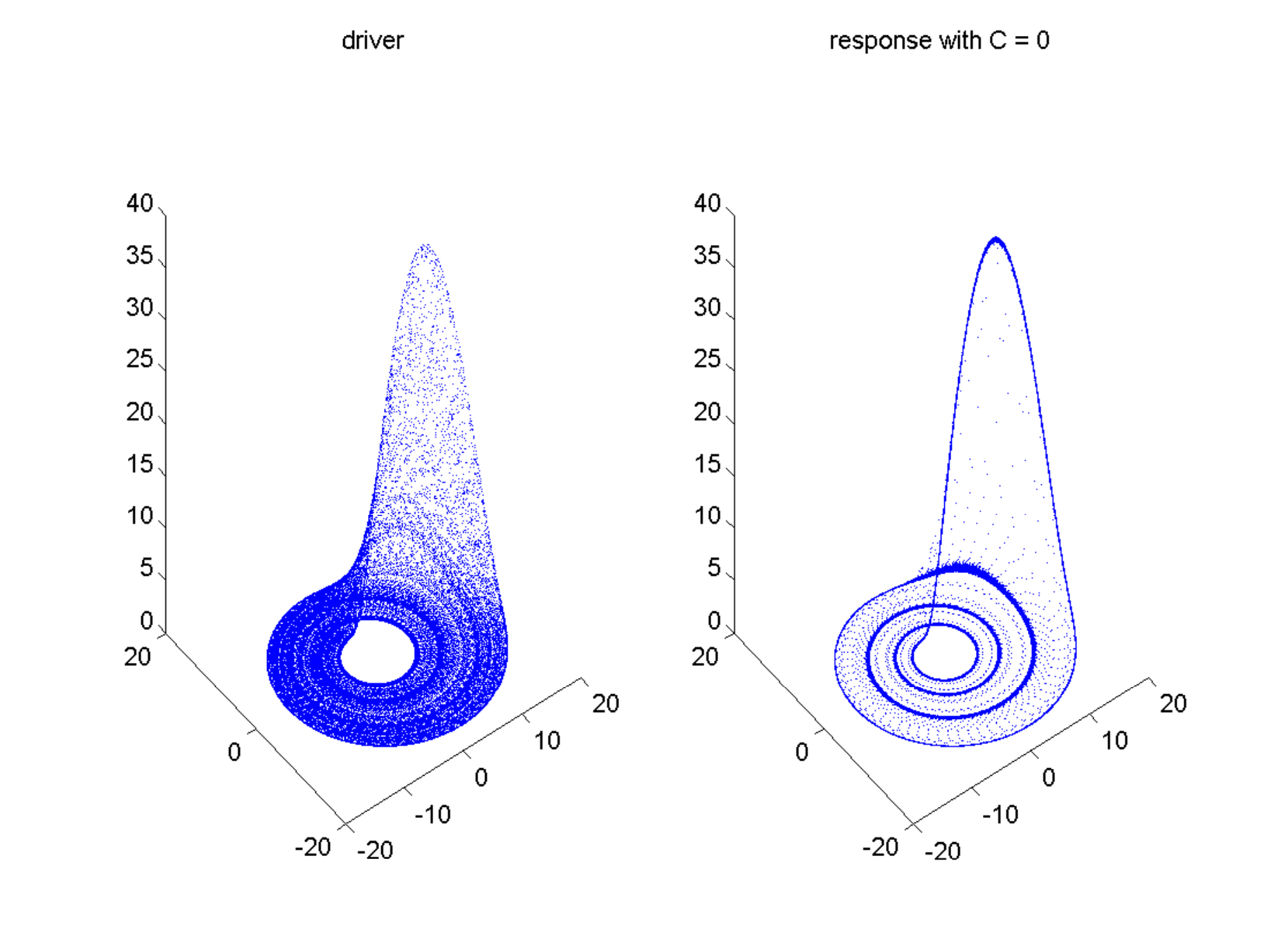}
\par\end{centering}

\centering{}\includegraphics[width=12cm,height=10.4cm]{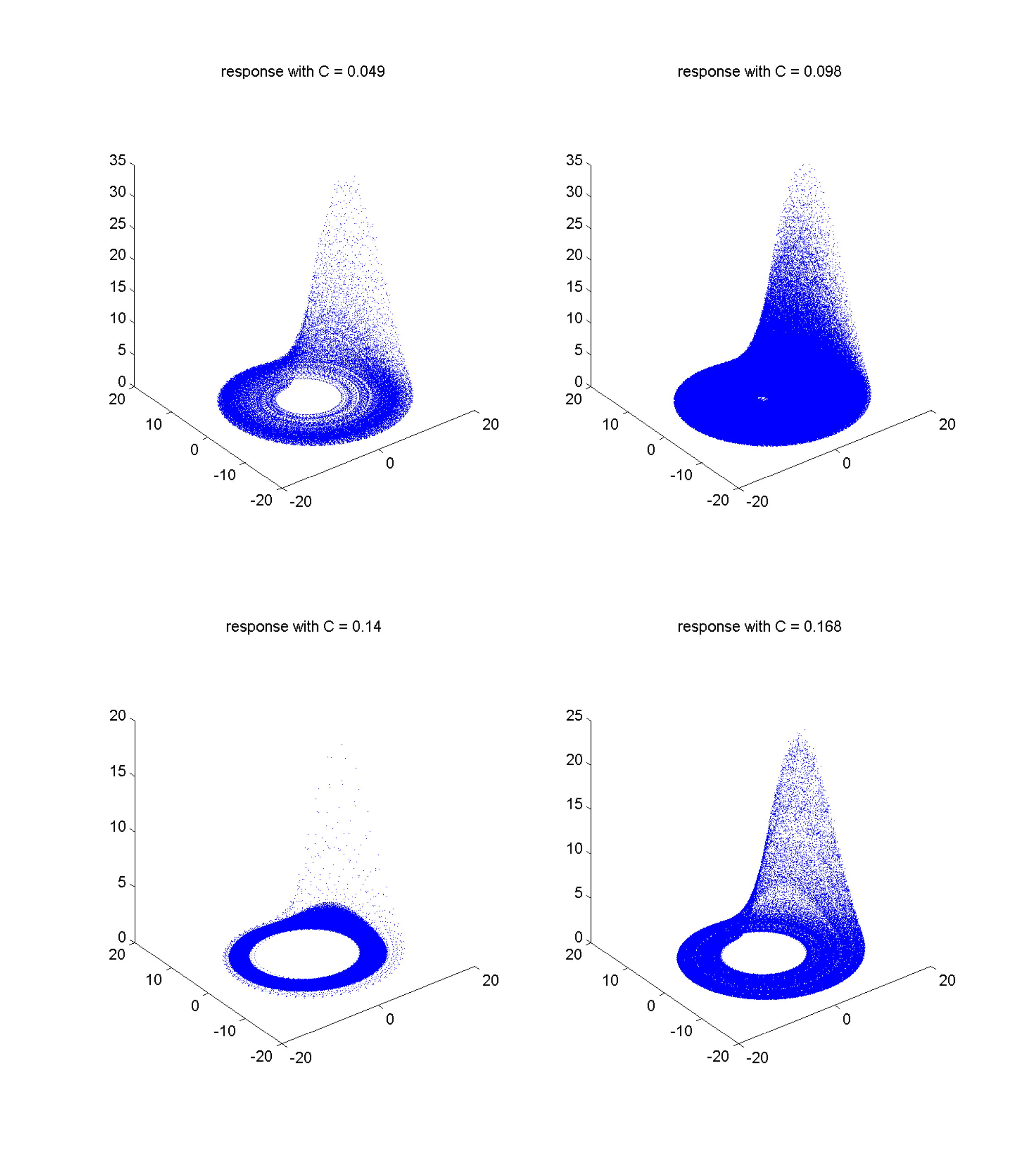}\protect\caption{Rössler system ($\omega_{1}=1.015$) driving another Rössler system
($\omega_{2}=0.985$). Attractors of the driver and of response system
for various couplings.}
\end{figure}

The next interaction graph shows that the two Rössler systems are
coupled through one-way driving relationship between variables $x_{1}$
and $y_{1}$. This causal link is what we would like to recover.

\usetikzlibrary{arrows,positioning} \newdimen\nodeDist \nodeDist=2cm
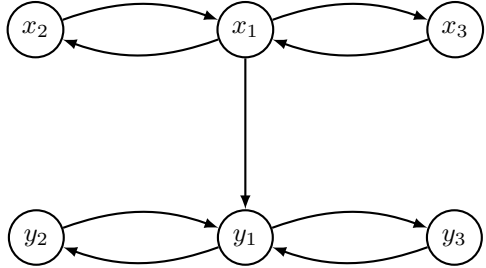
\begin{figure}[H] \centering \begin{tikzpicture}[->,>=latex,thick,node distance=\nodeDist,main node/.style={circle,draw} ]
   \node[main node] (x1) {$x_1$};    \node[main node] (x2) [left= of x1] {$x_2$};   \node[main node] (x3) [right= of x1] {$x_3$};   \node[main node] (y1) [below= of x1] {$y_1$};   \node[main node] (y2) [left= of y1] {$y_2$};   \node[main node] (y3) [right= of y1] {$y_3$};      \path[every node]     (x1) edge node {}  (y1)         edge [bend left=\edgeAngel ] node {} (x2)         edge [bend left=\edgeAngel ] node {} (x3)     (x2) edge [bend left=\edgeAngel ] node {} (x1)     (x3) edge [bend left=\edgeAngel ] node {} (x1)     (y1) edge [bend left=\edgeAngel ] node {} (y2)        edge [bend left=\edgeAngel ] node {} (y3)     (y2) edge [bend left=\edgeAngel ] node {} (y1)     (y3) edge [bend left=\edgeAngel ] node {} (y1); \end{tikzpicture}\caption{Interaction graph for the coupling of two Rössler systems.} \end{figure}The data were generated by Matlab solver of ordinary differential
equations ode45. The starting point was $[0,0,0.4,0,0,0.4].$ First
$1000$ data points were thrown away. The total number of obtained
data was $100000$ at an integration step size $0.1.$ This gives
about $60$ samples per one average orbit around the attractor. The
coupling strength $C$ was chosen from $0$ to $0.2$ with the step
$0.01.$ The same system was used in \cite{palus2007}, \cite{vejme2008},
\cite{palus2014}.

The plots of the conditional Lyapunov exponents for this Rössler-Rössler
system can be found in \cite{palus2007}. They show, similarly as
the following graph of $D_{2}$ estimates, that synchronization takes
place between couplings $0.11$ and $0.13$.

\begin{figure}[H]
\centering{}\includegraphics[width=10cm]{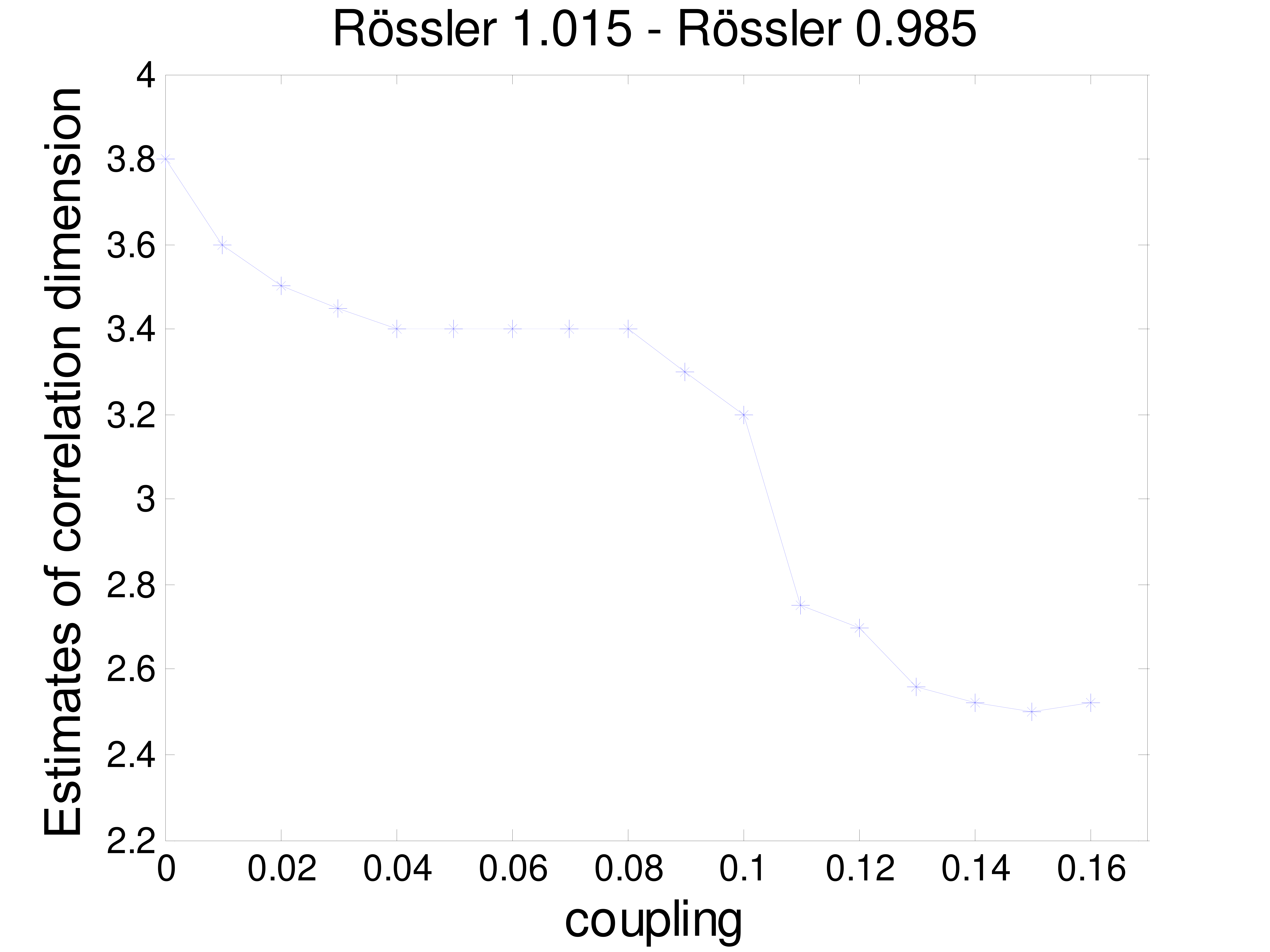}\protect\caption{Correlation dimension estimates for two Rössler systems (\ref{eq:rosros})
connected with different coupling strengths.}
\label{D2rosros}
\end{figure}

\subsubsection*{Results of causality detection using reconstructed manifolds}

Suppose that we know $50000$ data-points of variable $x_{1}$ of
the driving Rössler system and variable $y_{1}$ of the responsive
Rössler system and we would like to know whether there is a causal
relationship between the two systems. In order to use the state-space
based methods of search for causality we made reconstructions of the
state portraits. To this end, we used time-delayed vectors of $x_{1}$
and $y_{1}$ with time delay equal to $3$ and embedding dimension
of $7$. Methods for all three measures used $8$ nearest neighbors.

\begin{figure}[H]
\begin{centering}
\includegraphics[bb=50bp 190bp 580bp 620bp,clip,width=10cm,height=6.7cm]{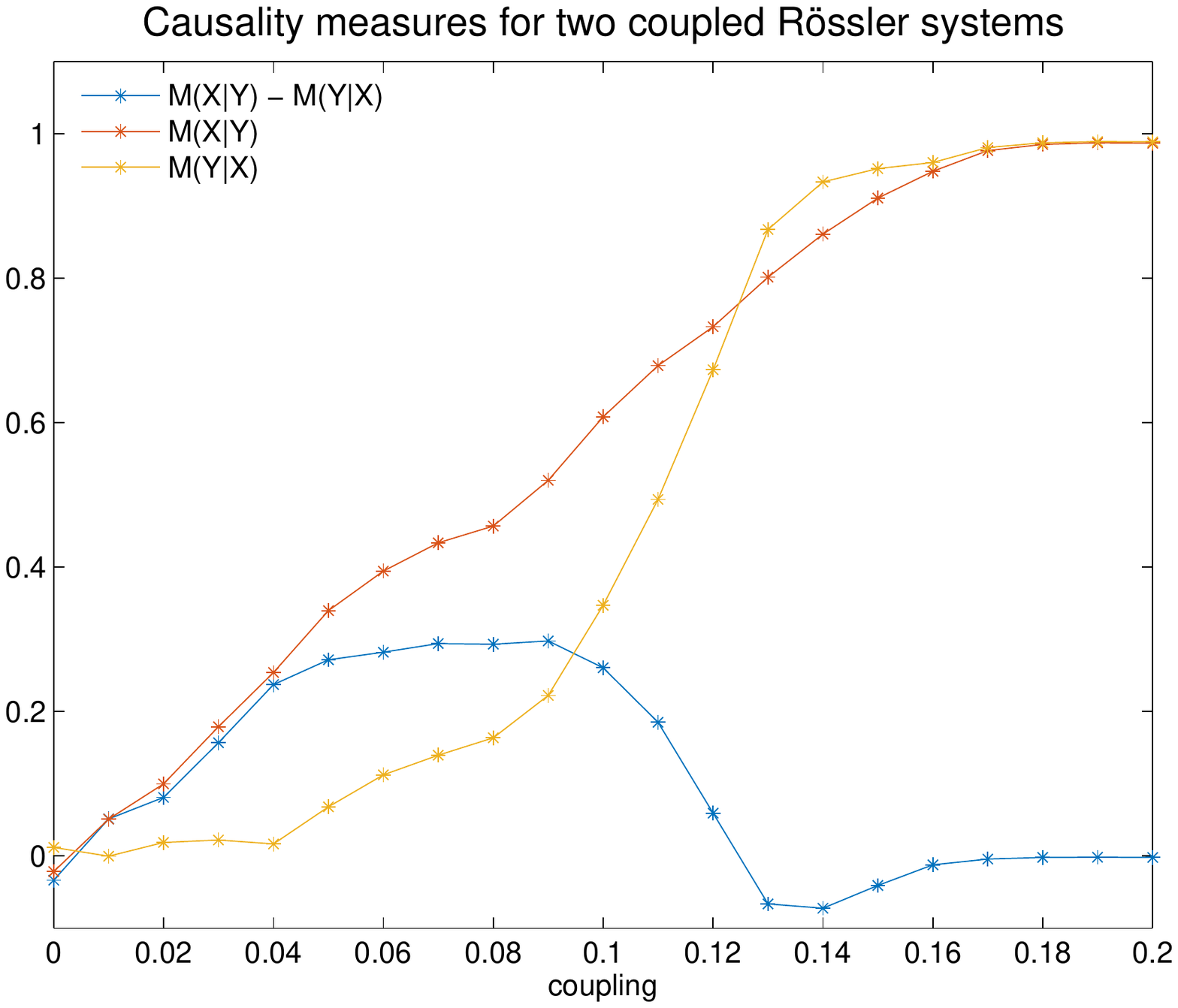}
\par\end{centering}

\medskip{}

\begin{centering}
\includegraphics[bb=50bp 190bp 580bp 620bp,clip,width=10cm,height=6.7cm]{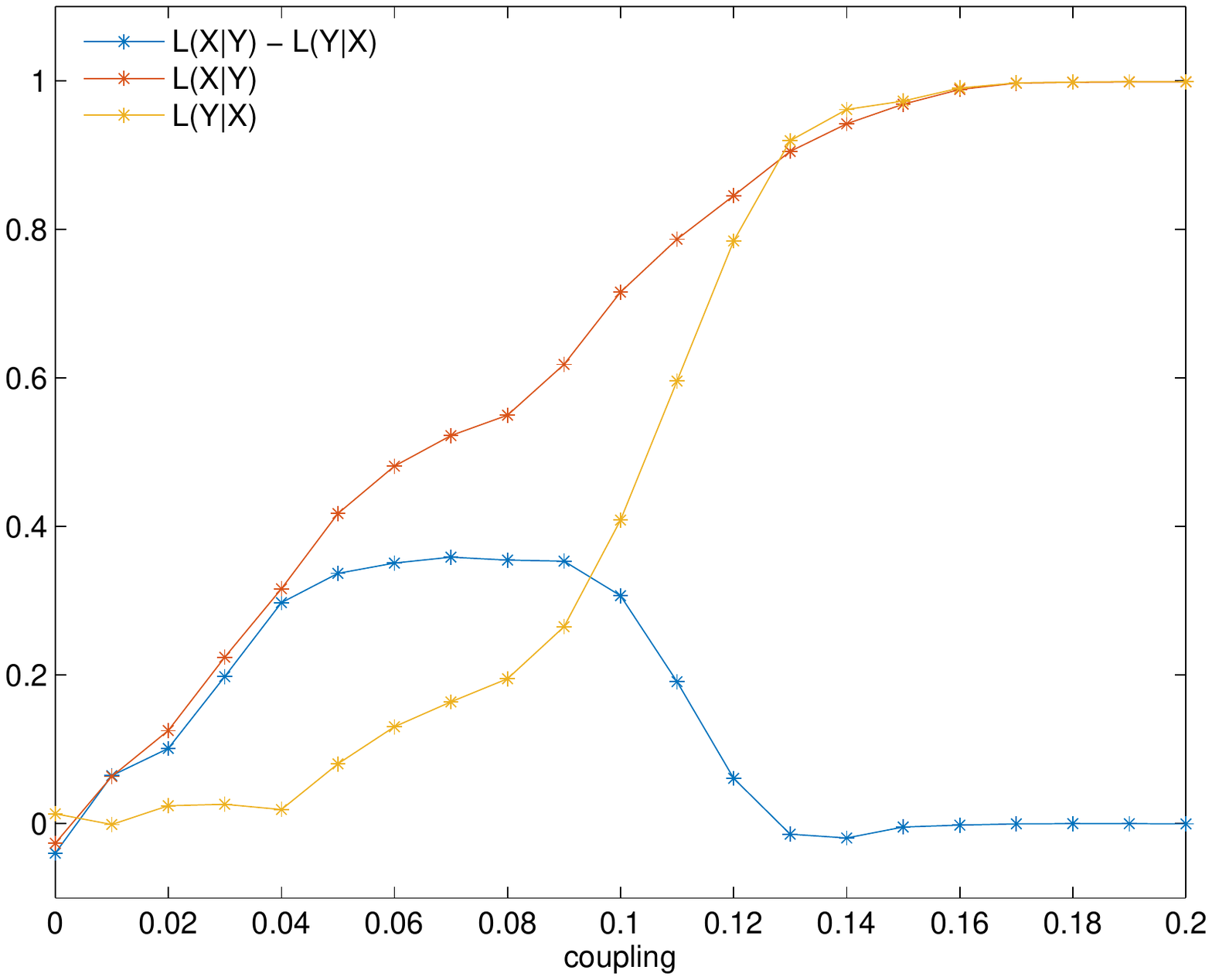}
\par\end{centering}

\medskip{}

\begin{centering}
\includegraphics[bb=50bp 190bp 580bp 620bp,clip,width=10cm,height=6.7cm]{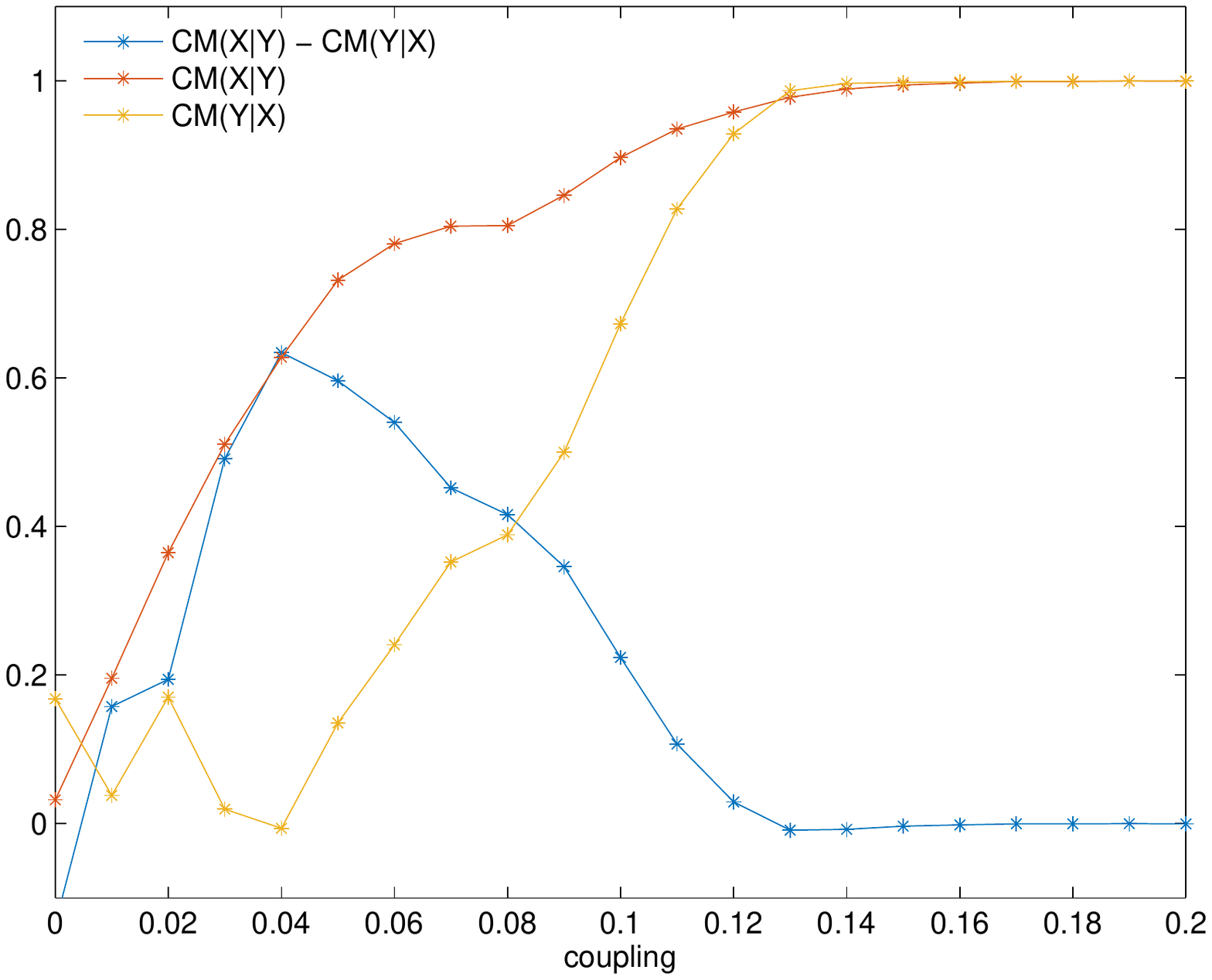}
\par\end{centering}

\protect\caption{Measures $M$, $L$, and $CM$ computed for two uni-directionally
coupled Rössler systems (\ref{eq:rosros}). The measures show that
$X$ drives $Y$ until the onset of synchronization between couplings
$0.11$ and $0.13$.}
\end{figure}

\newpage{}

\subsection{Rössler $0.5$ $\rightarrow$ Rössler $2.515$}

Another example of coupled Rössler systems:
\begin{eqnarray}
\dot{x}_{1} & = & -\omega_{1}x_{2}-x_{3}\nonumber \\
\dot{x}_{2} & = & \omega_{1}x_{1}+a_{1}x_{2}\nonumber \\
\dot{x}_{3} & = & 0.2+x_{3}(x_{1}-10)\label{eq:rosros15}\\
\dot{y}_{1} & = & -\omega_{2}y_{2}-y_{3}+C(x_{1}-y_{1})\nonumber \\
\dot{y}_{2} & = & \omega_{2}y_{1}+a_{2}y_{2}\nonumber \\
\dot{y}_{3} & = & 0.2+y_{3}(y_{1}-10)\nonumber 
\end{eqnarray}

The parameters are set to: 
\[
\omega_{1}=0.5,\quad\omega_{2}=2.515,\quad a_{1}=0.15,\quad a_{2}=0.72.
\]

\begin{figure}[H]

\begin{centering}
\includegraphics[bb=30bp 130bp 570bp 700bp,clip,width=13cm,height=15cm]{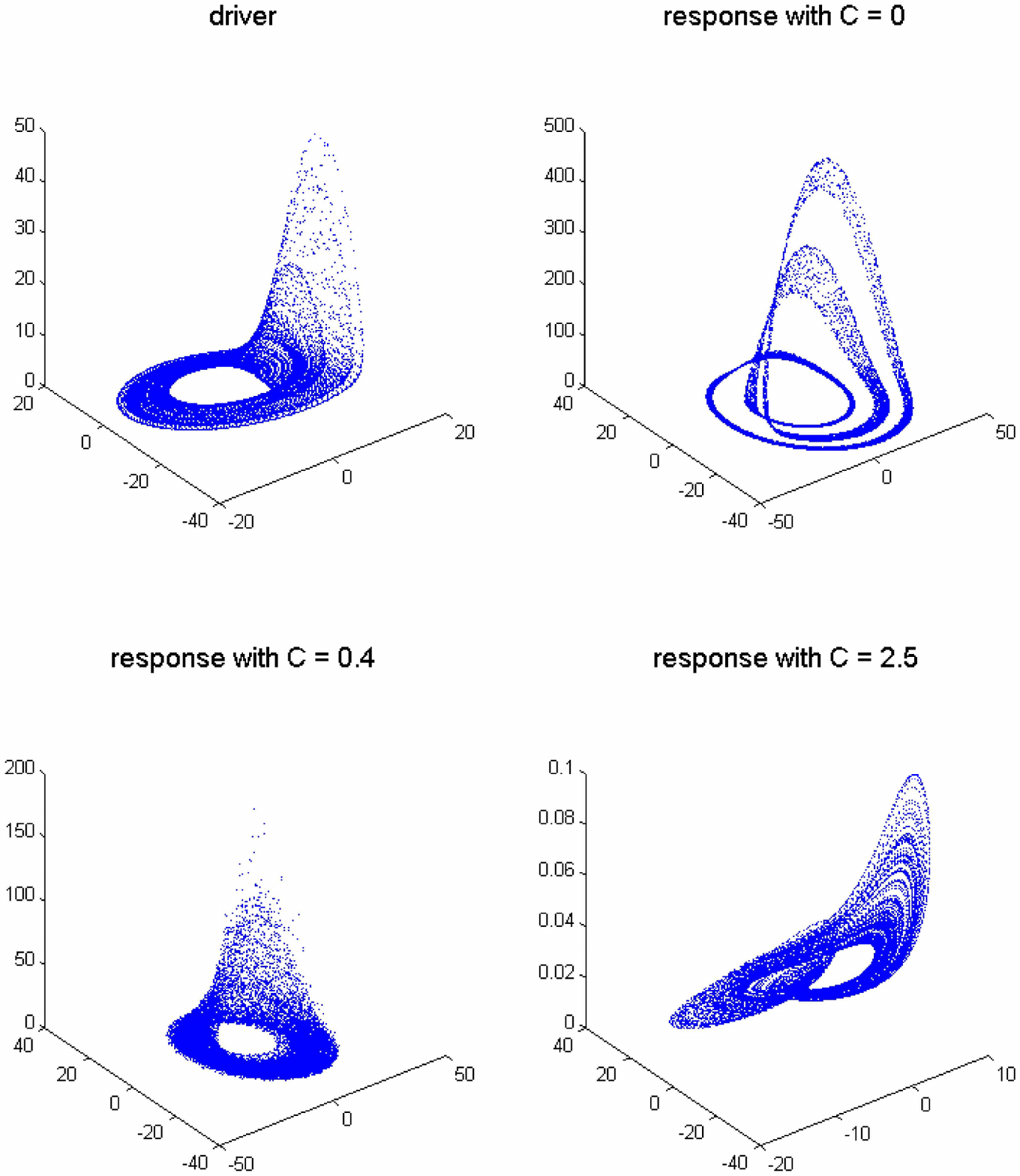}\protect\caption{Rössler system ($\omega_{1}=0.5$) driving Rössler system ($\omega_{2}=2.515$).
Attractors of the driver and of response system for various couplings.}

\par\end{centering}

\end{figure}

The frequency ratio of about $1:5$ used in this example reminds cardio-respiratory
interactions. The system was used in \cite{palus2007} and in \cite{vejme2008}
to show that in this case the problem of detecting directionality
is much more challenging than detecting directionality in two systems
with similar dynamics. 

The data were generated by Matlab solver of ordinary differential
equations ode45. The coupling strength $C$ was chosen from $0$ to
$2.5$ with step size $0.1.$ The starting point was $[0,0,0.4,0,0,0.4].$
First $1000$ data points were thrown away. The total number of obtained
data was $100000$ at an integration step size $0.1.$ 

The variables of the coupled systems may be arranged into the same
interaction graph as the previous example. The two connected Rössler
systems represent distinct dynamical subsystems coupled through one-way
driving relationship between variables $x_{1}$ and $y_{1}$. This
causal link is what we tried to uncover.

The plot of the $D_{2}$ estimates shows that synchronization takes
place at coupling of about $1$.

\begin{figure}[H]
\centering{}\includegraphics[bb=0bp 400bp 595bp 820bp,clip,width=10cm]{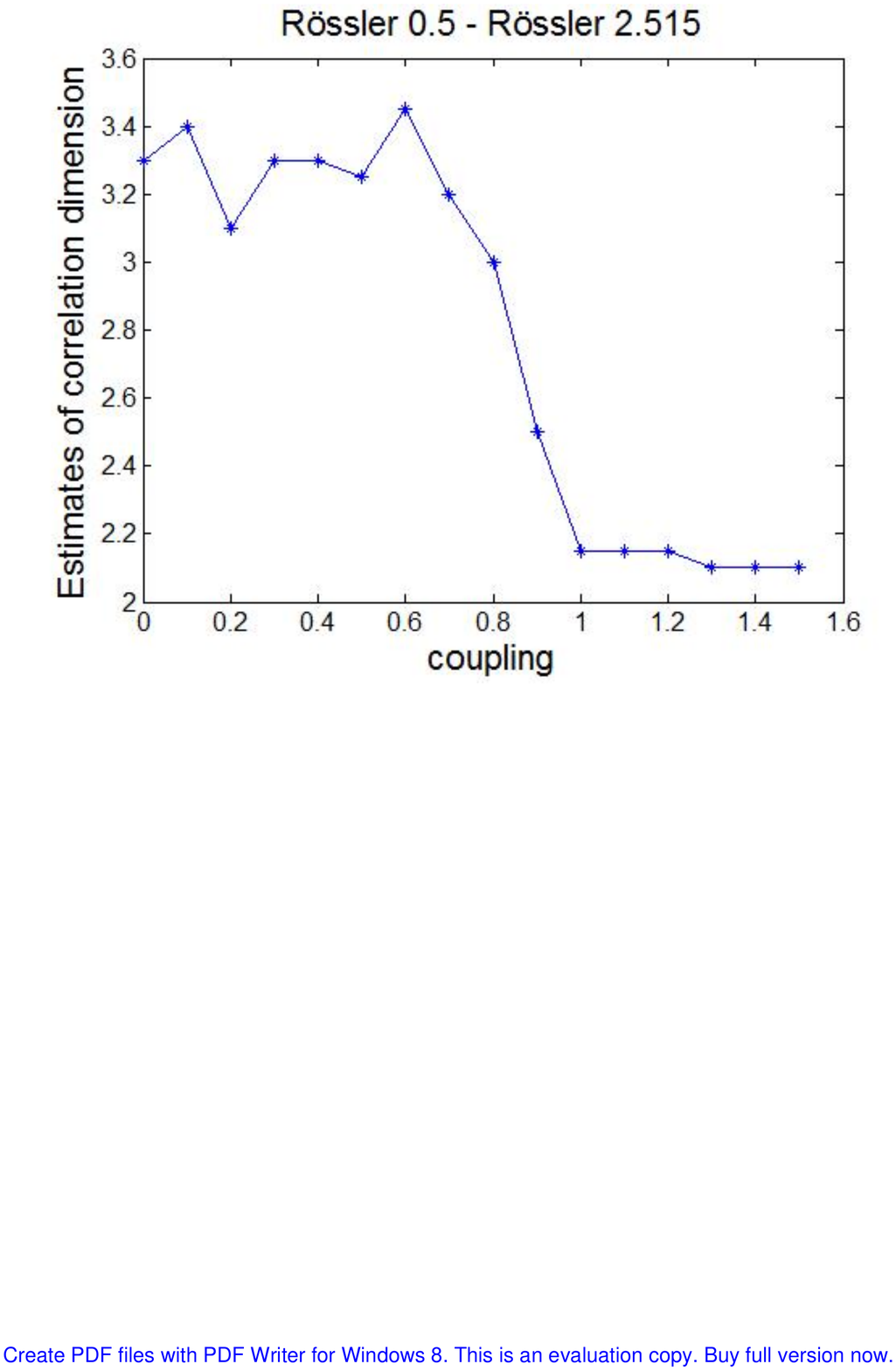}\protect\caption{Correlation dimension estimates for two Rössler systems (\ref{eq:rosros15})
connected with different coupling strengths.}
\label{D2rosros15}
\end{figure}

\subsubsection*{Results of causality detection using reconstructed manifolds}

Suppose that we know $50000$ data-points of variable $x_{1}$ of
the driving Rössler system and variable $y_{1}$ of the responsive
Rössler system and we would like to know whether there is a causal
relationship between the two systems. 

In order to use the state-space based methods of search for causality
we made reconstructions of the state portraits. To this end, we used
time-delayed vectors of $x_{1}$ and $y_{1}$ with time delay equal
to $1$ and embedding dimension of $7$. Methods for all three measures
used $8$ nearest neighbors.

\begin{figure}[H]
\begin{centering}
\includegraphics[bb=50bp 190bp 580bp 620bp,clip,width=10cm,height=6.7cm]{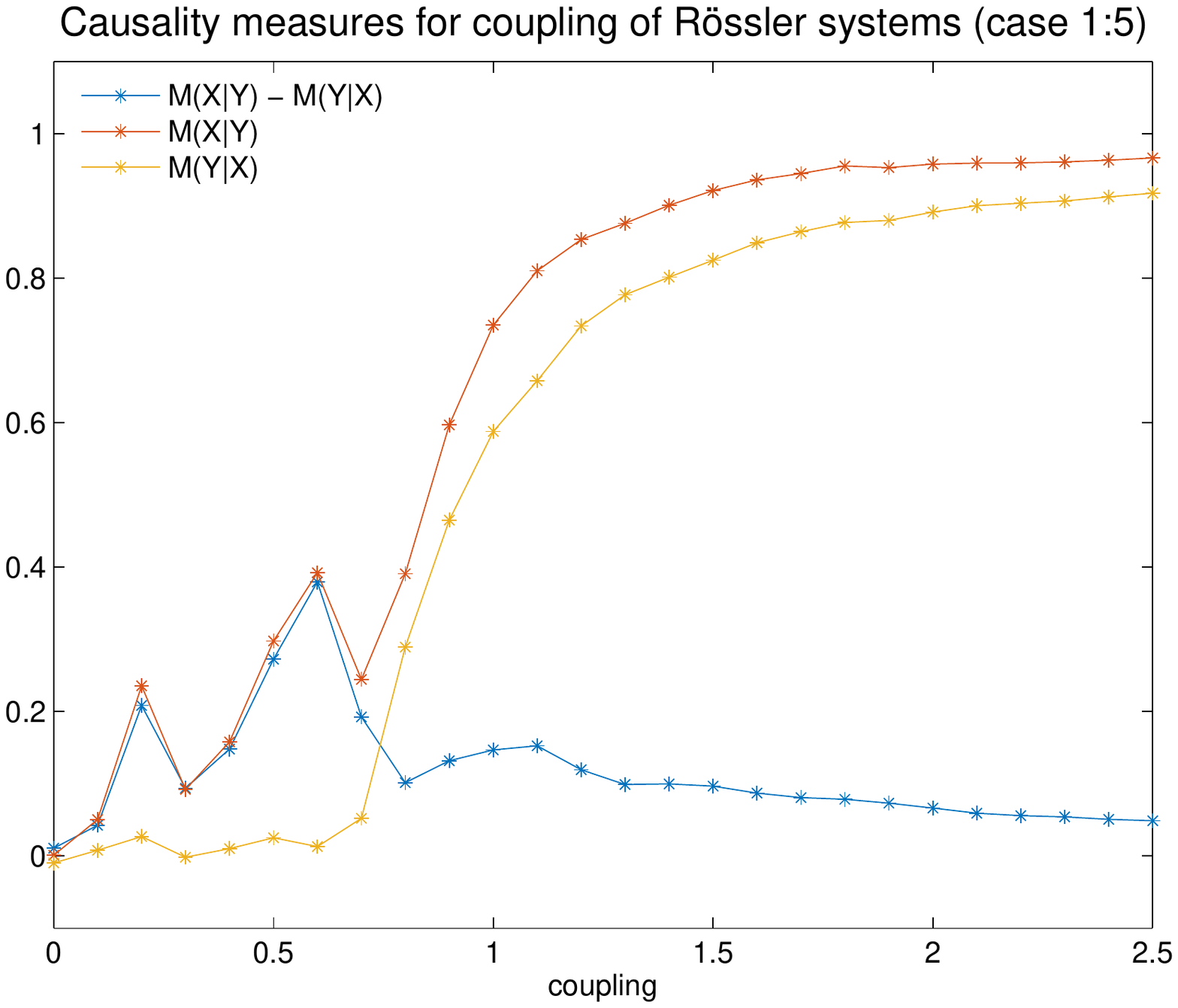}
\par\end{centering}

\medskip{}

\begin{centering}
\includegraphics[bb=50bp 190bp 580bp 620bp,clip,width=10cm,height=6.7cm]{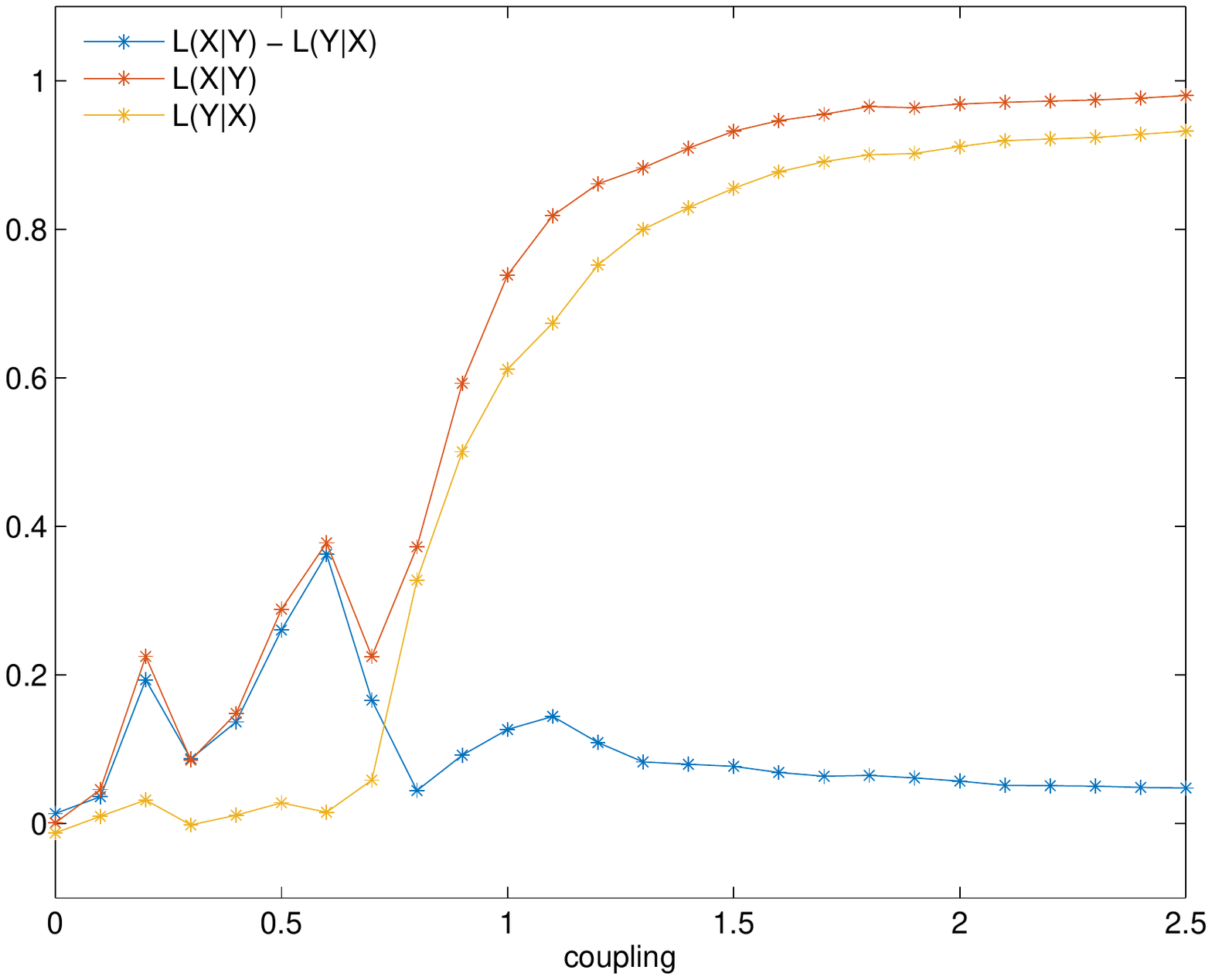}
\par\end{centering}

\medskip{}

\begin{centering}
\includegraphics[bb=50bp 190bp 580bp 620bp,clip,width=10cm,height=6.7cm]{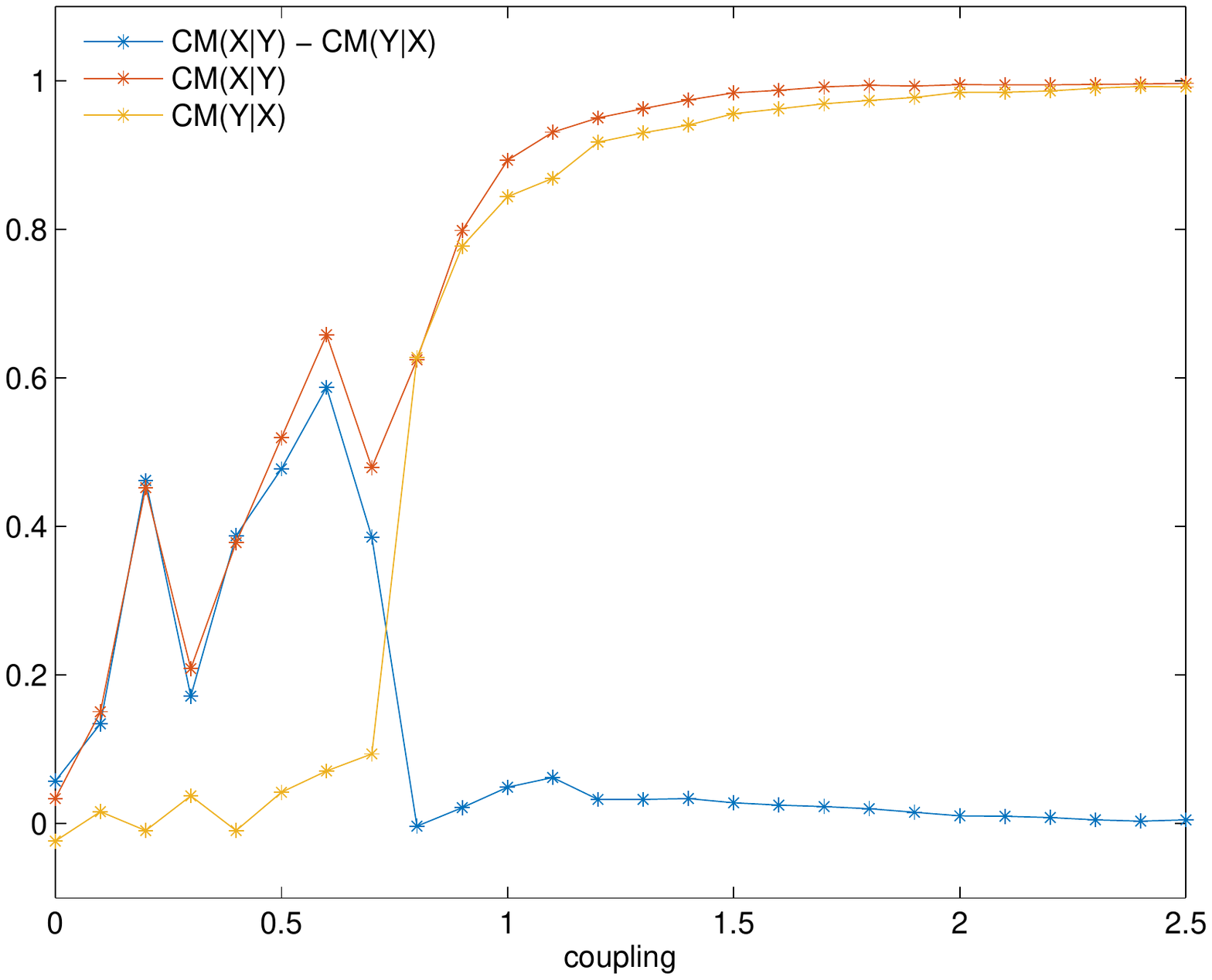}
\par\end{centering}

\protect\caption{Measures $M$, $L$, and $CM$ computed for two uni-directionally
coupled Rössler systems (\ref{eq:rosros15}). The measures show that
$X$ drives $Y$ until the onset of synchronization at coupling of
about $1$.}
\end{figure}

\newpage{}

\subsection{Rössler $2.515$ $\rightarrow$ Rössler $0.5$}

In this example, in reverse to the preceding case, the direction of
coupling is from the faster Rössler system to the slower system: 

\begin{eqnarray}
\dot{x}_{1} & = & -\omega_{1}x_{2}-x_{3}\nonumber \\
\dot{x}_{2} & = & \omega_{1}x_{1}+a_{1}x_{2}\nonumber \\
\dot{x}_{3} & = & 0.2+x_{3}(x_{1}-10)\label{eq:rosros51}\\
\dot{y}_{1} & = & -\omega_{2}y_{2}-y_{3}+C(x_{1}-y_{1})\nonumber \\
\dot{y}_{2} & = & \omega_{2}y_{1}+a_{2}y_{2}\nonumber \\
\dot{y}_{3} & = & 0.2+y_{3}(y_{1}-10)\nonumber 
\end{eqnarray}

The parameters were set to:
\[
\omega_{1}=2.515,\quad\omega_{2}=0.5,\quad a_{1}=0.72,\quad a_{2}=0.15.
\]

The data were generated by Matlab solver of ordinary differential
equations ode45. The values of coupling strength $C$ were chosen
from $0$ to $0.5$. The starting point was $[0,0,0.4,0,0,0.4].$
First $2000$ points were thrown away. $100000$ data points obtained
at an integration step size $0.1$ were saved. The same system was
used in \cite{palus2007} and in \cite{vejme2008}.

\begin{figure}[H]
\begin{centering}
\includegraphics[bb=30bp 180bp 570bp 700bp,clip,width=13cm,height=12cm]{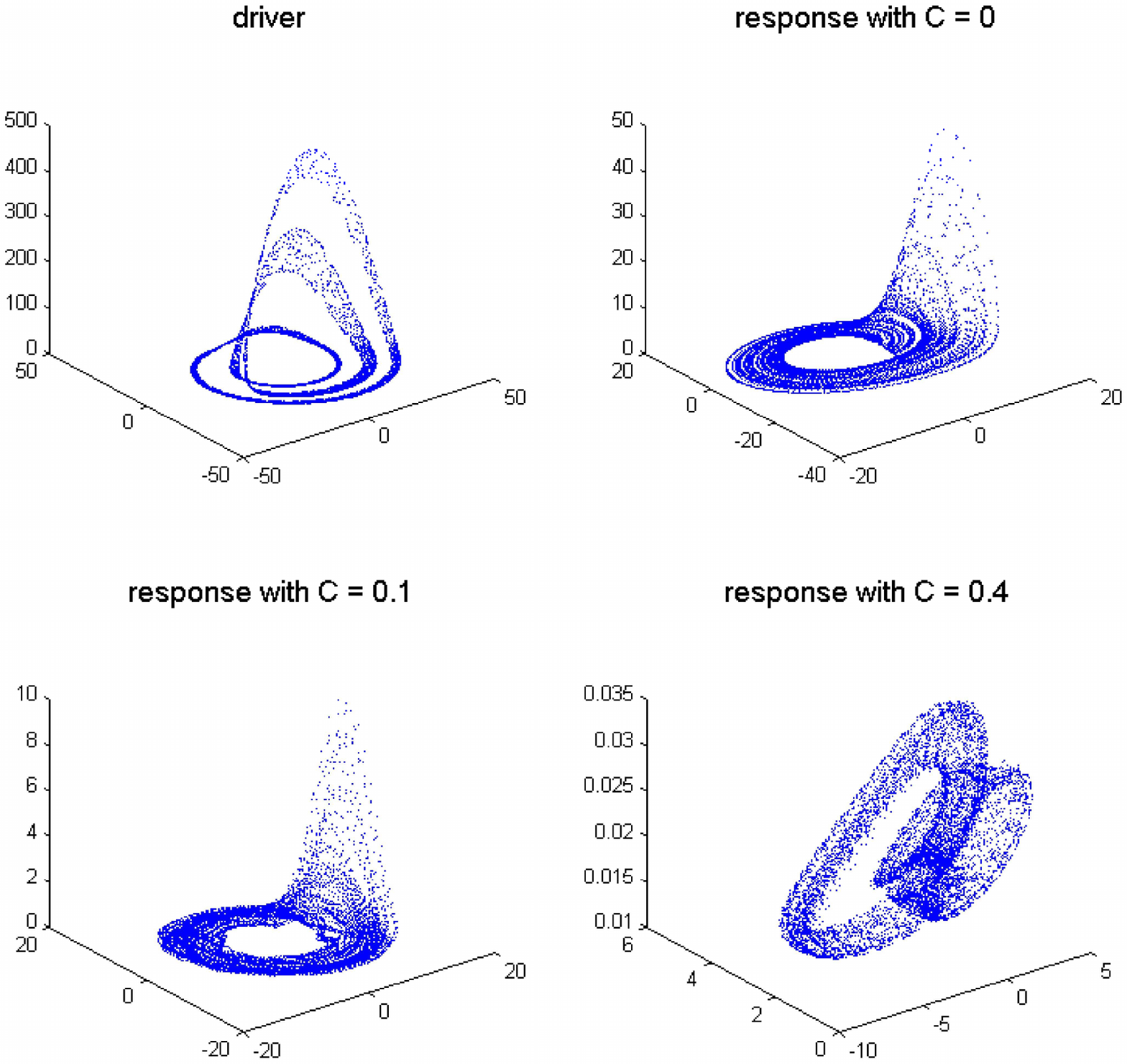}
\par\end{centering}

\protect\caption{Rössler ($\omega_{1}=0.5$) driving Rössler ($\omega_{2}=2.515$).
Attractors of the driver and of response system for various couplings.}
\end{figure}

The variables of the coupled systems may be once again arranged into
the interaction graph shown in Figure 17. It means that the two connected
Rössler systems represent distinct dynamical subsystems coupled through
one-way driving relationship between variables $x_{1}$ and $y_{1}$.
This causal link is what we would like to recover.

\begin{figure}[H]
\centering{}\includegraphics[bb=30bp 400bp 570bp 830bp,clip,width=10cm]{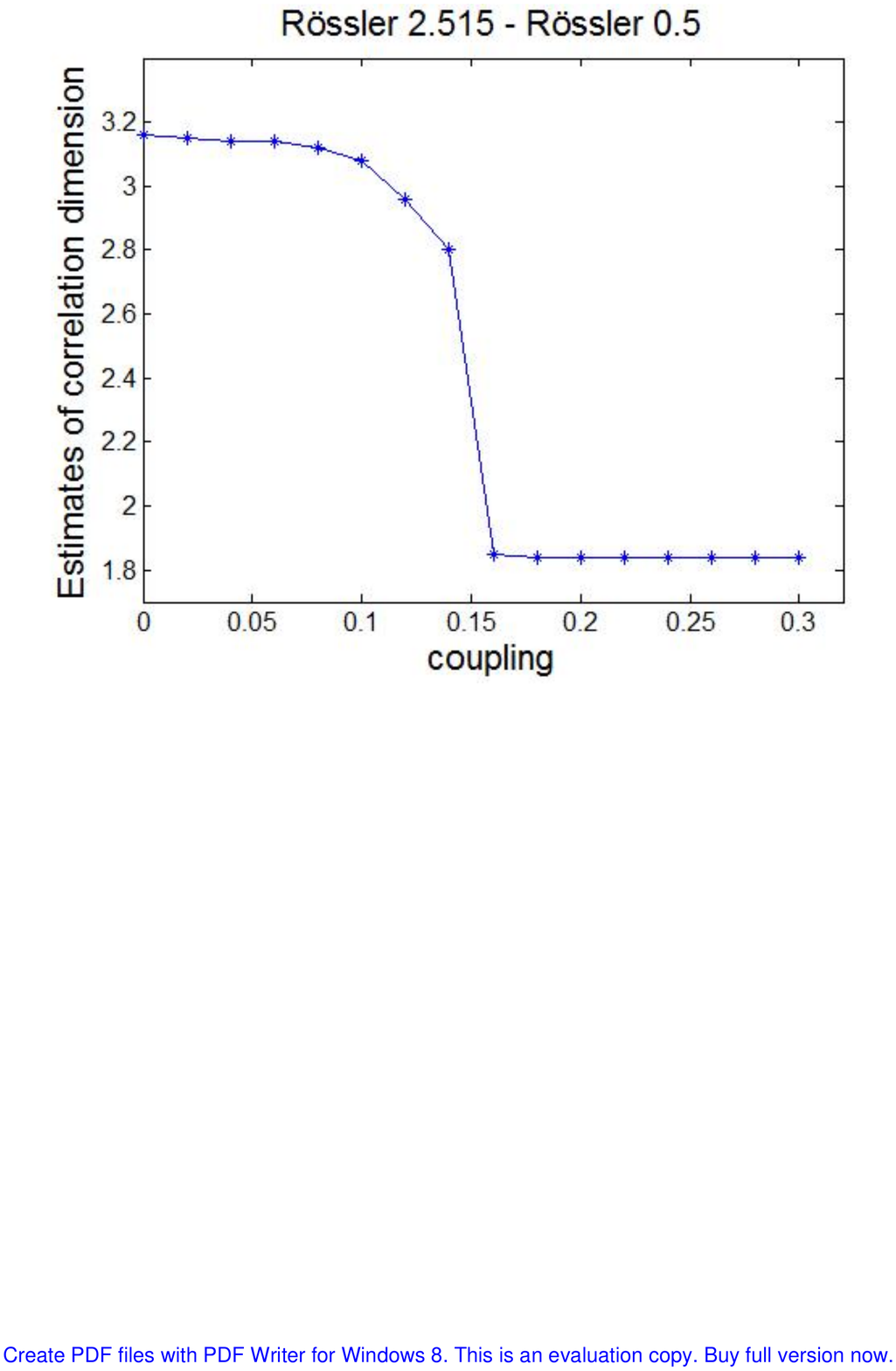}\protect\caption{Correlation dimension estimates for two Rössler systems (\ref{eq:rosros51})
connected with different coupling strengths.}
\label{D2rosros51}
\end{figure}

\subsubsection*{Results of causality detection using reconstructed manifolds}

Suppose that we know $50000$ data-points of variable $x_{1}$ of
the driving Rössler system and variable $y_{1}$ of the responsive
Rössler system and we would like to know whether there is a causal
relationship between the two systems. 

In order to use the state-space based methods of search for causality
we made reconstructions of the state portraits with the same parameters
as in the previous case. This means time-delayed vectors of $x_{1}$
and $y_{1}$ with delay equal to $1$ and embedding dimension of $7$.
Methods for all three measures used $8$ nearest neighbors.

\begin{figure}[H]
\begin{centering}
\includegraphics[bb=50bp 190bp 580bp 620bp,clip,width=10cm,height=6.7cm]{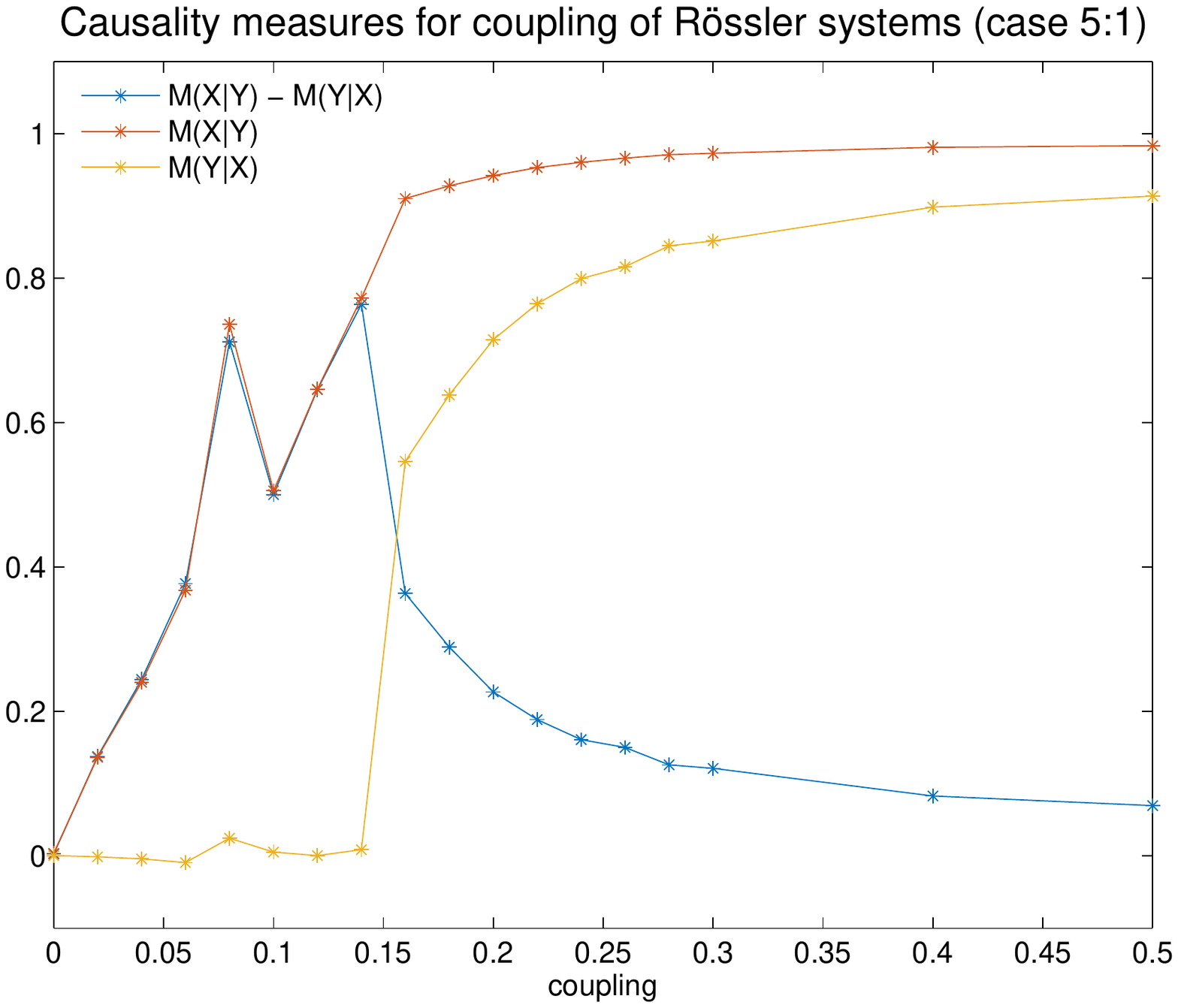}
\par\end{centering}

\medskip{}

\begin{centering}
\includegraphics[bb=50bp 190bp 580bp 620bp,clip,width=10cm,height=6.7cm]{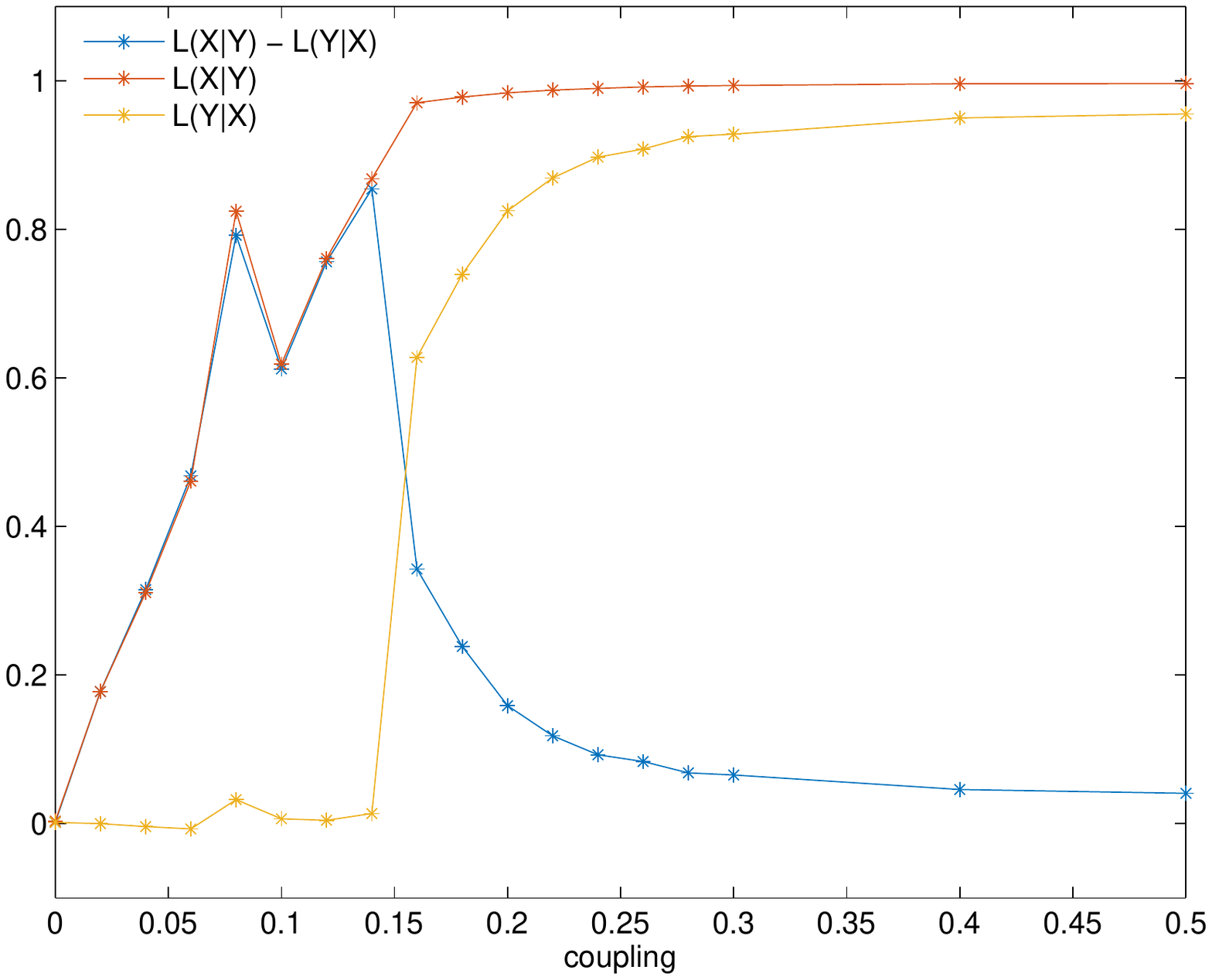}
\par\end{centering}

\medskip{}

\begin{centering}
\includegraphics[bb=50bp 190bp 580bp 620bp,clip,width=10cm,height=6.7cm]{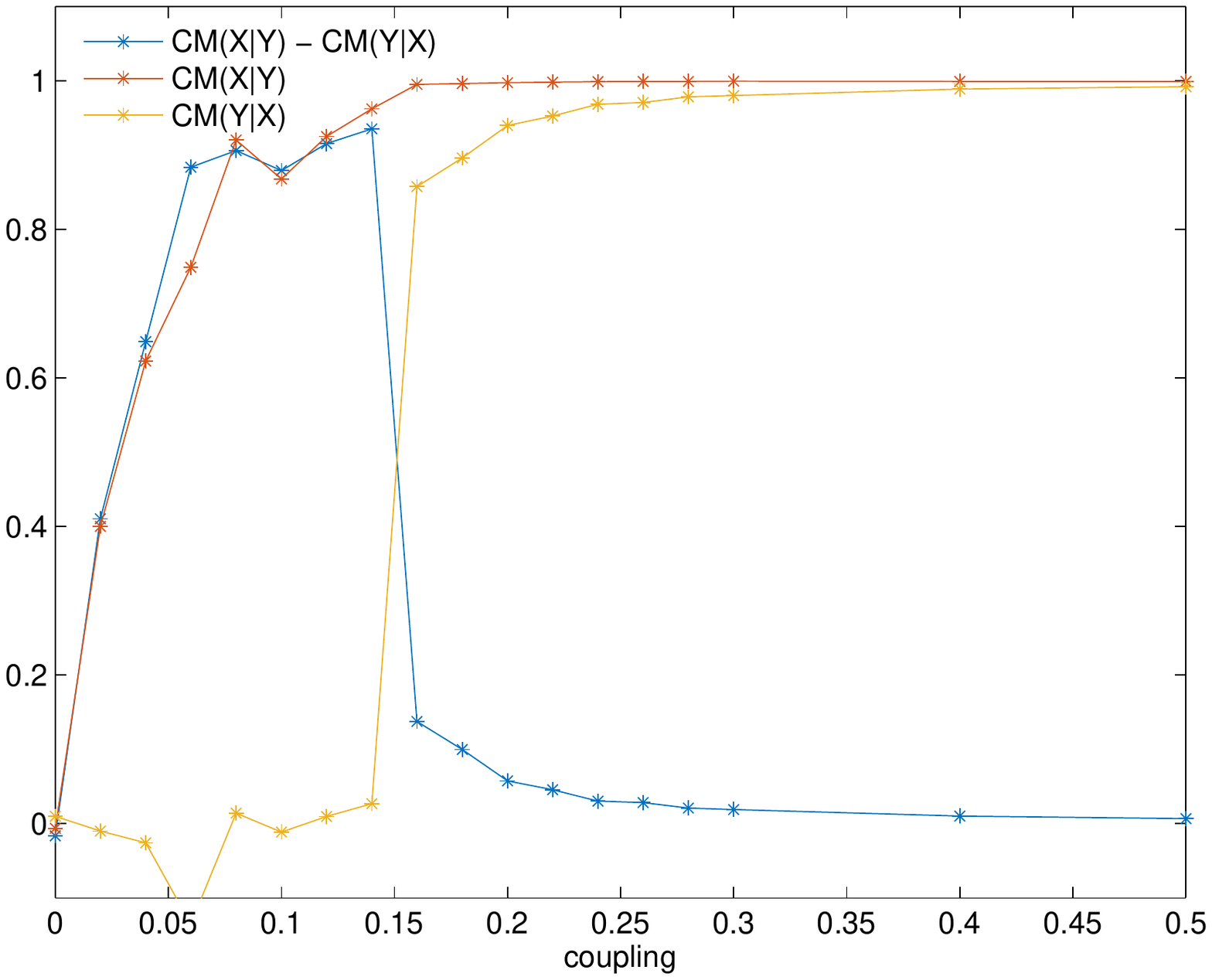}
\par\end{centering}

\protect\caption{Measures $M$, $L$, and $CM$ computed for two uni-directionally
coupled Rössler systems (\ref{eq:rosros51}). The measures show that
$X$ drives $Y$ until the onset of synchronization between couplings
$0.15$ and $0.2$.}
\end{figure}

\newpage{}

\subsection{Lorenz $28.5$ $\rightarrow$ Lorenz $27.5$}

Here the first three lines correspond to the driving Lorenz system
and the last three equations characterize the response Lorenz system:
\begin{eqnarray}
\dot{x}_{1} & = & 10(-x_{1}+x_{2})\nonumber \\
\dot{x}_{2} & = & 28.5x_{1}-x_{2}-x_{1}x_{3}\nonumber \\
\dot{x}_{3} & = & x_{1}x_{2}-\frac{8}{3}x_{3}\label{eq:lorlor}\\
\dot{y}_{1} & = & 10(-y_{1}+y_{2})+C(x_{1}-y_{1})\nonumber \\
\dot{y}_{2} & = & 27.5y_{1}-y_{2}-y_{1}y_{3}\nonumber \\
\dot{y}_{3} & = & y_{1}y_{2}-\frac{8}{3}y_{3}\nonumber 
\end{eqnarray}

The data were generated by Matlab solver of ordinary differential
equations ode45. The coupling strength $C$ was chosen from $0$ to
$10.08$ with step $0.42.$ The starting point was $[0.3,0.3,0.3,0.3,0.3,0.3].$
First $20000$ data points was thrown away. The total number of obtained
data was $100000$ at an integration step size $0.01.$ 

The same system was used in \cite{andrz2011} to show that for the
Lorenz dynamics both the flow waveforms and the events derived from
them enable detection of the coupling. 

\begin{figure}[H]
\begin{centering}
\includegraphics[bb=30bp 190bp 575bp 700bp,clip,width=14cm]{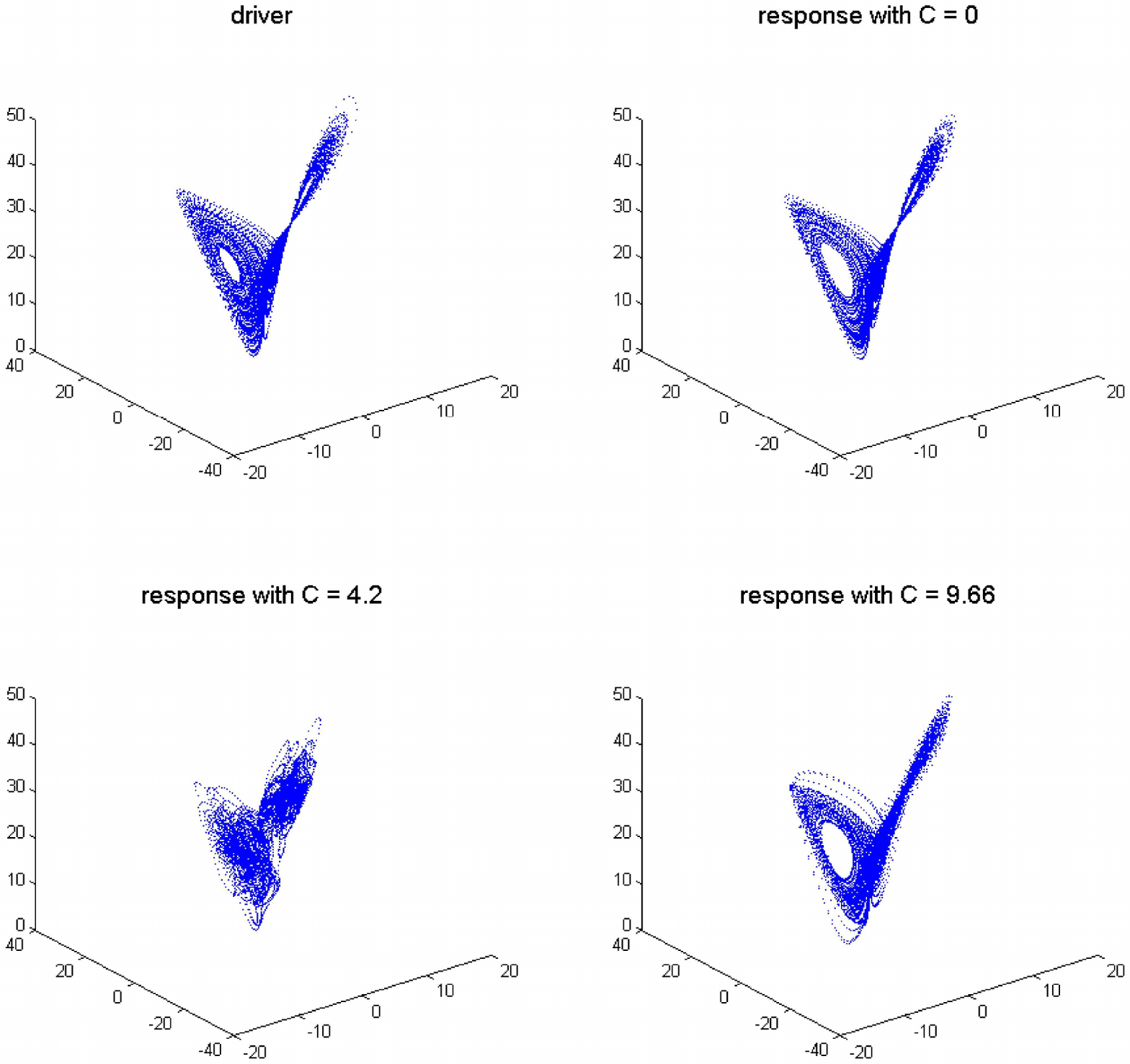}
\par\end{centering}

\protect\caption{Lorenz $28.5$ driving Lorenz $27.5$. Attractors of driver and response
system for various couplings.}
\end{figure}

The next interaction graph shows that the two Lorenz systems are coupled
through driving relationship between variables $x_{1}$ and $y_{1}$.
This causal link is what we would like to recover.

\usetikzlibrary{arrows,positioning} \newdimen\nodeDist \nodeDist=2cm
\begin{figure}[H] \centering \begin{tikzpicture}[->,>=latex,thick,node distance=\nodeDist,main node/.style={circle,draw} ]
   \node[main node] (x1) {$x_1$};    \node[main node] (x2) [right= of x1]{$x_2$};  \node[main node] (x3) [above right= sqrt(3)/2*\nodeDist  and .5*\nodeDist of x1] {$x_3$};   \node[main node] (y1) [right= of y2] {$y_1$};   \node[main node] (y2) [below= of x2] {$y_2$};   \node[main node] (y3) [below right= sqrt(3)/2*\nodeDist  and .5*\nodeDist of y1] {$y_3$};      \path[every node]     (x1) edge [bend left=\edgeAngel ] node {} (x2)     edge  node {} (x3)     edge node {}  (y1)     (x2) edge [bend left=\edgeAngel ] node {} (x3)     edge [bend left=\edgeAngel ] node {} (x1)     (x3) edge [bend left=\edgeAngel ] node {} (x2)     (y1) edge [bend left=\edgeAngel ] node {} (y2)     edge  node {} (y3)     (y2) edge [bend left=\edgeAngel ] node {} (y1)     edge [bend left=\edgeAngel ] node {} (y3)     (y3) edge [bend left=\edgeAngel ] node {} (y2); \end{tikzpicture}\caption{Interaction graph for the coupling of two Lorenz systems.} \end{figure}
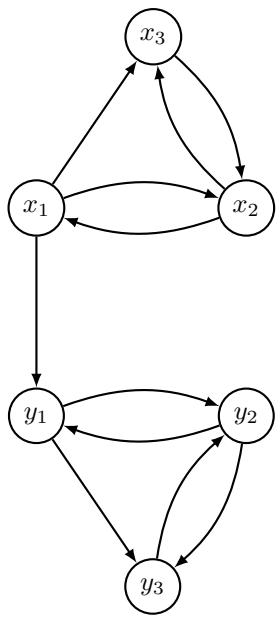

\begin{figure}[H]
\centering{}\includegraphics[bb=30bp 400bp 560bp 830bp,clip,width=10cm]{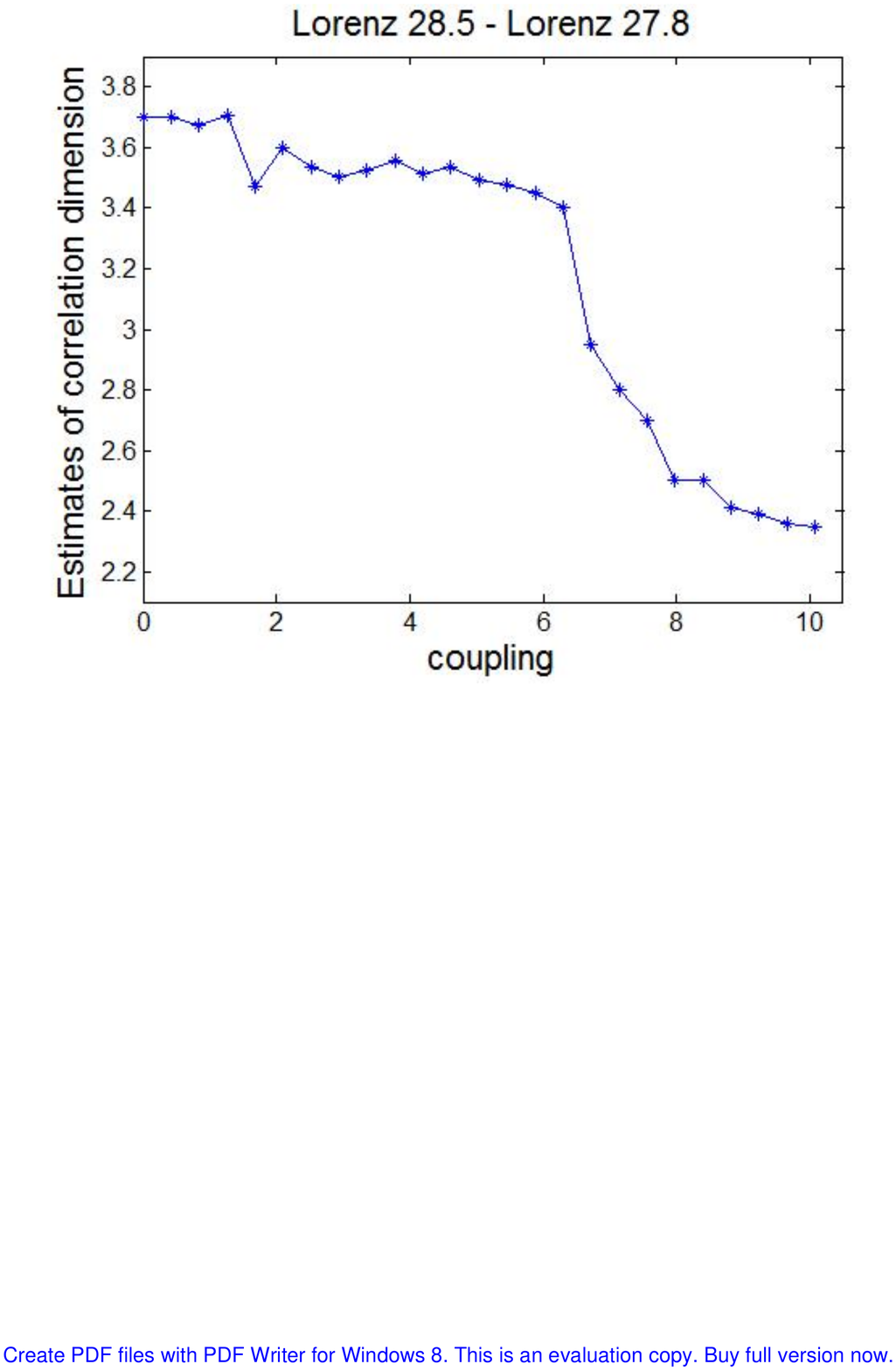}\protect\caption{Correlation dimension of connected Lorenz systems (\ref{eq:lorlor})
for different coupling strengths.}
\label{D2lorlor1}
\end{figure}

\subsubsection*{Results of causality detection using reconstructed manifolds}

Let us have $50000$ data-points of variable $x_{1}$ and variable
$y_{1}.$ In order to use the state-space based methods of search
for causality we made reconstructions of the state portraits. To this
end, we used time-delayed vectors of $x_{1}$ and $y_{1}$ with time
delay equal to $1$ and embedding dimension of $7$. Methods for all
three measures used $8$ nearest neighbors.

\begin{figure}[H]
\begin{centering}
\includegraphics[bb=50bp 190bp 580bp 620bp,clip,width=10cm,height=6.7cm]{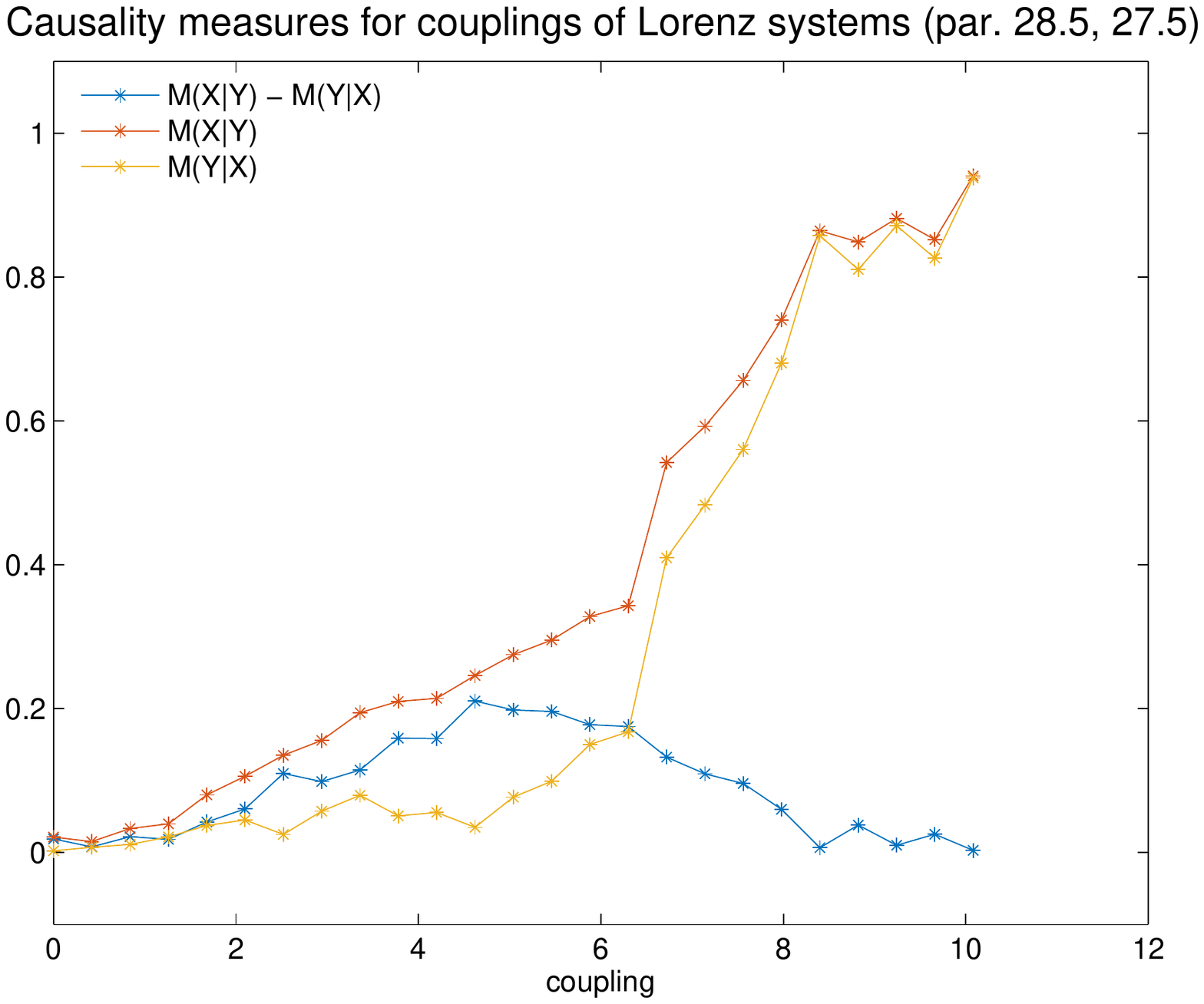}
\par\end{centering}

\medskip{}

\begin{centering}
\includegraphics[bb=50bp 190bp 580bp 620bp,clip,width=10cm,height=6.7cm]{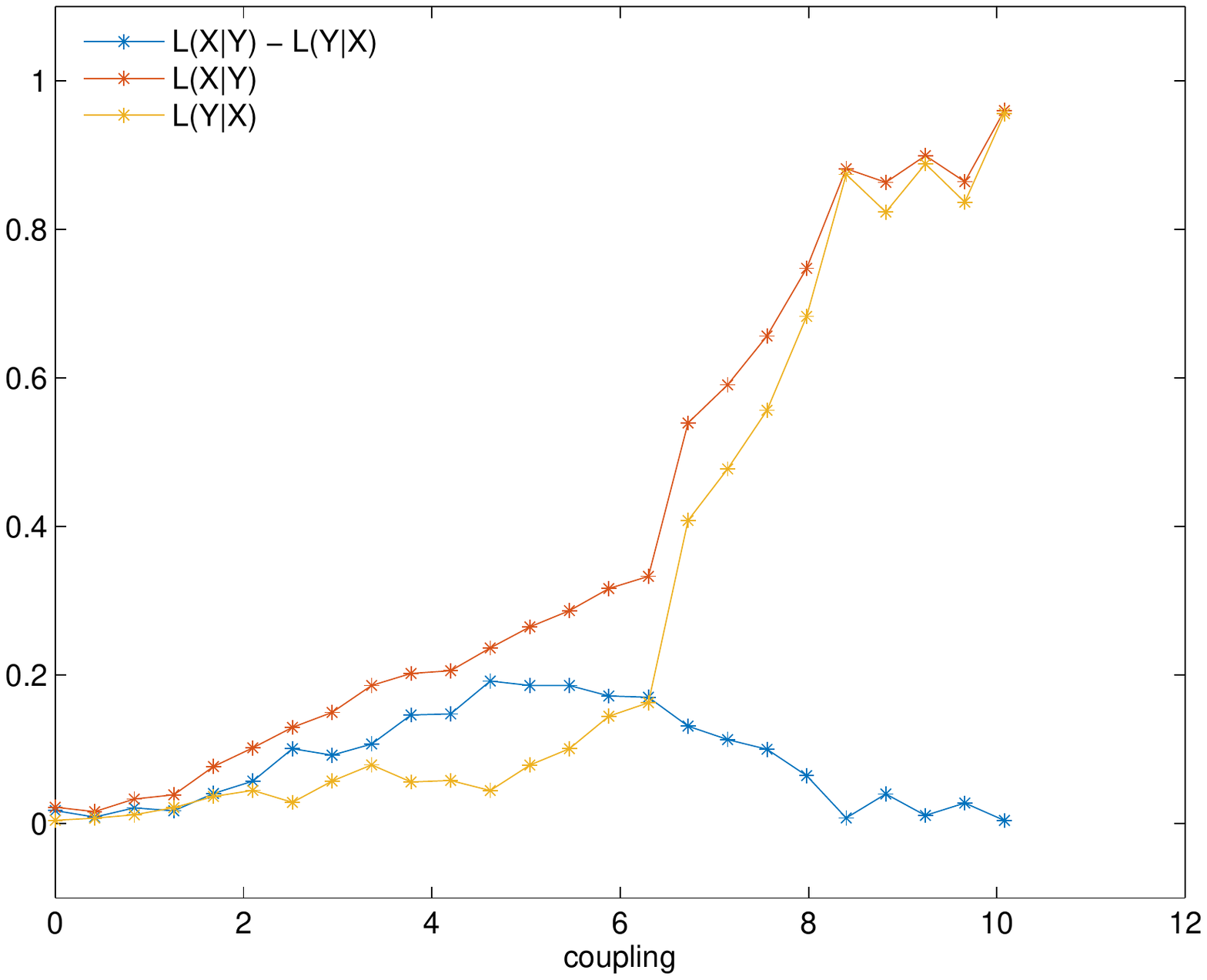}
\par\end{centering}

\medskip{}

\begin{centering}
\includegraphics[bb=50bp 190bp 580bp 620bp,clip,width=10cm,height=6.7cm]{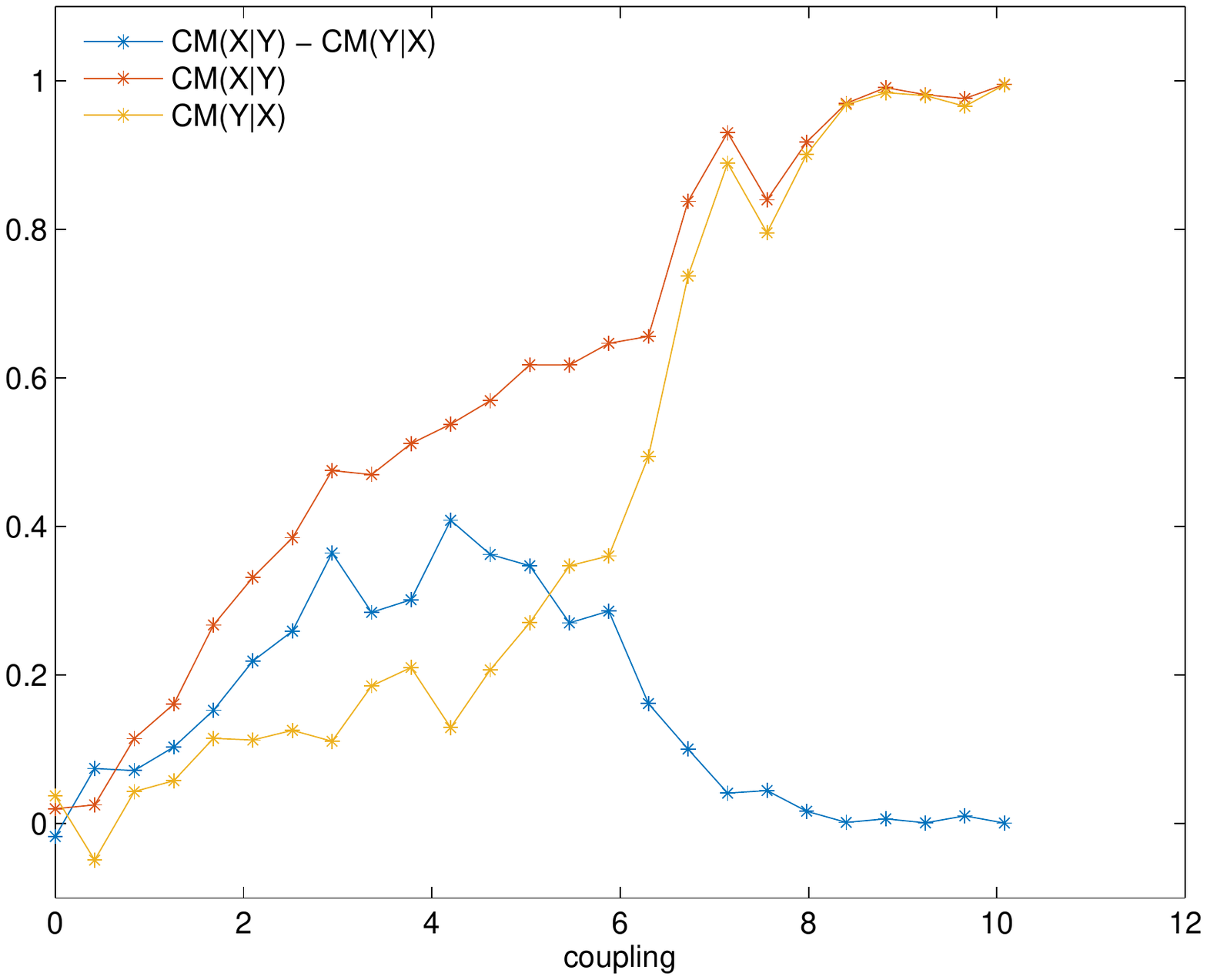}
\par\end{centering}

\protect\caption{Measures $M$, $L$, and $CM$ computed for two uni-directionally
coupled Lorenz systems (\ref{eq:lorlor}). The measures show that
$X$ drives $Y$ until the onset of synchronization between couplings
$8$ and $10$.}
\end{figure}

\newpage{}

\subsection{Lorenz $39$ $\rightarrow$ Lorenz $35$}

The last example is formed by another uni-directionally coupled nonidentical
Lorenz systems. Variables $x_{1}$, $x_{2}$, $x_{3}$ correspond
to the driver system and $y_{1}$, $y_{2}$, $y_{3}$ are the variables
of the response system: 
\begin{eqnarray}
\dot{x}_{1} & = & 10(-x_{1}+x_{2})\nonumber \\
\dot{x}_{2} & = & 39x_{1}-x_{2}-x_{1}x_{3}\nonumber \\
\dot{x}_{3} & = & x_{1}x_{2}-\frac{8}{3}x_{3}\label{eq:lorlor2}\\
\dot{y}_{1} & = & 10(-y_{1}+y_{2})+C(x_{1}-y_{1})\nonumber \\
\dot{y}_{2} & = & 35y_{1}-y_{2}-y_{1}y_{3}\nonumber \\
\dot{y}_{3} & = & y_{1}y_{2}-\frac{8}{3}y_{3}\nonumber 
\end{eqnarray}

The data were generated by Matlab solver of ordinary differential
equations ode45. The coupling strength $C$ was chosen from $0$ to
$14$ with the step $1.$ The starting point was $[0.3,0.3,0.3,0.3,0.3,0.3].$
First $2000$ data points was thrown away. The total number of obtained
data was $100\,000$ at an integration step size $0.01.$ Similar
system was used in \cite{Chich2009}.

\begin{figure}[H]
\begin{centering}
\includegraphics[width=16cm]{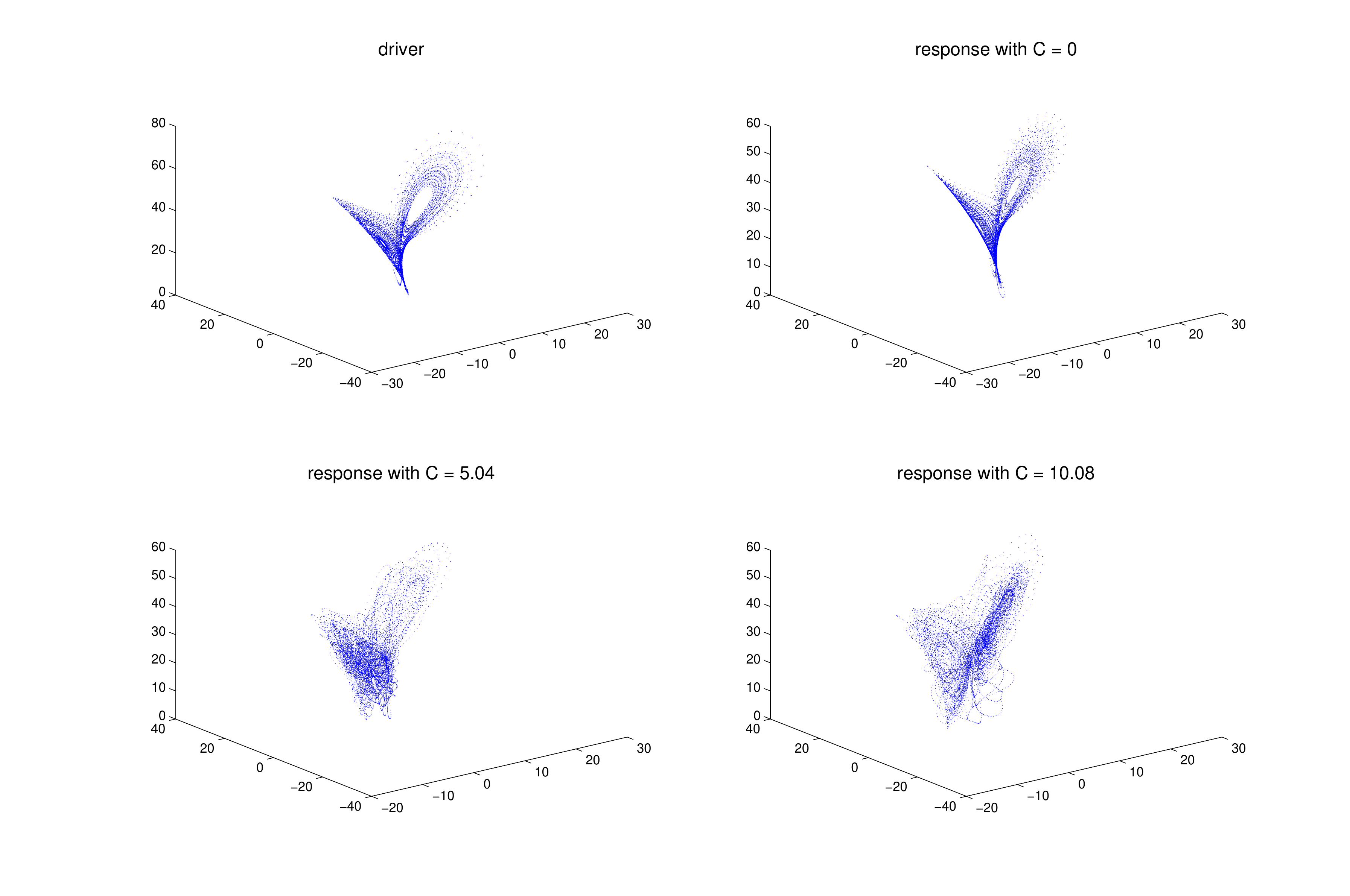}
\par\end{centering}

\protect\caption{Lorenz $39$ driving Lorenz $35$. Attractors of the driver and of
response system for various couplings.}

\end{figure}

The variables of the coupled systems may be arranged into the same
interaction graph as the previous example. The two connected Lorenz
systems represent distinct dynamical subsystems coupled through one-way
driving relationship between variables $x_{1}$ and $y_{1}$(see Figure
27). This causal link is what we tried to uncover.

Estimates of correlation dimension of the combined Lorenz-Lorenz system
(driver + response), computed for $100000$ data saturates to the
value which remains relatively unchanging for couplings somewhat higher
than $9$:

\begin{figure}[H]
\centering{}\includegraphics[width=10cm]{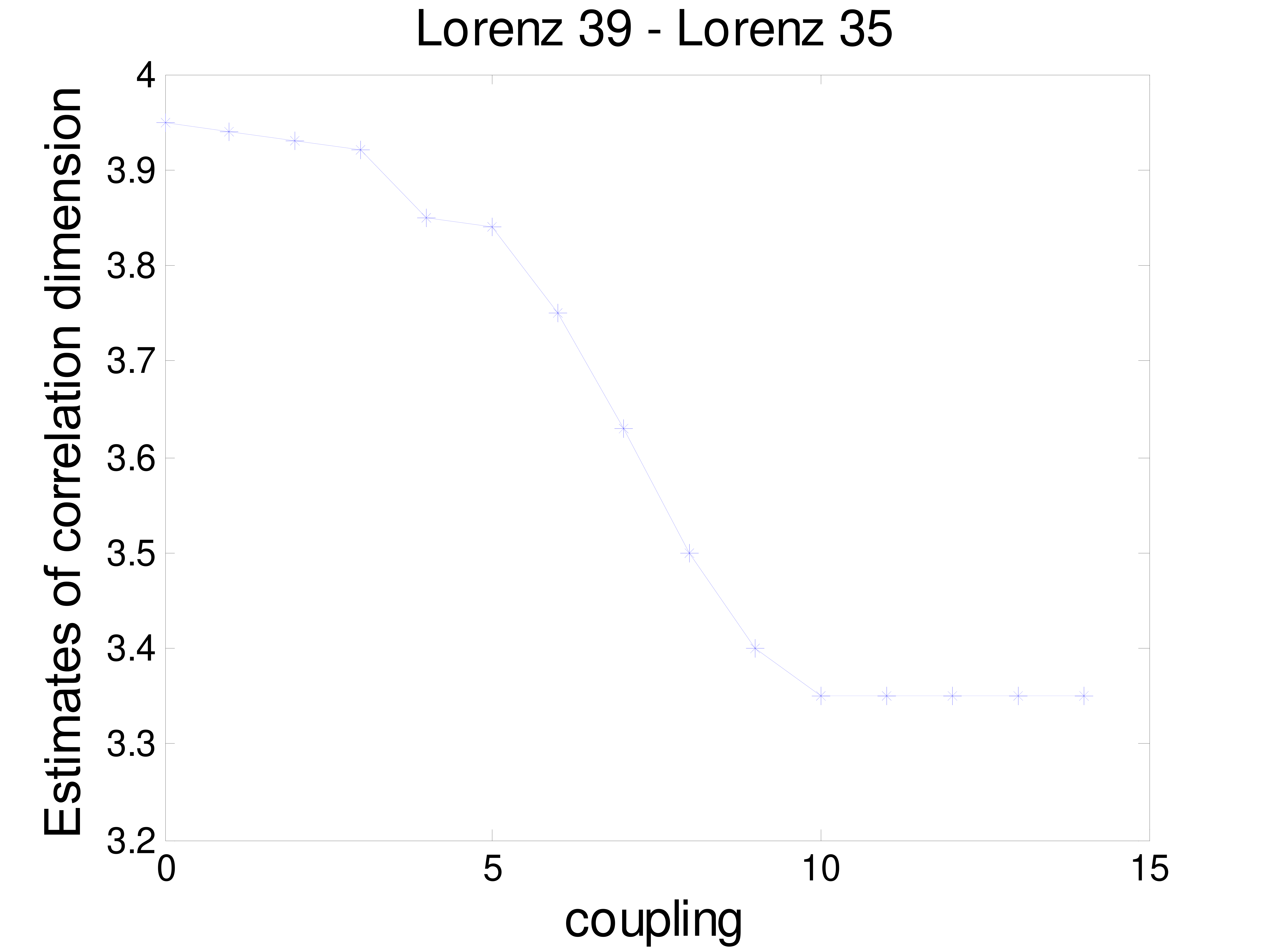}\protect\caption{Correlation dimension estimates for two Lorenz systems (\ref{eq:lorlor2})
connected with different coupling strengths.}
\label{D2lorlor2}
\end{figure}

\subsubsection*{Results of causality detection using reconstructed manifolds}

Suppose that we know $50000$ data-points of variable $x_{1}$ of
the driving Lorenz system and variable $y_{1}$ of the responsive
Lorenz system and we would like to know whether there is a causal
relationship between the two systems. 

In order to use the state-space based methods of search for causality
we made reconstructions of the state portraits with the same parameters
as in the previous case. This means time-delayed vectors of $x_{1}$
and $y_{1}$ with delay equal to $1$ and embedding dimension of $7$.
Methods for all three measures used $8$ nearest neighbors.

\begin{figure}[H]
\begin{centering}
\includegraphics[bb=50bp 190bp 580bp 620bp,clip,width=10cm,height=6.7cm]{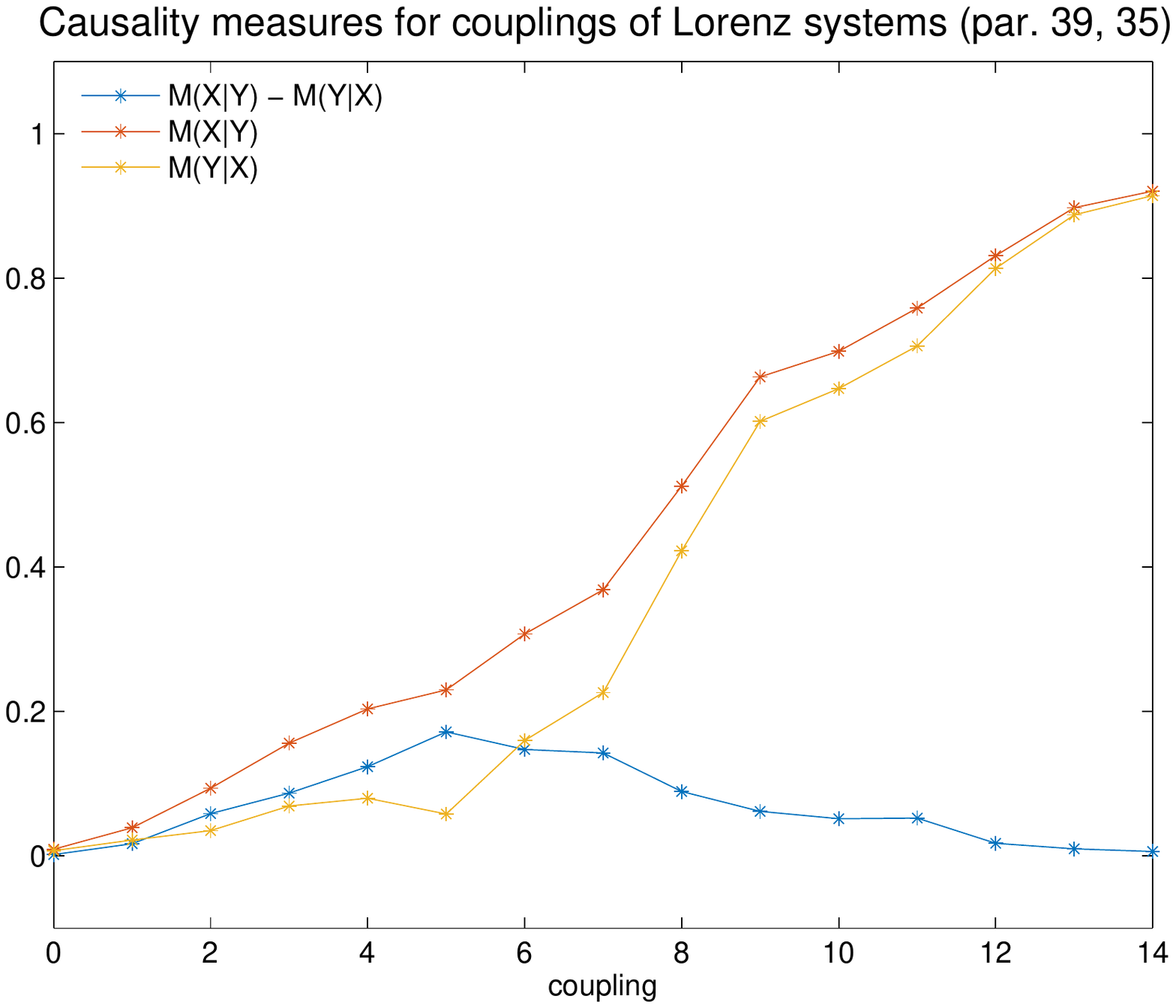}
\par\end{centering}

\medskip{}

\begin{centering}
\includegraphics[bb=50bp 190bp 580bp 620bp,clip,width=10cm,height=6.7cm]{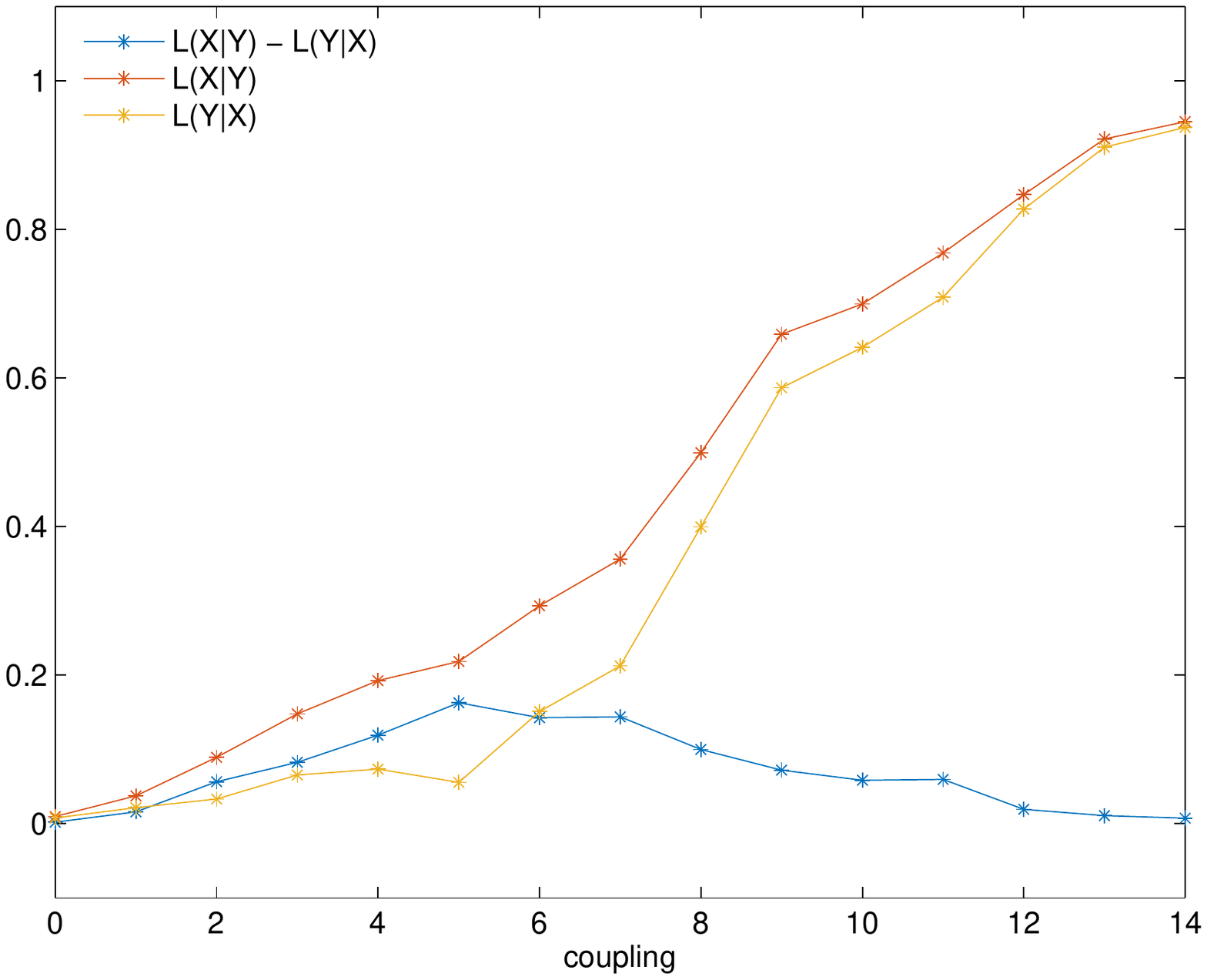}
\par\end{centering}

\medskip{}

\begin{centering}
\includegraphics[bb=50bp 190bp 580bp 620bp,clip,width=10cm,height=6.7cm]{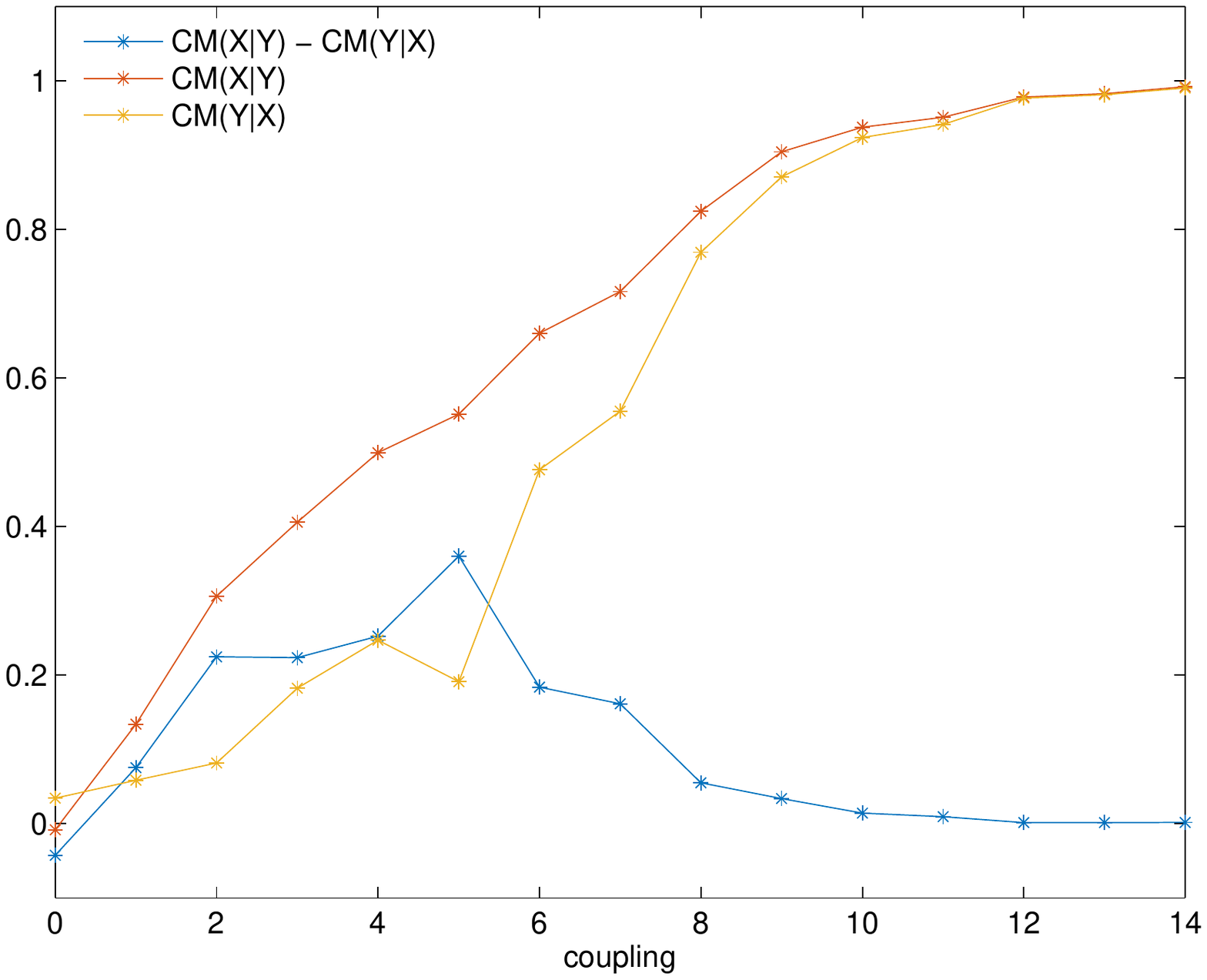}
\par\end{centering}

\protect\caption{Measures $M$, $L$, and $CM$ computed for two uni-directionally
coupled Lorenz systems. The measures show that $X$ drives $Y$ until
the onset of synchronization for couplings somewhat higher than $9$.}
\end{figure}

\newpage{}

\section{Conclusion}

In this study three methods to detect causality in reconstructed state
space were tested. The first method is based on measure named $M$
\cite{andrz2003}, the second one is based on measure $L$ \cite{Chich2009},
and the third one is a more recent variant of cross-mapping \cite{sugih2012}. 

The methods were compared on test examples of uni-directionally connected
chaotic systems of Hénon, Rössler and Lorenz type in relation to the
ability to detect unidirectional coupling and synchronization of interconnected
dynamical systems. 

Results show that each of the three examined methods managed to reveal
the presence and the direction of coupling and also to detect the
onset of full synchronization. Both the efficiency and the computational
complexity of the methods were comparable.

In case of real data, it may happen that the there is a correlation
between the systems and that is falsely declared as causality. Then
we can take any of the measures and look for performance improvement
with increasing number of used data. Lack of convergence means the
absence of causality. However, investigating the aspect of convergence
of the efficiency is left for further research.

\subsubsection*{Acknowledgment }

This work was supported by the Slovak Grant Agency for Science (Grant
no. 2/0043/13).

\section{Appendix}

\subsection{Computation of measures $M,$ $L$ and $CM$}

Matlab code prepared by Jozef Jakubík and retrieved in October 2015
from

{\footnotesize{}\href{https://www.mathworks.com/matlabcentral/fileexchange/52964-convergent-cross-mapping/content/SugiLM.m}{https://www.mathworks.com/matlabcentral/fileexchange/52964-convergent-cross-mapping/content/SugiLM.m}}{\footnotesize \par}

\medskip{}

\inputencoding{latin9}\begin{lstlisting}
function [ SugiCorr , SugiR , LM , SugiY , SugiX , origY , origX ] = SugiLM( X , Y , tau , E , LMN )

% Calculating Sugihara's CMM, L and M causality. 
%
% References:
% - for Sugihara's CCM method > Sugihara, George, et al., Detecting Causality in Complex Ecosystems, Science 26 October 2012, Vol. 338, no. 6106, pp. 496-500.   
% - for L method > Chicharro, Daniel, and Ralph G. Andrzejak, Reliable detection of directional couplings using rank statistics, Physical Review E 80.2 (2009): 026217.
% - for M method > Andrzejak, Ralph G., et al., Bivariate surrogate techniques: necessity, strengths, and caveats, Physical review E 68.6 (2003): 066202.	
% 
% Inputs:
% X,Y - time series with the same length
% tau - time step for the reconstruction
% E   - dimension of the reconstruction  
% LMN - number of neighborhoods for L and M methods 
%       the number of neighborhoods for Sugihara's CCM method is E+1  
%
% Outputs:
% SugiCorr - correlation between the CCM estimation of original data and original data
% SugiR    - sqrt((sum((originaldata-CCMestimaleddata).^2)/numel(originaldata)))/std(origaldata)
% LM - results for L and M methods
% SugiY, SugiX - the CCM estimate of original data 
% origY, origX - original data

switch nargin
    case 5
    case 4
        LMN = E+1;
    otherwise
        error('Bad input')
end

L=length(X);
T=1+(E-1)*tau;
Xm=zeros((L-T+1),E);
Ym=zeros((L-T+1),E);
SugiN=E+1;
N = L-T+1;

%% RECONTRUCTIONS OF ORIGINAL SYSTEMS

for t=1:(L-T+1)
    Xm(t,:)=X((T+t-1):-tau:(T+t-1-(E-1)*tau));
    Ym(t,:)=Y((T+t-1):-tau:(T+t-1-(E-1)*tau));
end
%%
LMj= zeros(2,2,N);

SugiX=zeros(N,1);
SugiY=zeros(N,1);

parfor j=1:N
%% neighborhood search 

[n1,d1]=knnsearch(Xm,Xm(j,:),'k',N);
[n2,d2]=knnsearch(Ym,Ym(j,:),'k',N);

%% LM

LMn1=n1(n1~=j);
LMn2=n2(n2~=j);
LMd1=d1(n1~=j);
LMd2=d2(n2~=j);
   
susXY=arrayfun(@(x) find(LMn1(:) == x,1,'first'), LMn2(1:LMN) );
susYX=arrayfun(@(x) find(LMn2(:) == x,1,'first'), LMn1(1:LMN) );
    
sum1=sum(LMd1(:))/(N-1);
sum2=sum(LMd2(:))/(N-1);
        
LMj(:,:,j) = [(N/2-sum(susXY)/LMN)/(N/2-(LMN+1)/2) , (sum1-sum(LMd1(susXY))/LMN)/(sum1-sum(LMd1(1:LMN))/LMN) ; (N/2-sum(susYX)/LMN)/(N/2-(LMN+1)/2) , (sum2-sum(LMd2(susYX))/LMN)/(sum2-sum(LMd2(1:LMN))/LMN)]; 
    
   % (GN-G(Y|X))/(GN-Gk) => L(Y|X)
   % (GN-G(X|Y))/(GN-Gk) => L(X|Y)
  
   %(RNY-Rcond(Y|X))/(RNY-RkY) => M(Y|X)
   %(RNX-Rcond(X|Y))/(RNX-RkX) => M(X|Y)
end

%% CMM

dat=floor((L-T+1)/2);

parfor ii=(dat+1):(L-T+1)
    [n1s,d1s]=knnsearch(Xm((ii-dat):(ii-1),:),Xm(ii,:),'k',SugiN);
    [n2s,d2s]=knnsearch(Ym((ii-dat):(ii-1),:),Ym(ii,:),'k',SugiN);
    
    u1s=exp(-d1s/d1s(1));
    w1s=u1s/sum(u1s);
    SugiY(ii)= w1s*Y(n1s+T-1+ii-(dat+1));
    
    u2s=exp(-d2s/d2s(1));
    w2s=u2s/sum(u2s);
    SugiX(ii)= w2s*X(n2s+T-1+ii-(dat+1));
end

origY=Y(T:end);
origY=origY((dat+1):(L-T+1));
SugiY=SugiY((dat+1):(L-T+1));
origX=X(T:end);
origX=origX((dat+1):(L-T+1));
SugiX=SugiX((dat+1):(L-T+1));

%%

SugiCorr1=corrcoef(origY,SugiY);
SugiCorr(2,1)=SugiCorr1(1,2);

SugiCorr2=corrcoef(origX,SugiX);
SugiCorr(1,1)=SugiCorr2(1,2);

SugiR(2,1)=sqrt((sum((origY-SugiY).^2)/numel(origY)))/std(origY);
SugiR(1,1)=sqrt((sum((origX-SugiX).^2)/numel(origX)))/std(origX);

LM = squeeze(mean(LMj,3));

end
\end{lstlisting}
\inputencoding{utf8}

\bibliographystyle{plain}
\bibliography{references}
 \bibliographystyle{plain}
\end{document}